# Physical modeling of shrink-swell cycles and cracking in a clayey vadose zone

V.Y. Chertkov*

Division of Environmental, Water, and Agricultural Engineering, Faculty of Civil and Environmental Engineering, Technion, Haifa 32000, Israel

**Abstract.** Physical understanding of the *crack* origin and quantitative physical prediction of the *crack* volume variation *far* from the clay soil surface are necessary to protect the underlying aquifers from pollutants. The *basis* of this work is an available physical model for predicting the shrinkage and swelling curves in the maximum water content range (the primary curves) and crack volume variation. The *objective* of the work is to generalize this model to the conditions of the deep layer of a clayey vadose zone with the overburden pressure, multiple shrinkage-swelling, and variation of water content in a small range. We aim to show that the scanning shrinkage and swelling curves, and steady shrink-swell cycles existing in such conditions, inevitably lead to the occurrence of cracks and a hysteretic crack volume. The *generalization* is based on the transition to the increasingly complex soil medium from the contributive clay, through the intra-aggregate matrix and aggregated soil with no cracking, to the soil with cracks. The *results* indicate the single-valued physical links between the scanning shrink-swell cycles and crack volume variation of the four soil media on the one hand, and primary shrinkage and swelling curves of the media on the other hand. The predicted cycles and crack volume hysteresis can be expressed through the *physical properties* and *conditions* of the soil at a given depth. The available observations of the cracks and crack volume variation in the clayey vadose zone give strong qualitative experimental *evidence* in favor of the feasibility of the model.

*Corresponding author. Tel.: 972-4829-2601.
*E-mail address:* agvictor@tx.technion.ac.il; vychert@ymail.com (V.Y. Chertkov).

**Contents**

## 1. Introduction

Shrinkage cracks close to the soil surface, their origin owing to the vertical water content gradient, and prediction of their volume, width, and spacing are well known [1-5]. Even though these cracks can extend to depths of several meters keeping a micro-width [4], their characteristics cannot explain the origin, volume, and volume variation of the cracks that are directly or indirectly observed at sufficiently large depths of a clayey vadose zone [6-10]. Contrary to the notion about the small permeability of clays, such cracks can essentially increase the hydraulic conductivity of the vadose zone and, correspondingly, transport different contaminant admixtures and solutes to the ground water [6]. Therefore, theoretical estimation of the crack volume, accompanying the natural multifold shrink-swell cycles far from the soil surface, as a function of soil structure and local hydrological conditions, is of theoretical as well as practical interest. In particular, the physical understanding of the origin of such cracks and quantitative physical prediction of the crack volume variation, are important to protect the underlying aquifers from different pollutants. The physical understanding and prediction should take into account the major *specific limitations* existing in a clay soil at large depths: (i) *overburden* action at a given depth; (ii) *limited* and frequently *small* range of water content variation (compared to the maximum range); and (iii) *multifold* drainage-wetting alternation (both transitive and steady).

The works that consider the explanation and prediction of shrink-swell cycling and accompanying cracking in a clayey vadose zone, based on the physical characteristics of soil texture and structure (with no fitting) and local hydrological characteristics, are absent. The contemporary approaches that combine the microphysics of water-vapor-solid interactions with different variants of transition to a clay soil continuum based on conservation laws and thermodynamics (see [11-14], among others), essentially consider a clay paste with no crack occurrence as well as no inter- and intra-aggregate structure leading to soil cracking at shrink-swell processes. The known approach to overburden effects in swelling soils [15] (for previous references see [15]) and subsequent discussion (see [16-19], among others) are based on the equilibrium thermodynamics relations and the use of the shrinkage curve slope as a function of water content and applied load, with the attraction of some form of empirical (fitted) shrinkage curve in the maximum possible water content range. The physical description (i.e., with no fitting) of overburden effects in connection with soil structure effects as well as cracking, swelling, and multifold drainage-wetting in an arbitrary range of water content variation, are beyond the scope of this approach. The same relates to a recent work [20] that modifies the approach [15] and its results using still another approximation of the fitted shrinkage curve [21]. Finally, it should be noted that available approaches to soil cracking, based on either a physical description (e.g., [22]) or description using a number of empirical (fitted) dependences (e.g., [23]), only relate to shrinkage in the maximum water content range close to the soil surface or in small samples, that is, with no effects of overburden, swelling, and multifold shrink-swell cycling in a small water content range.

A natural *reference background* for the development that accounts for the above limitations in a clay soil of large depths should include some description of the shrinkage-swelling curves and crack volume variation during a single drainage-wetting cycle in the maximum possible range of water content and without any soil loading. Such a description was recently suggested [24]. The *objective* of the present work is to extend the approach of the reference background [24] accounting for the three above natural factors (overburden, arbitrary water content range, and multifold transitive and steady drainage-wetting cycles) and keeping the *physical* character of the approach (i.e., without the use of fitting parameters). The methodology of the approach [24] is based on the recently suggested inter- and intra-aggregate structure of inorganic soils [25-28] (Fig.1) and successive consideration of the increasingly complex soil media: clay (paste), intra-aggregate soil matrix (including clay, silt, sand, and lacunar pores), aggregated soil without cracks (small samples), and aggregated soil with cracks (soil layer). The key point of the methodology is also a close connection between the swelling curves of these soil media as well as between the swelling and shrinkage curves for each of the media. Following the methodology [24], first, we consider the shrinkage-swelling of clay paste (contributing to the soil aggregates) in conditions of overburden pressure and multifold drainage-wetting with small variations of water content (Section 2). Then, we extend the results to be found for clay to the case of the intra-aggregate matrix (Section 3) and aggregated soil without and with cracks (Section 4). The theory content (Sections 2-4) is presented in the *concentrated* view in the beginning of the sections and their subsections as well as by their titles. Section 6 gives the theotetical results. In Sections 5 and 7 we analyze the limited available data on shrinkage under loading to check the major model aspects. The values that repeat in the text are summarized in the Notation.

**2. Shrinkage and swelling of a pure clay**

We successively consider the increasingly complex cases for a *clay* paste: a single shrink-swell cycle in the maximum range of water content and without loading (a brief review of the key points from [24] that are necessary in the following; Section 2.1.1), the scanning shrinkage and swelling curves in any possible range of water content without clay loading (Section 2.1.2); the multifold transitive and steady shrink-swell cycles in any possible range of water content and without clay loading (Section 2.1.3); and similar situations, but under loading (Sections 2.2.1-2.2.4).

*2.1. Clay shrinkage and swelling with no loading*

*2.1.1. Primary shrinkage and swelling curves*

We will refer to the shrinkage and swelling curves of a clay paste in the maximum possible range of water content as *primary* shrinkage and swelling curves (Fig.2, curves 1 and 2), by analogy with the soil-water retention curves at drying and wetting (e.g. [29]). The expressions of the primary shrinkage, $v(\zeta)$ [30,31] and swelling, $\hat{v}(\zeta)$ [24] curves of clay in *relative* coordinates are derived from considerations of the smallness of a number of values (Fig.2; see Notation) as

$$\zeta_n - \zeta_z \ll 1, \qquad v_n - v_z \ll 1, \qquad \zeta_h \cong 0.5 < 1, \qquad v_h - v_z \ll 1. \tag{1}$$

The primary shrinkage curve, $v(\zeta)$ [30] is presented as

$$v(\zeta) = \begin{cases} v_z, & 0 \leq \zeta \leq \zeta_z \\ v_z + \dfrac{(1-v_s)^2}{4(v_z - v_s)(1 - F_z)}(\zeta - \zeta_z)^2, & \zeta_z < \zeta \leq \zeta_n \\ v_s + (1 - v_s)\zeta, & \zeta_n < \zeta \leq \zeta_h. \end{cases} \tag{2}$$

where $F_z$ is the saturation degree $F(\zeta)$ as

$$F=(1-v_s)\zeta/(v-v_s) , \quad 0\leq\zeta\leq\zeta_h \quad (3)$$

at the shrinkage limit, $\zeta=\zeta_z$. $\zeta_z$ and the air-entry point, $\zeta_n$ (Fig.2) are

$$\zeta_z=(v_z-v_s)F_z/(1-v_s) \quad \text{and} \quad \zeta_n=(v_z-v_s)(2-F_z)/(1-v_s) . \quad (4)$$

$F_z$ is a function of $v_s$ and $v_z$ [31]. The $v_s$ and $v_z$ values totally determine $v(\zeta)$ (Eqs.(2)-(4)). The $v(\zeta)$ curve (Fig.2) consists of three parts (see Eq.(2)), two linear parts in the $0\leq\zeta\leq\zeta_n$ and $\zeta_n\leq\zeta\leq\zeta_h$ ranges and a curved part in the $\zeta_z\leq\zeta\leq\zeta_n$ range.

For the transition to primary clay swelling [24] and then to shrinkage and swelling in any water content range (Sections 2.1.2 and 2.1.3), it is also important to note the presentation of the clay matrix porosity, $P(v(\zeta))$ at primary shrinkage [30] as

$$P(v(\zeta))=1-v_s/v(\zeta) , \quad v_z\leq v\leq v_h , \quad 0\leq\zeta\leq\zeta_h \quad (5)$$

and the maximum and minimum *internal* sizes, $r_m(v(\zeta))$ and $r_o(v(\zeta))$, respectively, of clay matrix pores at primary shrinkage (excluding pore wall thickness) [30]

$$r_m(v(\zeta))=r_{mM} v(\zeta)^{1/3}(1-v_s/[Av(\zeta)]) , v_z\leq v\leq v_h , \quad 0\leq\zeta\leq\zeta_h \quad (6a)$$

$$r_o(v(\zeta))=r_{mM}v(\zeta)^{1/3}(\gamma v_s/A)(1-1/[\gamma v(\zeta)]), \quad (r_o\rightarrow 0 \text{ at } v_z\rightarrow 0.11) \quad v_z\leq v\leq v_h, \quad 0\leq\zeta\leq\zeta_h \quad (6b)$$

where $r_{mM}$ is the maximum external size of the clay pores (including pore wall thickness) at $\zeta=1$; $A\cong13.57$ and $\gamma\cong9$ are the characteristic constants of the clay matrix.

The primary swelling curve, $\hat{v}(\zeta)$ at slowest swelling [24] is

$$\hat{v}(\zeta)=v_h-\lambda(\zeta-\zeta_h)^2=v_h-[(v_h-v_z)/\zeta_h^2](\zeta-\zeta_h)^2 , \quad 0\leq\zeta\leq\zeta_h . \quad (7)$$

$\hat{v}(\zeta)$ is presented by one curved line (Fig.2). The maximum slope of $\hat{v}(\zeta)$ at $\zeta=0$ (Eq.(7); Fig.2), $2\lambda\zeta_h=2(v_h-v_z)/\zeta_h$ can be more and less than the maximum slope, $(1-v_s)$ of the primary shrinkage curve, $v(\zeta)$ (see Eq.(2)) in the basic shrinkage range, $\zeta_z\leq\zeta\leq\zeta_n$. The clay saturation degree, $\hat{F}(\zeta)$, porosity, $\hat{P}(\zeta)$, as well as the maximum and minimum internal pore sizes, $\hat{r}_m(\zeta)$ and $\hat{r}_o(\zeta)$ at primary swelling are obtained from Eqs.(3), (5), (6a) and (6b), respectively, by the replacement, $v(\zeta)\rightarrow\hat{v}(\zeta)$ [24]. That is, $\hat{F}(\zeta)=F(\hat{v}(\zeta))$, $\hat{P}(\zeta)=P(\hat{v}(\zeta))$, $\hat{r}_m(\zeta)=r_m(\hat{v}(\zeta))$, and $\hat{r}_o(\zeta)=r_o(\hat{v}(\zeta))$.

The coordinates $(\zeta,v)$ or $(\zeta,\hat{v})$ give the *customary* $(\overline{w},V)$ or $(\overline{w},\hat{V})$ (the specific volume $V$ or $\hat{V}$ vs. gravimetric water content, $\overline{w}$ of the clay) by [30,31,24]

$$V=v/(v_s\rho_s) , \quad \hat{V}=\hat{v}/(v_s\rho_s) , \quad \overline{w}=((1-v_s)/v_s)(\rho_w/\rho_s)\zeta . \quad (8)$$

Qualitatively, $V(\overline{w})$ and $\hat{V}(\overline{w})$ are similar to $v(\zeta)$ and $\hat{v}(\zeta)$, respectively, in Fig.2. For experimental validation of the results noted in this section see [30,31,25,24].

*2.1.2. Scanning swelling and shrinkage curves*

The swelling curve, $\hat{v}(\zeta,\zeta_o)$ (Fig.2, curve 3) that starts at a point $(\zeta_o,v(\zeta_o))$ of the primary shrinkage curve, $v(\zeta)$ (Fig.2, curve 1) will be referred to as the *scanning swelling* curve. The shrinkage curve, $\underline{v}(\zeta,\zeta_o)$ (Fig.3, curve 3) that starts at a point $(\zeta_o,$ $\hat{v}(\zeta_o)$) of the primary swelling curve, $\hat{v}(\zeta)$ (Fig.3, curve 2) will be referred to as the *scanning shrinkage* curve. The term "scanning" is introduced here by analogy to its use as applied to the soil-water retention curve at wetting and drainage (e.g. [29]). The possible initial $\zeta_o$ values are $0<\zeta_o<\zeta_h$ (Figs.2 and 3). One and only one pair of scanning swelling and shrinkage curves passes through any point of the $(\zeta,v)$ plane (here $v$ symbolizes the $v$ coordinate, but not a shrinkage curve) between the primary shrinkage and swelling curves (Figs.2 and 3). The simple analytical expressions for the scanning swell-shrink curves flow out of the *known* primary shrink-swell curves (Section 2.1.1) and the same goes for the smallness of a number of values (Eq.(1)).

The *scanning swelling curves*, $\hat{\underline{v}}(\zeta,\zeta_o)$ at $0<\zeta_o<\zeta_h$ (Fig.2, curve 3) can be divided into two subfamilies of *usual* scanning swelling curves, $\hat{\underline{v}}^u(\zeta,\zeta_o)$ at $0<\zeta_o<\zeta_n$ and *specific* ones, $\hat{\underline{v}}^s(\zeta,\zeta_o)$ at $\zeta_n<\zeta_o<\zeta_h$ (Fig.2). The $\hat{\underline{v}}^u(\zeta,\zeta_o)$ curves keep the general shape of the primary swelling curve, $\hat{v}(\zeta)$ (Fig.2; see below). The $\hat{\underline{v}}^s(\zeta,\zeta_o)$ curves coincide with the primary shrinkage curve, $v(\zeta)$ at $\zeta_n<\zeta_o<\zeta<\zeta_h$ (Fig.2) as

$$\hat{\underline{v}}^s(\zeta,\zeta_o)=v(\zeta)=v_h-(1-v_s)(\zeta_h-\zeta) , \qquad \zeta_n<\zeta_o<\zeta<\zeta_h . \qquad (9)$$

The $\hat{\underline{v}}^s(\zeta,\zeta_o)$ curves differ from the $\hat{\underline{v}}^u(\zeta,\zeta_o)$ curves (Fig.2) by their linear shape. The latter is connected with the similar linear shape of the $v(\zeta)$ curve in the basic shrinkage range, $\zeta_n<\zeta<\zeta_h$ (Fig.2). The physical reason for the linear $v(\zeta)$ curve at $\zeta_n<\zeta<\zeta_h$ is the lack of air bubbles in the water saturated clay matrix pores while they lose water and shrink (see, e.g. [31]). Correspondingly, the $\hat{\underline{v}}^s(\zeta,\zeta_o)$ curves at $\zeta_o>\zeta_n$ are not complicated by air entrapping when rewetting and retain the linear shape.

The *usual* scanning swelling curves, $\hat{\underline{v}}^u(\zeta,\zeta_o)$, similar to the primary swelling curve, $\hat{v}(\zeta)$ (Fig.2; Eq.(7)), can be approximated by the first terms of an expansion in powers of $(\zeta-\zeta_h)$ based on Eq.(1). However, in such a $\hat{\underline{v}}^u(\zeta,\zeta_o)$ presentation one should keep the linear term (unlike Eq.(7)) as

$$\hat{\underline{v}}^u(\zeta,\zeta_o)=v_h+\hat{a}(\zeta_o)(\zeta-\zeta_h)-\hat{b}(\zeta_o)(\zeta-\zeta_h)^2 , \qquad \zeta_o<\zeta<\zeta_h , \qquad 0<\zeta_o<\zeta_n . \qquad (10)$$

The $\hat{a}$ and $\hat{b}$ coefficients are linked by the condition that $\hat{\underline{v}}^u(\zeta,\zeta_o)$ and $v(\zeta)$ coincide at $\zeta=\zeta_o$ (Fig.2) (by the definition of the scanning swelling curves at $0<\zeta_o<\zeta_n$) as

$$\hat{\underline{v}}^u(\zeta_o,\zeta_o)=v(\zeta_o) , \qquad 0<\zeta_o<\zeta_n . \qquad (11)$$

Eqs.(10) and (11) give $\hat{b}(\zeta_o)$ as

$$\hat{b}(\zeta_o)=[v_h-v(\zeta_o)+\hat{a}(\zeta_o)(\zeta_o-\zeta_h)]/(\zeta_o-\zeta_h)^2 , \qquad 0<\zeta_o<\zeta_n . \qquad (12)$$

The $\hat{a}(\zeta_o)$ value in Eq.(10) determines the slope of the $\hat{\underline{v}}^u(\zeta,\zeta_o)$ curve at $\zeta=\zeta_h$ (see Fig.2) and a given $\zeta_o$. When $\zeta_o$ increases in the range, $0<\zeta_o<\zeta_n$ (Eq.(10)) the $\hat{a}(\zeta_o)$ slope grows in the range, $0<\hat{a}(\zeta_o)<(1-v_s)$ (between the slopes of the primary swelling and shrinkage curves at $\zeta=\zeta_h$; Fig.2), i.e., $\hat{a}(\zeta_n)=(1-v_s)$. Accounting for this condition and the smallness of $\zeta_n$ ($\zeta_n<\zeta_h<0.5$; Eq.(1)) we present $\hat{a}(\zeta_o)$ at $0<\zeta_o<\zeta_n$ as

$$\hat{a}(\zeta_o)=(1-v_s)\zeta_o/\zeta_n \ . \qquad\qquad 0<\zeta_o<\zeta_n \ . \qquad\qquad (13)$$

The replacement of $\hat{a}(\zeta_o)$ in Eq.(12) from Eq.(13) gives $\hat{b}$ as a function of $\zeta_o$. Note that at $\zeta_o\to\zeta_n$ $\hat{a}(\zeta_o)\to(1-v_s)$ (Eq.(13)), $\hat{b}(\zeta_o)\to 0$ (Eq.(12)), and $\hat{v}^u(\zeta,\zeta_o)\to\hat{v}^s(\zeta,\zeta_n)$ (Eqs.(9) and (10)). Hence, the transition between the *usual scanning swelling* curves, $\hat{v}^u(\zeta,\zeta_o)$ and *specific* ones, $\hat{v}^s(\zeta,\zeta_o)$ at $\zeta_o=\zeta_n$ (Fig.2) is continuous, and the $\hat{v}^s(\zeta,\zeta_n)$ curve simultaneously plays the part of the *boundary* curve between the two subfamilies. Thus, the expressions for the *scanning swelling* curves, $\hat{v}(\zeta,\zeta_o)$ (Eqs.(9), (10), (12), and (13)) in the *single-valued* manner, flow out of the above prerequisites: (i) the primary shrink-swell curves (Section 2.1.1), and (ii) conditions from Eq.(1).

Now we consider the *scanning shrinkage curves*, $\underline{v}(\zeta,\zeta_o)$ at $0<\zeta_o<\zeta_h$ (Fig.3, curve 3). First, we assume that they also can be divided into two subfamilies. In other words each curve, $\underline{v}(\zeta,\zeta_o)$ (Fig.3) can be related to the usual scanning shrinkage curves, $\underline{v}^u(\zeta,\zeta_o)$ at $\zeta_{oB}<\zeta_o<\zeta_h$ that are adjacent to the primary shrinkage curve, $v(\zeta)$ (Fig.3) and repeat its structure, i.e., consist of two linear parts and curved part between them [30] (Section 2.1.1). Alternatively, $\underline{v}(\zeta,\zeta_o)$ can be related to the specific scanning shrinkage curves, $\underline{v}^s(\zeta,\zeta_o)$ at $0<\zeta_o<\zeta_{oB}$ that have a simpler structure and are only presented by one curved line. However, the examination of this assumption (accounting for the primary shrinkage and swelling curves (Section 2.1.1) and Eq.(1)) showed that the boundary point of the continuous transition between the two subfamilies, $\zeta_o=\zeta_{oB}$ can only have the value, $\zeta_{oB}=0$. In other words, in all the range $0<\zeta_o<\zeta_h$ (Fig.3) the scanning shrinkage curves, $\underline{v}(\zeta,\zeta_o)$ only exist that keep the same general shape as the primary shrinkage curve, $v(\zeta)$ (Section 2.1.1) of the three parts (see above). These parts change with variation of $\zeta_o$. Let us consider the presentation of the scanning shrinkage curve of a clay, $\underline{v}(\zeta,\zeta_o)$ at a given $0<\zeta_o<\zeta_h$ within the limits of each of the three ranges (Fig.3): (i) $0<\zeta\leq\underline{\zeta}_z(\zeta_o)$; (ii) $\underline{\zeta}_n(\zeta_o)\leq\zeta\leq\zeta_o$; and (iii) $\underline{\zeta}_z(\zeta_o)<\zeta<\underline{\zeta}_n(\zeta_o)$. The $\underline{\zeta}_z(\zeta_o)$ and $\underline{\zeta}_n(\zeta_o)$ values of $\zeta$ (Fig.3) correspond to the points of the smooth transition between the linear and curved parts of the $\underline{v}(\zeta,\zeta_o)$ curve, and are similar to the $\zeta_z$ and $\zeta_n$ values relating to the primary shrinkage curve, $v(\zeta)$ [30]. $\underline{\zeta}_z(\zeta_o)$ and $\underline{\zeta}_n(\zeta_o)$ will be estimated below.

The linear presentation of $\underline{v}(\zeta,\zeta_o)$ in the range, $0<\zeta\leq\underline{\zeta}_z(\zeta_o)$ (Fig.3) is obvious as

$$\underline{v}(\zeta,\zeta_o)=v_z+a(\zeta_o)\zeta \ , \qquad 0<\zeta\leq\underline{\zeta}_z(\zeta_o) \ , \qquad 0<\zeta_o<\zeta_h \qquad (14)$$

and similar to the linear presentation of the primary shrinkage curve, $v(\zeta)$ (Fig.3) at $0<\zeta\leq\zeta_z$. However, in Eq.(14) unlike the $v(\zeta)$ case (Fig.3; Eq.(2)), the line slope $a(\zeta_o)>0$. If $\zeta_o\to 0$ the scanning shrinkage curve, $\underline{v}(\zeta,\zeta_o)$ shortens up to the $(0,v_z)$ point (Fig.3). With that the slope, $a(\zeta_o)$ strives to the slope of the primary swelling curve at $\zeta=0$ (see the line after Eq.(7)), i.e., $a(\zeta_o)\to 2\lambda\zeta_h$ (for $\lambda$ see Eq.(7)) and $\underline{\zeta}_z(\zeta_o)\to 0$. If $\zeta_o\to\zeta_h$ the scanning shrinkage curve, $\underline{v}(\zeta,\zeta_o)$ eventually coincides with the primary shrinkage curve (Fig.3). With that $a(\zeta_o)\to 0$ and $\underline{\zeta}_z(\zeta_o)\to\zeta_z$ (Fig.3). Thus, $a(\zeta_o)$ and $\underline{\zeta}_z(\zeta_o)$ are in the ranges, $2\lambda\zeta_h>a(\zeta_o)>0$ and $0<\underline{\zeta}_z(\zeta_o)<\zeta_z$ when $0<\zeta_o<\zeta_h$. Accounting for the smallness of the values in Eq.(1) and for the above ranges of the $a(\zeta_o)$ and $\underline{\zeta}_z(\zeta_o)$ values, one can, with sufficient accuracy, approximate $a(\zeta_o)$ and $\underline{\zeta}_z(\zeta_o)$ with simple linear functions of $\zeta_o$ at $0<\zeta_o<\zeta_h$ as

$$a(\zeta_o)=2\lambda\zeta_h(1-\zeta_o/\zeta_h)\,, \qquad \underline{\zeta}_z(\zeta_o)=\zeta_z\zeta_o/\zeta_h\,, \qquad 0<\zeta_o<\zeta_h\,. \qquad (15)$$

The general form of the linear presentation of the $\underline{v}(\zeta,\zeta_o)$ curve in the range, $\underline{\zeta}_n(\zeta_o)\leq\zeta\leq\zeta_o$ [Fig.3; close to the intersection between $\underline{v}(\zeta,\zeta_o)$ and the primary swelling curve, $\hat{v}(\zeta)$ in the point $(\zeta_o, \hat{v}(\zeta_o))$] is

$$\underline{v}(\zeta,\zeta_o)=d(\zeta_o)+e(\zeta_o)\zeta\,, \qquad \underline{\zeta}_n(\zeta_o)\leq\zeta\leq\zeta_o\,, \qquad 0<\zeta_o<\zeta_h\,. \qquad (16)$$

This presentation is similar to that of the primary shrinkage curve, $v(\zeta)$ in the range, $\zeta_n<\zeta<\zeta_h$ (Fig.3; Eq.(2)) by changing $v_s$ and $(1-v_s)$ with more general expressions, $d(\zeta_o)$ and $e(\zeta_o)$, respectively. We should find three functions, $d(\zeta_o)$, $e(\zeta_o)$, and $\underline{\zeta}_n(\zeta_o)$. $d(\zeta_o)$ and $e(\zeta_o)$ are linked by the intersection condition between $\underline{v}(\zeta,\zeta_o)$ and $\hat{v}(\zeta)$ (Fig.3) as

$$\underline{v}(\zeta,\zeta_o)|_{\zeta=\zeta_o}=\hat{v}(\zeta_o)\,, \qquad 0<\zeta_o<\zeta_h \qquad (17)$$

or (see Eqs.(7), (16), and (17))

$$d(\zeta_o)+e(\zeta_o)\zeta_o=v_h-\lambda(\zeta_o-\zeta_h)^2\,, \qquad (\lambda=(v_h-v_z)/\zeta_h^2) \qquad 0<\zeta_o<\zeta_h\,. \qquad (18)$$

It follows from Eq.(18) that $e(\zeta_o)$ should be a linear function of $\zeta_o$ and $d(\zeta_o)$ a square function of $\zeta_o$. For this reason, first, we find $e(\zeta_o)$ from its variation range at $0<\zeta_o<\zeta_h$. As noted above if $\zeta_o\to 0$ the scanning shrinkage curve, $\underline{v}(\zeta,\zeta_o)$ shortens up to the $(0,v_z)$ point (Fig.3). Then $e(\zeta_o)$ (Eq.(16)) and $a(\zeta_o)$ (Eq.(14)) coincide, i.e., at $\zeta_o\to 0$ $e(\zeta_o)\to a(\zeta_o)\to 2\lambda\zeta_h$ (see Eq.(15); for $\lambda$ see Eq.(7) or Eq.(18)). If $\zeta_o\to\zeta_h$ the scanning shrinkage curve, $\underline{v}(\zeta,\zeta_o)$ coincides with the primary shrinkage curve (Fig.3), i.e., $e(\zeta_o)\to(1-v_s)$ (cf. Eq.(2)). In general, it can be both $2\lambda\zeta_h>(1-v_s)$ and $2\lambda\zeta_h<(1-v_s)$ [24]. Thus, depending on that, either $2\lambda\zeta_h>e(\zeta_o)>(1-v_s)$ or $2\lambda\zeta_h<e(\zeta_o)<(1-v_s)$ when $0<\zeta_o<\zeta_h$. In both the cases the linear $e(\zeta_o)$ is

$$e(\zeta_o)=2\lambda\zeta_h-[2\lambda\zeta_h-(1-v_s)]\zeta_o/\zeta_h\,, \qquad 0<\zeta_o<\zeta_h\,. \qquad (19)$$

Replacing $e(\zeta_o)$ in Eq.(18) from Eq.(19) and rearranging we obtain the square $d(\zeta_o)$ as

$$d(\zeta_o)=v_z+[\lambda-(1-v_s)/\zeta_h]\zeta_o^2\,, \qquad 0<\zeta_o<\zeta_h\,. \qquad (20)$$

Note that at $\zeta_o\to 0$ $d(\zeta_o)\to v_z$ and at $\zeta_o\to\zeta_h$ $d(\zeta_o)\to v_s$ as it should be. The $\underline{\zeta}_n(\zeta_o)$ function will be found below. Finally, the general form of the intermediate curved line, $\underline{v}(\zeta,\zeta_o)$ in the range, $\underline{\zeta}_z(\zeta_o)<\zeta<\underline{\zeta}_n(\zeta_o)$ (Fig.3) is

$$\underline{v}(\zeta,\zeta_o)=\underline{v}_z(\zeta_o)+a(\zeta_o)(\zeta-\underline{\zeta}_z(\zeta_o))+b(\zeta_o)(\zeta-\underline{\zeta}_z(\zeta_o))^2\,, \quad \underline{\zeta}_z(\zeta_o)<\zeta<\underline{\zeta}_n(\zeta_o),\ 0<\zeta_o<\zeta_h\,. \quad (21)$$

This presentation is similar to that of the primary shrinkage curve, $v(\zeta)$ in the range, $\zeta_z<\zeta<\zeta_n$ (Fig.3; Eq.(2)) [30] with a number of differences: (1) $v_z\rightarrow \underline{v}_z(\zeta_o)$; (2) a linear term, $a(\zeta_o)(\zeta-\underline{\zeta}_z(\zeta_o))$ appears; (3) coefficients $a$ and $b$ become functions of $\zeta_o$; (4) $\zeta_z\rightarrow \underline{\zeta}_z(\zeta_o)$; and (5) $\zeta_n\rightarrow \underline{\zeta}_n(\zeta_o)$. The physical conditions that determine $\underline{v}_z(\zeta_o)$, $a(\zeta_o)$, $b(\zeta_o)$, and $\underline{\zeta}_n(\zeta_o)$ in Eq.(21) are also similar to those in the $v(\zeta)$ case [30], namely, *continuity* and *smoothness* of $\underline{v}(\zeta,\zeta_o)$ at $\zeta=\underline{\zeta}_z(\zeta_o)$ and $\zeta=\underline{\zeta}_n(\zeta_o)$ (Fig.3). The $\underline{v}_z(\zeta_o)$ function flows out of the *continuity* at $\zeta=\underline{\zeta}_z(\zeta_o)$ (Fig.3) $[\underline{v}(\zeta,\zeta_o)|_{\substack{\zeta=\underline{\zeta}_z-\varepsilon \\ \varepsilon\rightarrow 0}}=\underline{v}(\zeta,\zeta_o)|_{\substack{\zeta=\underline{\zeta}_z+\varepsilon \\ \varepsilon\rightarrow 0}}]$ accounting for Eqs.(14), (15) and (21) as

$$\underline{v}_z(\zeta_o)=v_z+a(\zeta_o)\underline{\zeta}_z(\zeta_o)=v_z+2\lambda(1-\zeta_o/\zeta_h)\zeta_z\zeta_o \ , \qquad 0<\zeta_o<\zeta_h \ . \qquad (22)$$

The $a(\zeta_o)$ function in Eq.(21) naturally coincides with $a(\zeta_o)$ from Eqs.(14) and (15) and flows out of the *smoothness* condition at $\zeta=\underline{\zeta}_z(\zeta_o)$ (Fig.3) $[\partial\underline{v}(\zeta,\zeta_o)/\partial\zeta|_{\substack{\zeta=\underline{\zeta}_z-\varepsilon \\ \varepsilon\rightarrow 0}}=\partial\underline{v}(\zeta,\zeta_o)/\partial\zeta|_{\substack{\zeta=\underline{\zeta}_z+\varepsilon \\ \varepsilon\rightarrow 0}}]$ with $\underline{v}(\zeta,\zeta_o)$ in the left part from Eqs.(14) and (15) and $\underline{v}(\zeta,\zeta_o)$ in the right part from Eq.(21). Finally, $b(\zeta_o)$, and $\underline{\zeta}_n(\zeta_o)$ in Eq.(21) are jointly found from the *continuity and smoothness* conditions at $\zeta=\underline{\zeta}_n(\zeta_o)$ that are obtained from the above continuity and smoothness conditions after replacement of $\zeta=\underline{\zeta}_z(\zeta_o)\pm\varepsilon$ with $\zeta=\underline{\zeta}_n(\zeta_o)\pm\varepsilon$. Substitution for $\underline{v}(\zeta,\zeta_o)$ in the left part of the conditions from Eq.(21) and in the right part of the conditions from Eq.(16) leads to the equation system as

$$\underline{v}_z(\zeta_o)+a(\zeta_o)[\underline{\zeta}_n(\zeta_o)-\underline{\zeta}_z(\zeta_o)]+b(\zeta_o)[\underline{\zeta}_n(\zeta_o)-\underline{\zeta}_z(\zeta_o)]^2=d(\zeta_o)+e(\zeta_o)\underline{\zeta}_n(\zeta_o) \ , \ 0<\zeta_o<\zeta_h \ , \ (23)$$

$$a(\zeta_o)+2b(\zeta_o)[\underline{\zeta}_n(\zeta_o)-\underline{\zeta}_z(\zeta_o)]=e(\zeta_o) \ , \qquad 0<\zeta_o<\zeta_h \ , \qquad (24)$$

relative to the unknown, $\underline{\zeta}_n(\zeta_o)$ and $b(\zeta_o)$ since $\underline{v}_z(\zeta_o)$ (Eq.(22)), $a(\zeta_o)$ and $\underline{\zeta}_z(\zeta_o)$ (Eq.(15)), $d(\zeta_o)$ (Eq.(20)), and $e(\zeta_o)$ (Eq.(19)) are already known. Solution of Eqs.(23) and (24) after some term rearrangement gives $\underline{\zeta}_n(\zeta_o)$ and $b(\zeta_o)$ as

$$\underline{\zeta}_n(\zeta_o)=\underline{\zeta}_z(\zeta_o)+2[\underline{v}_z(\zeta_o)-d(\zeta_o)-e(\zeta_o)\underline{\zeta}_z(\zeta_o)]/[e(\zeta_o)-a(\zeta_o)] \ , \qquad 0<\zeta_o<\zeta_h \ , \qquad (25)$$

$$b(\zeta_o)=[e(\zeta_o)-a(\zeta_o)]^2/\{4[\underline{v}_z(\zeta_o)-d(\zeta_o)-e(\zeta_o)\underline{\zeta}_z(\zeta_o)]\} \ , \qquad 0<\zeta_o<\zeta_h \ . \qquad (26)$$

According to their physical meaning $b(\zeta_o)>0$ (see Eq.(21)), $\underline{\zeta}_n(\zeta_o)>\underline{\zeta}_z(\zeta_o)$ (see Fig.3), and $e(\zeta_o)>a(\zeta_o)$ (see the slopes of $\underline{v}(\zeta,\zeta_o)$ in Fig.3 at $\zeta=0$ and $\zeta=\zeta_o$) if $\zeta_o>0$. For this reason $[\underline{v}_z(\zeta_o)-d(\zeta_o)-e(\zeta_o)\underline{\zeta}_z(\zeta_o)]$ in Eqs.(25) and (26) should be more than zero if $\zeta_o>0$. It is confirmed by substitution for $\underline{v}_z(\zeta_o)$, $d(\zeta_o)$, $e(\zeta_o)$, and $\underline{\zeta}_z(\zeta_o)$ the above found expressions. Thus, the expressions of the *scanning shrinkage* curves, $\underline{v}(\zeta,\zeta_o)$ also flow out of the above prerequisites in the *single-valued* manner (as $\hat{\underline{v}}(\zeta,\zeta_o)$).

The expressions for the maximum and minimum internal pore sizes, $r_m(\zeta)$, $r_o(\zeta)$, $\hat{r}_m(\zeta)$, and $\hat{r}_o(\zeta)$, saturation degree, $F(\zeta)$ and $\hat{F}(\zeta)$, and clay matrix porosity, $P(\zeta)$ and $\hat{P}(\zeta)$ along the *scanning* shrinkage ($\underline{v}(\zeta,\zeta_o)$) and swelling ($\underline{\hat{v}}(\zeta,\zeta_o)$) curves are obtained from the known expressions along the *primary* shrinkage ($v(\zeta)$) and swelling ($\hat{v}(\zeta)$) curves [30,24] (Section 2.1.1; Eqs.(3), (5), and (6) as they are and after replacement, $v(\zeta) \to \hat{v}(\zeta)$) after replacements, $v(\zeta) \to \underline{v}(\zeta,\zeta_o)$ and $\hat{v}(\zeta) \to \underline{\hat{v}}(\zeta,\zeta_o)$ as

$\underline{r}_m(\zeta,\zeta_o) = r_m(\underline{v}(\zeta,\zeta_o))$,   $\underline{r}_o(\zeta,\zeta_o) = r_o(\underline{v}(\zeta,\zeta_o))$,   $\underline{\hat{r}}_m(\zeta,\zeta_o) = r_m(\underline{\hat{v}}(\zeta,\zeta_o))$,

$\underline{\hat{r}}_o(\zeta,\zeta_o) = r_o(\underline{\hat{v}}(\zeta,\zeta_o))$,   $\underline{F}(\zeta,\zeta_o) = F(\underline{v}(\zeta,\zeta_o))$,   $\underline{\hat{F}}(\zeta,\zeta_o) = F(\underline{\hat{v}}(\zeta,\zeta_o))$,

$\underline{P}(\zeta,\zeta_o) = P(\underline{v}(\zeta,\zeta_o))$,   $\underline{\hat{P}}(\zeta,\zeta_o) = P(\underline{\hat{v}}(\zeta,\zeta_o))$ . (27)

Finally, the transition from the *relative* to *customary* coordinates as applied to *scanning* curves of the clay matrix is realized by the same relations (Section 2.1.1; Eq.(8)) after the obvious replacements, $v(\zeta) \to \underline{v}(\zeta,\zeta_o)$, $\hat{v}(\zeta) \to \underline{\hat{v}}(\zeta,\zeta_o)$, $V(\overline{w}) \to \underline{V}(\overline{w},\overline{w}_o)$, $\hat{V}(\overline{w}) \to \underline{\hat{V}}(\overline{w},\overline{w}_o)$, and addition, $\overline{w}_o = ((1-v_s)/v_s)(\rho_w/\rho_s)\zeta_o$. Qualitatively, $\underline{V}(\overline{w},\overline{w}_o)$ and $\underline{\hat{V}}(\overline{w},\overline{w}_o)$ are as $\underline{v}(\zeta,\zeta_o)$ and $\underline{\hat{v}}(\zeta,\zeta_o)$ in Figs.2 and 3.

*2.1.3. Transitive and steady scanning shrink-swell cycles*

Usually, at a given depth of soil profile, the water content periodically varies in a range, $W_1 < W < W_2$ where $W_1 > 0$ and $W_2 < W_h$. With that $W_1$ and $W_2$ can be approximately considered as constant because usually possible variations of $W_1$ and $W_2$, $\delta W_1 \ll W_2 - W_1$ and $\delta W_2 \ll W_2 - W_1$. Thus, we should estimate the shrink-swell cycles of a soil volume in an arbitrary range, $W_1 < W < W_2$ inside the maximum range, $0 < W < W_h$ of water content. First, we consider the shrink-swell cycles of a *clay* volume in a range, $\zeta_1 < \zeta < \zeta_2$. Since $\zeta_1 > 0$ and $\zeta_2 < \zeta_h$ (Fig.4) the *shrinkage* and *swelling* branches of a cycle should coincide with the *known scanning shrinkage* and *swelling* curves (Section 2.1.2) or their parts. For instance, let us consider two consecutive cycles starting from the point, $\zeta_o = \zeta_1 \equiv \zeta_{o1}$ on the primary shrinkage curve (Fig.4, curve $v(\zeta)$). The *swelling* branch of the *first* cycle coincides with the scanning swelling curve, $\underline{\hat{v}}_1(\zeta,\zeta_1)$ (Fig.4, curve 1) at $\zeta_1 \equiv \zeta_{o1} < \zeta < \zeta_2$. The *shrinkage* branch of the *first* cycle coincides with the scanning shrinkage curve, $\underline{v}_1(\zeta,\zeta_{o2})$ (Fig.4, curve 2) at $\zeta_1 < \zeta < \zeta_2$. The $\zeta_{o2}$ parameter that determines the position of this scanning shrinkage curve is found from the obvious condition (Fig.4) at $\zeta = \zeta_2$ to be

$\underline{v}_1(\zeta_2,\zeta_{o2}) = \underline{\hat{v}}_1(\zeta_2,\zeta_1)$ . (28)

The *swelling* branch of the *second* cycle coincides with the scanning swelling curve, $\underline{\hat{v}}_2(\zeta,\zeta_{o3})$ (Fig.4, curve 3) at $\zeta_1 < \zeta < \zeta_2$. The $\zeta_{o3}$ parameter that determines the position of this scanning swelling curve is found from the condition (Fig.4) at $\zeta = \zeta_1$ as

$\underline{\hat{v}}_2(\zeta_1,\zeta_{o3}) = \underline{v}_1(\zeta_1,\zeta_{o2})$ , (29)

($\zeta_{o2}$ is preliminarily found from Eq.(28)). Further, the *shrinkage* branch of the *second* cycle coincides with the scanning shrinkage curve, $\underline{v}_2(\zeta,\zeta_{o4})$ (Fig.4, curve 4) at $\zeta_1<\zeta<\zeta_2$. The $\zeta_{o4}$ parameter that determines the position of this scanning shrinkage curve is found from the condition (Fig.4) at $\zeta=\zeta_2$ to be

$$\underline{v}_2(\zeta_2,\zeta_{o4})=\hat{\underline{v}}_2(\zeta_2,\zeta_{o3}) \ . \tag{30}$$

This process can be continued. However, even from the above graphical construction (Fig.4) it is clear that, eventually (through a number of steps), we come to a *steady* multifold shrink-swell cycle (Fig.5) with repeating shrinkage and swelling branches in the $\zeta_1<\zeta<\zeta_2$ range. The above first, second, and a number of following cycles can be referred to as *transitive* ones. In the above consideration we started at the $\zeta=\zeta_1$ point of the *primary shrinkage* curve, $v(\zeta)$ (Fig.4). One can be convinced that starting at the $\zeta=\zeta_2$ point of the *primary swelling* curve, $\hat{v}(\zeta)$ and conducting the similar consideration we come to the *same* steady multifold shrink-swell cycle (Fig.5). Thus, the $\zeta_1<\zeta<\zeta_2$ range (within the limits of a *given* primary ($v(\zeta),\hat{v}(\zeta)$) cycle) in the *single-valued* manner determines the corresponding *steady* shrink-swell cycle of a clay (Fig.5). Its two branches, $\underline{v}(\zeta,\zeta_{o1})$ and $\hat{\underline{v}}(\zeta,\zeta_{o2})$, i.e., the corresponding $\zeta_{o1}$ and $\zeta_{o2}$ values that determine the branches (Section 2.1.2) are immediately found from two of the evident conditions at $\zeta=\zeta_1$ and $\zeta=\zeta_2$ (Fig.5) as

$$\underline{v}(\zeta_1,\zeta_{o1})=\hat{\underline{v}}(\zeta_1,\zeta_{o2}) \ , \qquad \text{and} \tag{31}$$

$$\underline{v}(\zeta_2,\zeta_{o1})=\hat{\underline{v}}(\zeta_2,\zeta_{o2}) \ . \tag{32}$$

Note that at possible erratic movements of the boundary $\zeta_1$ and $\zeta_2$ points, the steady cycle (Fig.5) can be subject to corresponding migration.

### 2.2. Clay shrinkage and swelling under loading
#### 2.2.1. Remarks of the loading effect on the primary shrinkage and swelling curves

Hereafter the loading, $L$ is considered to be approximately *constant* at a given depth during the shrinkage and swelling processes. The primary shrinkage and swelling curves of clay at $L=0$ are only expressed through two clay characteristics, the relative volume of clay solids, $v_s$ and relative minimum clay volume, $v_z$ [30,31,24] (see Section 2.1.1). In particular, characteristic points of the primary shrinkage and swelling curves without loading (Fig.2), ($\zeta_h$, $v_h$), ($\zeta_n$, $v_n$), and ($\zeta_z$, $v_z$) are determined by $v_s$ and $v_z$. By its physical meaning [30,31] $v_s$ cannot change with $L$ increase and retains its value. Unlike that, $v_z$ by its physical meaning [30,31], in general, should depend on $L$. In addition, at $L>0$ $\zeta_h$ and $v_h$ become functions of $L$ independent of $v_z(L)$. It follows that the above characteristic points and primary clay shrinkage and swelling curves depend on $L$ through $v_z(L)$ and $v_h(L)$ (or $\zeta_h(L)$). $v_z(L)$ is considered in Section 2.2.2. $\zeta_h(L)$ and $v_h(L)$ are considered in Section 2.2.3. The resulting transformation of different shrinkage and swelling curves under loading is regarded in Section 2.2.4. In the case of the primary shrinkage curve (Fig.6) the point of maximum swelling, ($\zeta_h$, $v_h$) is displaced under loading to the point, ($\zeta_h(L),v_h(L)$) along the saturation line since loading keeps the saturation state of clay matrix at a given water content, $\zeta>\zeta_n(L)$. The $\zeta_n$ and $\zeta_z$ points are, in general, also transformed to $\zeta_n(L)$ and $\zeta_z(L)$, and $v_n$ and $v_z$ decrease to $v_n(L)<v_n$ and $v_z(L)<v_z$ (Fig.6). The primary swelling curve is also subject to the corresponding change, $\hat{v}(\zeta)\to\hat{v}(\zeta,L)$ where $\hat{v}(\zeta)\equiv\hat{v}(\zeta,L=0)$.

*2.2.2. The minimum clay volume ($v_z$) as a function of loading ($L$)*

The minimum clay volume, $v_z$ corresponds to a *rigid* clay particle network between the oven-dried state ($\zeta=0$) and shrinkage limit ($\zeta=\zeta_z$) (Fig.2). For this reason, to describe the $v_z$ variation under loading, we can take advantage of the *elastic* model and characterize a dry clay matrix by Young's modulus, $E$ and Poisson's ratio, $\sigma$. We consider the state of the dry clay under loading, $L$ as the result of homogeneous compression of a thick isotropic elastic horizontal layer by pressure $L$ that is applied to its surfaces. The deformation of the layer is written as $u_{zz}=\Delta v_z/v_z$ (z is a vertical axis). $v_z$ is the initial value (Fig.6). By definition and from the elasticity theory [33]:

$$\Delta v_z = v_z - v_z(L) , \tag{33}$$

$$\Delta v_z = [L/(\alpha E)]v_z , \tag{34}$$

$$\alpha = (1-\sigma)/[(1+\sigma)(1-2\sigma)] . \tag{35}$$

Eqs.(33) and (34) give $v_z(L)$ as

$$v_z(L) = v_z[1-L/(\alpha E)] . \tag{36}$$

For different clays the $L/(\alpha E)$ factor in Eq.(36) is always small. Indeed, available data on Young's modulus, $E$ and Poisson's ratio, $\sigma$ of dry clays that were obtained by different methods (see [34-37], among others) are characterized by the spread, approximately, in the ranges, $E \sim 10^2\text{-}10^3$MPa and $\sigma \sim 0.1\text{-}0.3$ (i.e., $\alpha \sim 1\text{-}1.4$; see Eq.(35)). However, in any case at $L<L_{max} \sim 1\text{-}1.5$MPa (for the maximum soil depth $\sim 50$m) one has the estimate, $L/(\alpha E) \leq 10^{-3}\text{-}10^{-2}$. The smallness of $L/(\alpha E)$ justifies the use of the elastic model in estimating the $v_z(L)$ dependence of clay. At sufficiently small loading, $L </\cong 100$kPa Eq.(36) gives $v_z(L) \cong v_z$ since $L/(\alpha E) \leq 10^{-4}\text{-}10^{-3}$. The corresponding available data are analyzed in Sections 5 and 7.

*2.2.3. The maximum swelling point ($\zeta_h, v_h$) variation as a function of loading ($L$)*

For all clays at $L=0$ $\zeta_h \cong 0.5$ and $v_h \cong 0.5(1+v_s) \cong 0.57\text{-}0.6$ [25,32]. At $L>0$ we need explicit dependences, $\zeta_h(L)$ and $v_h(L)$. They are linked by Eq.(2) at $\zeta=\zeta_h(L)$ as

$$v_h(L) = v_s + (1-v_s)\zeta_h(L) . \tag{37}$$

With the $L$ growth $\zeta_h(L)$ varies between the maximum, $\zeta_h=0.5$ at $L=0$ (Eq.(1)) and some minimum. The latter follows from Eq.(37) since $v_h(L)$ also decreases with the $L$ increase starting from $v_h=0.5(1+v_s)$, but cannot be less than $v_z$ at $L\rightarrow\infty$ (Fig.6). Replacing $v_h(L)$ in Eq.(37) with $v_z$ we come to the minimum of $\zeta_h(L\rightarrow\infty)=(v_z-v_s)/(1-v_s)$. By its physical meaning, $(v_z-v_s)/(1-v_s)$ is the ratio of the minimum to the potential maximum (at the liquid limit) pore volumes of the clay (i.e., its major swelling characteristic). This physical meaning is in the agreement with the physical meaning of $\zeta_h(L\rightarrow\infty)$. Fig.7 shows the qualitative view of the $\zeta_h(L/L^*)$ dependence where $L^*$ is some characteristic loading of the clay that determines the velocity of $\zeta_h$ decrease when $L$ increases. We assume that the negative increment, $d\zeta_h$ per unit increment of loading, $L$, i.e., $d\zeta_h/dL$, is proportional to the part of the relative maximum water content, $\zeta_h(L/L^*)$ that can change at a given $L$, i.e., to the difference $[\zeta_h(L/L^*)-(v_z-v_s)/(1-v_s)]$ (see Fig.7). Then ($1/L^*$ is a proportionality coefficient)

$$d\zeta_h = -[\zeta_h(L/L^*) - (v_z - v_s)/(1-v_s)]dL/L^* . \tag{38}$$

Integration of Eq.(38) with condition, $\zeta_h(L=0)=0.5$ (Fig.7) gives $\zeta_h(L/L^*)$ as

$$\zeta_h(L/L^*) = (v_z - v_s)/(1-v_s) + [0.5 - (v_z - v_s)/(1-v_s)]\exp(-L/L^*) . \tag{39}$$

Substitution for $\zeta_h(L/L^*)$ in Eq.(37) from Eq.(39) gives $v_h(L/L^*)$ as (at $v_h=0.5(1+v_s)$)

$$v_h(L/L^*) = v_z + (v_h - v_z)\exp(-L/L^*) . \tag{40}$$

The characteristic loading of a clay, $L^*$ can only depend on the ratio, $(v_z-v_s)/(1-v_s)$ (see Fig.7). According to Eq.(39), with the increase of $(v_z-v_s)/(1-v_s)$ to 0.5 $L^*$ should decrease to 0 because at any $L>0$ $\zeta_h(L/L^*)\to 0$. In addition $(v_z-v_s)/(1-v_s)<<1$, but $L^*$, in general, is not small. Thus, $L^*$ is the quickly decreasing and non-small function of the small ratio, $(v_z-v_s)/(1-v_s)$ when it increases. The simplest variants of such function are

$$L^* = L_u/[(v_z-v_s)/(1-v_s)] \quad \text{or} \quad L^* = L_u/[(v_z-v_s)/(1-v_s)]^2 \tag{41}$$

where $L_u$ is a constant (with dimension of loading) that does not depend on the properties $v_s$ and $v_z$ of a particular clay and in this meaning is *universal*. To choose a variant and estimate $L_u$ one should turn to available data (see Sections 5.2.3 and 7.1).

*2.2.4. Different clay shrinkage and swelling curves under loading*

The known dependences, $v_z(L)$ (Eq.(36)), $v_h(L)$ and $\zeta_h(L)$ (Eqs.(40) and (39)), enable the physical *quantitative* prediction of different shrinkage and swelling curves *under loading* based on their expressions for the case with *no loading* in Section 2.1. With this end in all equations of Section 2.1 one should just replace $v_z$, $v_h$, and $\zeta_h=0.5$ with $v_z(L)$, $v_h(L)$, and $\zeta_h(L)$. As applied to the *primary* shrinkage and swelling curves, it immediately follows from what is stated in Section 2.1.1. For calculations of $v(\zeta,L)$ and $\hat{v}(\zeta,L)$ (Fig.6) the usual Eqs.(2) and (7) can be utilized with the above expressions for $F_z$, $\zeta_z$, $\zeta_n$ through $v_z(L)$, $v_h(L)$, and $\zeta_h(L)$. Relations between the *primary* curves, $v(\zeta,L)$ and $\hat{v}(\zeta,L)$ (Fig.6) on the one side and *scanning* shrinkage ($\underline{v}(\zeta,\zeta_o,L)$) and swelling ($\hat{\underline{v}}(\zeta,\zeta_o,L)$) curves under loading, $L$ on the other side, are totally similar to such relations in a no loading case (Figs.2 and 3) since the *scanning* shrinkage and swelling curves are in the single-valued manner connected with *primary* ones (Section 2.1.2). Therefore, the *scanning* shrinkage and swelling curves under loading, $\underline{v}(\zeta,\zeta_o,L)$ and $\hat{\underline{v}}(\zeta,\zeta_o,L)$ are *quantitatively* predicted using the same Eqs.(9), (10), (12)-(16), (19)-(22), (25) and (26) (Section 2.1.2) with the above expressions, $F_z(v_s,v_z(L))$, $\zeta_z(v_s,v_z(L))$, $\zeta_n(v_s,v_z(L))$, $v_h(L)$, and $\zeta_h(L)$. At the known *scanning* curves under loading, $\underline{v}(\zeta,\zeta_o,L)$ and $\hat{\underline{v}}(\zeta,\zeta_o,L)$ the issue of finding the *transitive* and *steady* shrink-swell cycles under loading in the range, $0<\zeta_1<\zeta<\zeta_2<\zeta_h(L)$ (Fig.6; $\zeta_1$ and $\zeta_2$ are not shown), is similar to the analogous issue in the case without loading (Section 2.1.3; Figs.4 and 5). Now Eqs.(28)-(32) are utilized to the curves, $\underline{v}(\zeta,\zeta_o,L)$ and $\hat{\underline{v}}(\zeta,\zeta_o,L)$.

In spite of the *always* small change of the *clay* volume, $v_z(L)$ with the $L$ growth (Eq.(36)), the volume variation of a *soil*, that includes this clay, can be quite appreciable at small water contents even at $L<10$kPa when $v_z(L)\cong v_z=$const. The reason for this is the non-clay porosity variation with $L$ (Sections 5 and 7). The two last paragraphs of Section 2.1.2 (about the maximum and minimum internal pore sizes, saturation degree, and clay porosity as well as transition to the customary coordinates) are extended to the loading case by the replacement of any relative or specific clay volume with relevant volume under loading (e.g., $\underline{v}(\zeta,\zeta_o) \to \underline{v}(\zeta,\zeta_o,L)$ and $\underline{\hat{v}}(\zeta,\zeta_o) \to \underline{\hat{v}}(\zeta,\zeta_o,L)$, $\underline{V}(\overline{w},\overline{w}_o) \to \underline{V}(\overline{w},\overline{w}_o,L)$ and $\underline{\hat{V}}(\overline{w},\overline{w}_o) \to \underline{\hat{V}}(\overline{w},\overline{w}_o,L)$) and by accounting for $\zeta_h \to \zeta_h(L)$ while calculating $\overline{w}_h(L)$ using Eq.(8).

## 3. Shrinkage and swelling of the intra-aggregate matrix of a soil

In this section we consider the transformation of the results from Section 2 for a clay paste to the case of a soil intra-aggregate matrix that consists of the clay (as a shrink-swell agent), silt and sand grains, and lacunar pores (Fig.1). In other words, we consider the expressions for the *primary* shrink-swell curves (Section 3.1.1; a part of the section is a brief review of necessary points from [24]) as well as for the *scanning* shrink-swell curves and multifold shrink-swell *cycles* (Section 3.1.2) of an *intra-aggregate* matrix with no loading through the similar curves and cycles of the contributive *clay*, and then the similar relations for the loading case (Section 3.2).

### *3.1. Shrinkage and swelling of intra-aggregate matrix with no loading*
### *3.1.1. Primary shrinkage and swelling curves*

The relative volumes of the intra-aggregate matrix at shrinkage, $u(\zeta)$ and swelling, $\hat{u}(\zeta)$ without loading in the maximum range of water content, $0<\zeta\leq\zeta_h$, i.e., the *primary* shrink-swell curves, $u(\zeta)$ and $\hat{u}(\zeta)$ are linked with the *primary* shrink-swell curves of the contributive clay, $v(\zeta)$ and $\hat{v}(\zeta)$ (Section 2.1.1; Fig.2) as [25,26,24]

$$v(\zeta)=(u(\zeta)-u_{lp}(\zeta)-u_S)/(1-u_S), \qquad 0<\zeta\leq\zeta_h \tag{42a}$$

$$\hat{v}(\zeta)=(\hat{u}(\zeta)-\hat{u}_{lp}(\zeta)-u_S)/(1-u_S), \qquad 0<\zeta\leq\zeta_h \tag{42b}$$

where $u_S$ is the relative volume of the non-clay solids in the intra-aggregate matrix (Fig.1; silt and sand grains); $u_{lp}(\zeta)$ and $\hat{u}_{lp}(\zeta)$ are the relative volumes of the lacunar pores (Fig.1) in the intra-aggregate matrix at shrinkage [25-27] and swelling [24], respectively. Relations given by Eqs.(42a) and (42b) flow out of definitions of the values that enter them. To estimate $u(\zeta)$ and $\hat{u}(\zeta)$ (at the known $v(\zeta)$ and $\hat{v}(\zeta)$) one should first find $u_{lp}(\zeta)$ and $\hat{u}_{lp}(\zeta)$. To this end the lacunar factor, $k$ was introduced [26,27]; it only depends on the internal characteristics of the intra-aggregate matrix [38] (Fig.1), does not depend on water content, and coincides at shrinkage and swelling [24]. By definition of $k$, $du_{lp}(\zeta)$ and $d\hat{u}_{lp}(\zeta)$ are presented as [26,27,24]

$$du_{lp}(\zeta)=-k du_{cp}(\zeta), \qquad 0<\zeta\leq\zeta_h \tag{43a}$$

$$d\hat{u}_{lp}(\zeta)=-k d\hat{u}_{cp}(\zeta), \qquad 0<\zeta\leq\zeta_h \tag{43b}$$

where $u_{cp}(\zeta)$ and $\hat{u}_{cp}(\zeta)$ are the relative volumes of clay pores in the intra-aggregate matrix (along the primary shrinkage and swelling curves, respectively). $u_{cp}(\zeta)$ and $\hat{u}_{cp}(\zeta)$ can be written as [26,27,24]

$$u_{cp}(\zeta)=(v(\zeta)-v_s)(1-u_S), \qquad 0<\zeta\leq\zeta_h \tag{44a}$$

$$\hat{u}_{cp}(\zeta)=(\hat{v}(\zeta)-v_s)(1-u_S), \qquad 0<\zeta\leq\zeta_h . \tag{44b}$$

Eqs.(44a) and (44b) flow out of definitions of the values that enter them. In addition, we can use the conditions at the *shrinkage end* [26,27] as

$$u_{lp}(0)=u_{lpz} \tag{45a}$$

($u_{lpz}$ is the $u_{lp}(\zeta)$ maximum) and at the *swelling start* [24] as

$$\hat{u}_{lp}(0)=u_{lpz} \tag{45b}$$

(at the evident condition, $\hat{u}_{lpz}=u_{lpz}$ [24]). Replacing $u_{cp}(\zeta)$ in Eq.(43a) from Eq.(44a) and $\hat{u}_{cp}(\zeta)$ in Eq.(43b) from Eq.(44b) and integrating the relations to be obtained with the use of Eqs.(45a) and (45b), we find the relations being sought to be

$$u_{lp}(\zeta)=u_{lpz}-k(1-u_S)(v(\zeta)-v_z), \qquad 0<\zeta\leq\zeta_h \tag{46a}$$

$$\hat{u}_{lp}(\zeta)=u_{lpz}-k(1-u_S)(\hat{v}(\zeta)-v_z), \qquad 0<\zeta\leq\zeta_h. \tag{46b}$$

According to [26,27]: (i) if lacunar pores already exist at maximum swelling, $\zeta=\zeta_h$ then $k>0$ and $u_{lpz}>0$; (ii) if lacunar pores only appear at $\zeta=\zeta_l<\zeta_h$, then $k=0$ at $\zeta_l<\zeta<\zeta_h$ and $k>0$ at $0<\zeta\leq\zeta_l$ and $u_{lpz}>0$; and (iii) if lacunar pores do not appear at shrinkage, then $k=0$ and $u_{lpz}=0$. Substituting for $u_{lp}(\zeta)$ and $\hat{u}_{lp}(\zeta)$ in Eqs.(42a) and (42b) the expressions found from Eqs.(46a) and (46b), we obtain the *primary* shrinkage and swelling curves of the intra-aggregate matrix to be

$$u(\zeta)=u_{lpz}+k(1-u_S)v_z+u_S+(1-k)(1-u_S)v(\zeta), \qquad 0<\zeta\leq\zeta_h, \tag{47a}$$

$$\hat{u}(\zeta)=u_{lpz}+k(1-u_S)v_z+u_S+(1-k)(1-u_S)\hat{v}(\zeta), \qquad 0<\zeta\leq\zeta_h, \tag{47b}$$

where $v(\zeta)$ is from Eq.(2) and $\hat{v}(\zeta)$ from Eq.(7). The primary shrink-swell cycle in the intra-aggregate matrix, $(u(\zeta),\hat{u}(\zeta))$ (according to Eqs.(47a) and (47b) $u(0)=\hat{u}(0)$ and $u(\zeta_h)=\hat{u}(\zeta_h)$) is qualitatively similar to the primary shrink-swell cycle of the contributive clay, $(v(\zeta),\hat{v}(\zeta))$ (cf. Eqs.(47a) and (47b) and Fig.2 or 3). However, in general, unlike the case of clay (Figs.2 or 3), the *slope* of the *primary shrinkage* curve of the *intra-aggregate matrix*, $du/d\zeta$ in the *basic* shrinkage range, $\zeta_n<\zeta\leq\zeta_h$ is *less* then the *slope* of the *saturation* line (Fig.8), and the initial point of the curve ($\zeta_h,u_h$) is on the pseudo saturation line (Fig.8) [26,27]. The reason for that is the presence of lacunar pores in the intra-aggregate matrix (Fig.1). In this case $0<k<1$ and in Eqs.(47a) and (47b) $(1-k)<1$ [26,27].

To transit to the customary coordinates $(w,U)$ and $(w,\hat{U})$ (specific volume vs. gravimetric water content of the soil intra-aggregate matrix at shrinkage and swelling; $w\equiv\hat{w}$) one can use [24-26]

$$w=((1-u_s)/u_s)(\rho_w/\rho_s)\zeta; \quad U=u/(u_s\rho_s); \quad \hat{U}=\hat{u}/(u_s\rho_s), \quad 0<w\leq w_h \quad (w=c\overline{w}) \tag{48}$$

where $w_h=w(\zeta_h)$; $c$ being the clay content. The cycle $(U(w),\hat{U}(w))$ is qualitatively similar to the cycles $(u(\zeta),\hat{u}(\zeta))$ and $(v(\zeta),\hat{v}(\zeta))$ (Figs.2 or 3).

*3.1.2. Scanning shrinkage and swelling curves as well as transitive and steady shrink-swell cycles without loading*

Eqs.(42a) and (42b) that connect the *primary* shrinkage and swelling curves of the intra-aggregate matrix ($u(\zeta)$ and $\hat{u}(\zeta)$) with those of the contributive clay ($v(\zeta)$ and $\hat{v}(\zeta)$), reflect a *particular* case of the more general link between a point of ($\zeta,u$) plane (here $u$ symbolizes the $u$ coordinate, but not a shrinkage curve) that is between the primary shrinkage and swelling curves of the intra-aggregate matrix (Fig.8) and the *corresponding* point of ($\zeta,v$) plane that is between the primary shrinkage and swelling curves of the contributive clay (Figs. 2 or 3). In other words the relations that are similar to Eqs.(42a) and (42b) should also be fulfilled at possible water contents, $\zeta$ for the volumes of the intra-aggregate matrix and contributive clay along *any scanning* shrinkage ($\underline{v}(\zeta,\zeta_o)$ and $\underline{u}(\zeta,\zeta_o)$) and swelling ($\hat{v}(\zeta,\zeta_o)$ and $\hat{u}(\zeta,\zeta_o)$) curves as

$$\underline{v}(\zeta,\zeta_o)=(\underline{u}(\zeta,\zeta_o)-\underline{u}_{lp}(\zeta,\zeta_o)-u_S)/(1-u_S), \qquad 0<\zeta<\zeta_o\leq\zeta_h \tag{49a}$$

$$\hat{v}(\zeta,\zeta_o)=(\hat{u}(\zeta,\zeta_o)-\hat{u}_{lp}(\zeta,\zeta_o)-u_S)/(1-u_S), \qquad 0<\zeta_o<\zeta\leq\zeta_h . \tag{49b}$$

Here $\underline{u}_{lp}(\zeta,\zeta_o)$ and $\hat{u}_{lp}(\zeta,\zeta_o)$ designate the relative lacunar pore volume along the corresponding scanning shrinkage and swelling curves of the intra-aggregate matrix; $\zeta_o$ indicates the initial point of the scanning shrinkage or swelling curve (as in Figs.3 or 2 for clay). $\zeta_o$ can be used for "marking" the scanning curves.

The relation between the relative volumes of lacunar pores and clay pores in the intra-aggregate matrix that was indicated in Eqs.(43a) and (43b) along the *primary* curves is also more general and can be written along the *scanning* curves as

$$d\underline{u}_{lp}(\zeta,\zeta_o)=-k\,d\underline{u}_{cp}(\zeta,\zeta_o), \qquad 0<\zeta<\zeta_o\leq\zeta_h \tag{50a}$$

$$d\hat{u}_{lp}(\zeta,\zeta_o)=-k\,d\hat{u}_{cp}(\zeta,\zeta_o), \qquad 0<\zeta_o<\zeta\leq\zeta_h \tag{50b}$$

where $\underline{u}_{cp}(\zeta,\zeta_o)$ and $\hat{u}_{cp}(\zeta,\zeta_o)$ are the relative volumes of the clay pores (in the intra-aggregate matrix) along the *scanning* shrinkage and swelling curves marked by $\zeta_o$.

Eqs.(44a) and (44b) that relate to the *primary* curves are also fulfilled along the *scanning* shrinkage and swelling curves as

$$\underline{u}_{cp}(\zeta,\zeta_o)=(\underline{v}(\zeta,\zeta_o)-v_s)(1-u_S), \qquad 0<\zeta<\zeta_o\leq\zeta_h \tag{51a}$$

$$\hat{u}_{cp}(\zeta,\zeta_o)=(\hat{v}(\zeta,\zeta_o)-v_s)(1-u_S), \qquad 0<\zeta_o<\zeta\leq\zeta_h . \tag{51b}$$

For the lacunar pore volume, $\underline{u}_{lp}(\zeta,\zeta_o)$ along the *scanning shrinkage* curve, we can use the same condition at the *shrinkage end* as in Eq.(45a)

$$\underline{u}_{lp}(0,\zeta_o)=u_{lpz}, \qquad 0<\zeta_o\leq\zeta_h . \tag{52a}$$

However for the lacunar pore volume, $\hat{u}_{lp}(\zeta,\zeta_o)$ along the *scanning swelling* curve, unlike Eq.(45b) it is more convenient to use the condition at the *swelling end* as

$\hat{u}_{lp}(\zeta_h,\zeta_o)=u_{lph}$ ,     $0<\zeta_o\leq\zeta_h$                           (52b)

($\hat{u}_{lph}=u_{lph}$ is the relative volume of lacunar pores at maximum swelling [24]).

Replacing $\underline{u}_{cp}(\zeta,\zeta_o)$ in Eq.(50a) from Eq.(51a) and $\hat{u}_{cp}(\zeta,\zeta_o)$ in Eq.(50b) from Eq.(51b), and integrating the relations to be obtained with the use of Eqs.(52a) and (52b) we find (cf. Eqs.(46a) and (46b))

$\underline{u}_{lp}(\zeta,\zeta_o)=u_{lpz}-k(1-u_S)(\underline{v}(\zeta,\zeta_o)-v_z)$ ,     $0<\zeta<\zeta_o\leq\zeta_h$           (53a)

$\hat{u}_{lp}(\zeta,\zeta_o)=u_{lph}+k(1-u_S)(v_h-\hat{v}(\zeta,\zeta_o))$ ,     $0<\zeta_o<\zeta\leq\zeta_h$ .         (53b)

Replacing $u_{lph}$ in Eq.(53b) as $u_{lph}=u_{lpz}-k(1-u_S)(v_h-v_z)$ [26] (see Eq.(46a) at $\zeta=\zeta_h$) one can rewrite Eq.(53b) as (cf. Eq.(46b))

$\hat{u}_{lp}(\zeta,\zeta_o)=u_{lpz}-k(1-u_S)(\hat{v}(\zeta,\zeta_o)-v_z)$ ,     $0<\zeta_o<\zeta\leq\zeta_h$ .         (53c)

Note that $\underline{u}_{lp}(\zeta,\zeta_o)$ from Eq.(53a) *automatically* meets the necessary condition, $\underline{u}_{lp}(\zeta,\zeta_o)|_{\zeta=\zeta_o}=\hat{u}_{lp}(\zeta_o)$ at the *scanning shrinkage start* (Fig.3 after the symbol change $v\to u$) with $\hat{u}_{lp}(\zeta_o)$ from Eq.(46b) since $\underline{v}(\zeta,\zeta_o)|_{\zeta=\zeta_o}=\hat{v}(\zeta_o)$ (Eq.(17); see Fig.3). Similarly, $\hat{u}_{lp}(\zeta,\zeta_o)$ from Eq.(53c) *automatically* meets the necessary condition, $\hat{u}_{lp}(\zeta,\zeta_o)|_{\zeta=\zeta_o}=u_{lp}(\zeta_o)$ at the *scanning swelling start* (see Fig.2 after the symbol change $v\to u$) with $u_{lp}(\zeta_o)$ from Eq.(46a) since $\hat{v}(\zeta_o,\zeta_o)=v(\zeta_o)$ (Eq.(11); see Fig.2).

Substituting for $\underline{u}_{lp}(\zeta,\zeta_o)$ and $\hat{u}_{lp}(\zeta,\zeta_o)$ in Eqs.(49a) and (49b) the expressions found from Eqs.(53a) and (53c), we obtain the *scanning* shrinkage and swelling curves of the intra-aggregate matrix to be (cf. Eqs.(47a) and (47b))

$\underline{u}(\zeta,\zeta_o)=u_{lpz}+k(1-u_S)v_z+u_S+(1-k)(1-u_S)\underline{v}(\zeta,\zeta_o)$ ,     $0<\zeta<\zeta_o\leq\zeta_h$        (54a)

$\hat{u}(\zeta,\zeta_o)=u_{lpz}+k(1-u_S)v_z+u_S+(1-k)(1-u_S)\hat{v}(\zeta,\zeta_o)$ ,     $0<\zeta_o<\zeta\leq\zeta_h$ ,      (54b)

where $\underline{v}(\zeta,\zeta_o)$ is from Eqs.(14)-(16), (19)-(22), (25), and (26), and $\hat{v}(\zeta,\zeta_o)$ is from Eqs.(9), (10), (12), and (13). The expressions for $\underline{u}(\zeta,\zeta_o)$ and $\hat{u}(\zeta,\zeta_o)$ (Eqs.(54a) and (54b)) show that the *scanning* shrinkage and swelling curves of the intra-aggregate matrix are arranged inside the *primary* shrink-swell cycle, $(u(\zeta),\hat{u}(\zeta))$ (Eqs.(47a) and (47b)) similar to the arrangement of the clay scanning curves, $\underline{v}(\zeta,\zeta_o)$ and $\hat{v}(\zeta,\zeta_o)$ inside the primary cycle, $(v(\zeta),\hat{v}(\zeta))$ (Figs.2 and 3).

Using the found $\underline{u}(\zeta,\zeta_o)$ and $\hat{u}(\zeta,\zeta_o)$ scanning curves (Eqs.(54a) and (54b)) in Eqs.(28)-(32) instead of $\underline{v}(\zeta,\zeta_o)$ and $\hat{v}(\zeta,\zeta_o)$, one can construct the *transitive* and *steady* shrink-swell cycles for the intra-aggregate matrix without loading in the *total analogy* to the case of the contributive clay in Section 2.1.3 (Figs.4 and 5). Possible differences in the slope of the primary shrinkage curves (in the basic shrinkage range) of the clay and intra-aggregate matrix (see the paragraph after Eq.(47b) and Fig.8), are not essential for the analogy.

The transition from the relative to customary coordinates as applied to the scanning curves of the intra-aggregate matrix, is realized by the same relations (Eq.(48)) after the evident replacements, $u(\zeta) \rightarrow \underline{u}(\zeta,\zeta_o)$, $\hat{u}(\zeta) \rightarrow \underline{\hat{u}}(\zeta,\zeta_o)$, $U(w) \rightarrow \underline{U}(w,w_o)$, $\hat{U}(w) \rightarrow \underline{\hat{U}}(w,w_o)$, and addition, $w_o=((1-u_s)/u_s)(\rho_w/\rho_s)\zeta_o$. Qualitatively, the $\underline{U}(w,w_o)$ and $\underline{\hat{U}}(w,w_o)$ curves are similar to the $\underline{u}(\zeta,\zeta_o)$ and $\underline{\hat{u}}(\zeta,\zeta_o)$ or $\underline{v}(\zeta,\zeta_o)$ and $\hat{v}(\zeta,\zeta_o)$ curves, respectively, in Figs.2 and 3.

### 3.2. Shrinkage and swelling of intra-aggregate matrix under loading
*3.2.1. The lacunar factor and maximum lacunar pore volume under loading*

Transformations of the lacunar factor, $k$ (entering Eqs.(47a), (47b), (54a), and (54b)) and maximum lacunar pore volume in the intra-aggregate matrix at the end of shrinkage, $u_{lpz}$ (in Eqs.(47a), (47b), (54a), and (54b) $u_{lpz} \equiv u_{lpz}(L=0)$) under loading, $L$ have so far not been considered. Note that $u_S$ entering indicated equations, by its definition [26], cannot vary with $L$ (since $u_S=v_s(1-c)[c+v_s(1-c)]^{-1}$ [26]).

The lacunar factor, $k$ as a function of only internal characteristics of intra-aggregate matrix -clay content, $c$, clay type, and soil texture- is presented [38] as

$$k(c/c^*)=[1-(c/c^*)^3]^{1/3}, \qquad 0<c/c^*<1 \qquad (55a)$$

$$k(c/c^*)=0, \qquad 1<c/c^*<1/c^* \qquad (55b)$$

$$c^*=[1+(v_z/v_s)(1/p-1)]^{-1} \qquad (55c)$$

where $c^*$ is the critical clay content and $p$ is the porosity of the contributive silt and sand grains when they are in the state of *imagined* contact [25]. $k(c/c^*(L))$ is determined by substitution for $v_z$ in Eq.(55c) the $v_z(L)$ dependence (Eq.(36)).

To estimate $u_{lpz}(L)$, first we estimate $u_{lph}(L)$. By definition, $u_{lph}$ is the lacunar pore volume in the intra-aggregate matrix (Fig.1) at $\zeta=\zeta_h$ (Eq.(46a)). One can consider that at the maximum swelling point, $\zeta=\zeta_h$ the intra-aggregate matrix is in the *visco-plastic* state or close to that. In *such* state the pore volumes of different type approximately change with loading, $L$ proportionally to each other. In other words, it is natural to consider that the *lacunar* pore volume, $u_{lph}(L)$ (Fig.1) varies with loading, $L$ proportionally to the volume, $(v_h(L)-v_s)$ of the *clay matrix* pores (in relative units) in the intra-aggregate matrix at $\zeta=\zeta_h$ (Fig.1). Then, taking $u_{lph}(L=0) \equiv u_{lph}$ and $v_h(L=0) \equiv v_h$ we estimate the sought $u_{lph}(L)$ dependence as ($v_h(L)$ being from Eq.(40))

$$u_{lph}(L)=u_{lph}(v_h(L)-v_s)/(v_h-v_s), \qquad L>0 \ . \qquad (56a)$$

Then, $u_{lpz}(L)$ follows from Eqs.(46a), (55a)-(55c), (56a), (40), and (36) as

$$u_{lpz}(L)=u_{lph}(L)+k(c/c^*(L))(1-u_S)(v_h(L)-v_z(L)), \qquad L>0 \ . \qquad (56b)$$

Thus, $k(L)$ and $u_{lpz}(L)$ are determined by the same major dependences, $v_z(L)$ (Eq.(36)) and $v_h(L)$ (Eq.40). According to Eq.(56a) $u_{lph}(L)=0$ follows $u_{lph}=0$, as it should be (if lacunar pores at $\zeta=\zeta_h$ are absent before loading, all the more they cannot appear after loading). However, if in this case ($u_{lph}=0$) $k(c/c^*(L))$ (from Eqs.(55a)-(55c)) is more than zero the lacunar pores can appear at shrinkage under loading (cf. Sections 5 and 7). If $u_{lph}>0$, according to Eq.(56a) $u_{lph}(L)<u_{lph}$, as it should be. With that $k(c/c^*(L))>k(c/c^*(L=0))$ (see Eqs.(55a)-(55c), and (36)). Therefore the *fraction*, $k(L)$ of the clay pore volume increment at shrinkage-swelling that transforms into the lacunar pore volume increment [38] (Eqs.(43a) and (43b)) can increase with loading. However, the shrinkage-swelling itself becomes weaker since the slope (1-$k(L)$) of the shrinkage curve in the basic shrinkage range [26] decreases at loading. Note that at $L</\cong 100$kPa when $v_z(L)\cong v_z$=const (Section 2.2.2) $k(L)$ is also constant (see Eqs.(55)).

### 3.2.2. Different shrink-swell curves of intra-aggregate matrix under loading

The *quantitative* prediction of the *primary* shrinkage ($u(\zeta,L)$) and swelling ($\hat{u}(\zeta,L)$) curves of an intra-aggregate matrix *under loading* immediately follows from Eqs.(47a) and (47b) as

$$u(\zeta,L)=u_{lpz}(L)+k(L)(1-u_S)v_z(L)+u_S+(1-k(L))(1-u_S)v(\zeta,L), \quad 0<\zeta\leq\zeta_h(L), \quad L\geq 0, \quad (57a)$$

$$\hat{u}(\zeta,L)=u_{lpz}(L)+k(L)(1-u_S)v_z(L)+u_S+(1-k(L))(1-u_S)\hat{v}(\zeta,L), \quad 0<\zeta\leq\zeta_h(L), \quad L\geq 0, \quad (57b)$$

with the preliminarily calculated basic dependences, $v_z(L)$ (Eq.(36)), $\zeta_h(L)$ (Eq.(39)), $v_h(L)$ (Eq.(40)), and then the primary shrinkage ($v(\zeta,L)$) and swelling ($\hat{v}(\zeta,L)$) curves for the contributive clay under loading (see Sections 2.2.1-2.2.4) as well as $u_{lpz}(L)$ and $k(L)$ (Section 3.2.1). Fig.8 shows the transformation geometry of the *maximum* (i.e., *primary*) shrink-swell cycle of the intra-aggregate matrix, ($u(\zeta),\hat{u}(\zeta)$) *without* loading to the similar cycle, ($u(\zeta,L), \hat{u}(\zeta,L)$) *under* loading (cf. Fig.6 for clay).

The *scanning* shrinkage ($\underline{u}(\zeta,\zeta_o,L)$) and swelling ($\underline{\hat{u}}(\zeta,\zeta_o,L)$) curves *inside* the maximum shrink-swell cycle of the intra-aggregate matrix under loading, ($u(\zeta,L), \hat{u}(\zeta,L)$) (Fig.8) follow from Eqs.(54a) and (54b) as ($L\geq 0$)

$$\underline{u}(\zeta,\zeta_o,L)=u_{lpz}(L)+k(L)(1-u_S)v_z(L)+u_S+(1-k(L))(1-u_S)\underline{v}(\zeta,\zeta_o,L), \quad 0<\zeta<\zeta_o\leq\zeta_h(L), \quad (58a)$$

$$\underline{\hat{u}}(\zeta,\zeta_o,L)=u_{lpz}(L)+k(L)(1-u_S)v_z(L)+u_S+(1-k(L))(1-u_S)\underline{\hat{v}}(\zeta,\zeta_o,L), \quad 0<\zeta_o<\zeta\leq\zeta_h(L), \quad (58b)$$

with the preliminarily calculated $v_z(L)$, $\zeta_h(L)$, $v_h(L)$, $u_{lpz}(L)$, and $k(L)$, as above, as well as the scanning shrinkage ($\underline{v}(\zeta,\zeta_o,L)$) and swelling ($\underline{\hat{v}}(\zeta,\zeta_o,L)$) curves for the contributive clay under loading (see Sections 2.2.1-2.2.4).

At the known *scanning* curves, $\underline{u}(\zeta,\zeta_o,L)$ and $\underline{\hat{u}}(\zeta,\zeta_o,L)$ the finding of the *transitive* and *steady* shrink-swell cycles in the intra-aggregate matrix under loading in the range, $0<\zeta_1<\zeta<\zeta_2<\zeta_h(L)$ (Fig.8; $\zeta_1$ and $\zeta_2$ are not shown) is totally similar to the analogous issue in the case of contributive clay under loading (section 2.2.4; Figs.4 and 5 with replacements, $\underline{v}(\zeta,\zeta_o)\to\underline{v}(\zeta,\zeta_o,L)$ and $\underline{\hat{v}}(\zeta,\zeta_o)\to\underline{\hat{v}}(\zeta,\zeta_o,L)$). Eqs.(28)-(32) are utilized to the scanning curves, $\underline{u}(\zeta,\zeta_o,L)$ and $\underline{\hat{u}}(\zeta,\zeta_o,L)$ as indicated above.

The last paragraph of Section 3.1.2, about the transition to the customary coordinates, is extended to the loading case by the replacement of any relative or specific volume of the intra-aggregate matrix with the relevant volume under loading (e.g., $\underline{u}(\zeta,\zeta_o)\to\underline{u}(\zeta,\zeta_o,L)$ and $\underline{\hat{u}}(\zeta,\zeta_o)\to\underline{\hat{u}}(\zeta,\zeta_o,L)$; $\underline{U}(w,w_o)\to\underline{U}(w,w_o,L)$ and $\underline{\hat{U}}(w,w_o)\to\underline{\hat{U}}(w,w_o,L)$) and using $\zeta_h\to\zeta_h(L)$ in calculation of $w_h(L)$ from Eq.(48).

Finally, it should be stressed that all the different shrinkage and swelling curves of the contributive *clay* (Section 2) in the *single-valued* manner determine the *corresponding* curves of the *soil intra-aggregate matrix* both without and under loading through $c$, $v_s$, $v_z(L)$, and $v_h(L)$ or $\zeta_h(L)$.

## 4. Shrinkage-swelling and cracking of a soil

Based on [24,28,39] and results of the previous sections, we consider shrinkage-swelling of a real soil with no cracking and loading (Section 4.1), with cracking and no loading (Section 4.2), and with cracking and under loading (Section 4.3).

### *4.1. Reference shrinkage and swelling of a soil*

Proceeding to a real soil (Fig.1) that includes the intra-aggregate matrix (Section 3), interface layer of aggregates, and inter-aggregate (structural) pores, we, first, consider the *reference* shrinkage and swelling of the soil. By definition *reference* shrinkage and swelling occurs *without cracking and loading*. Such shrinkage and swelling in terms of the primary and, hence, scanning shrink-swell curves can be realized using sufficiently *small* soil samples [28,24]. The *reference primary* shrinkage and swelling curves (i.e., for small samples) are necessary to present the *primary* shrinkage and swelling curves with *crack volume* contribution (i.e., in the case of a layer and large sample) [28,24] (see the brief necessary points in Section 4.2.1). Similarly, the *reference scanning* shrink-swell curves are necessary to present the *scanning* shrink-swell curves with *crack* contribution (Section 4.2.2).

### *4.1.1. Reference multifold primary shrinkage and swelling of a soil*

By definition, *primary* shrinkage and swelling of a soil occurs in the *maximum possible* water content range [24]. Such shrink-swell cycle leads to aggregate destruction (see [40,41,21], among others). For this reason, after the cycle the soil does not return to the initial state. Its volume and water content at the new maximum swelling point decrease (Fig.9) (see observations in [21] and a physical explanation and prediction in [24]) unlike the volume and water content of clay paste or intra-aggregate matrix (see Figs.2 and 8). A brief exposition, we will rely on, of the necessary points of the *first reference primary* shrink-swell cycle is as follows [24].

(1) The soil specific volume of the *first reference primary* shrinkage ($Y_r$) and swelling ($\hat{Y}_r$) is presented as (Fig.9, curves 1 and 2)

$$Y_r(w') = U'(w') + U_i + U_s, \qquad 0 \leq w' \leq w'_h \qquad (w' = w/K), \qquad (59a)$$

$$\hat{Y}_r(\hat{w}') = \hat{U}'(\hat{w}') + \hat{U}_i + U_s, \qquad 0 \leq \hat{w}' \leq \hat{w}'_h \qquad (\hat{w}' = w/\hat{K}), \qquad (59b)$$

with $U_s$ being the constant (at reference shrinkage and swelling) contribution of the inter-aggregate (structural) pores (Fig.1); $U_i$ and $\hat{U}_i$ being the constant (for a given soil) contributions of the interface layer (Fig.1) at shrinkage and swelling, respectively; $U'(w')$ and $\hat{U}'(\hat{w}')$ being the contributions of the intra-aggregate matrix (Fig.1) at shrinkage and swelling, respectively; $w'$ and $\hat{w}'$ being the contributions of the intra-aggregate matrix (Fig.1) to the total gravimetric water content, $W$ and $\hat{W}$ (see below) at shrinkage and swelling, respectively; and $w'_h$ and $\hat{w}'_h$ corresponding to shrinkage start and swelling finish, respectively, in the first primary shrink-swell cycle (for $K$ and $\hat{K}$ see below). $U'(w')$ and $\hat{U}'(\hat{w}')$ are expressed through the specific volume of the intra-aggregate matrix, $U(w)$ and $\hat{U}(\hat{w})$ at primary shrinkage and swelling, respectively (that were considered in Section 3.1.1; see Eq.(48)), as

$$U'(w') = U(w'K)/K, \qquad 0 \leq w' \leq w'_h \qquad (w'K = w), \qquad (60a)$$

$$\hat{U}'(\hat{w}') = \hat{U}(\hat{w}'\hat{K})/\hat{K} , \quad 0 \le \hat{w}' \le \hat{w}'_h \quad (\hat{w}'\hat{K}=\hat{w}=w) , \tag{60b}$$

where $K$ and $\hat{K}$ are aggregate/intra-aggregate mass ratio at shrinkage and swelling, respectively. In turn, $K$ and $\hat{K}$ are expressed through $U_i$ [39] and $\hat{U}_i$ [24] as

$$K=(1-U_i/U_h)^{-1} \quad (U_h=u_h/(u_s\rho_s), \text{ see Eq.(48)}) , \tag{61a}$$

$$\hat{K}=K[1-K(\hat{U}_i-U_i)/U_z]^{-1} \quad (U_z=u_z/(u_s\rho_s), \text{ see Eq.(48)}) . \tag{61b}$$

The $\hat{U}_i-U_i$ difference gives the increment, $\Delta\hat{U}_i$ of the interface layer volume (Fig.1) at the expense of the aggregate destruction while the transition from shrinkage to swelling occurs at $W=\hat{W}\cong 0$. The $\hat{U}_i-U_i$ difference originates the decrease of the maximum specific soil volume after the reference primary shrink-swell cycle, $\Delta Y_{rh} \equiv Y_{rh} - \hat{Y}_{rh} \equiv Y_r(w'_h) - \hat{Y}_r(\hat{w}'_h) > 0$ [24] (Fig.9). With that (see Eqs.(60a) and (59a), Eqs.(60b) and (59b), and Eq.(48) for $u_h=\hat{u}_h$ )

$$Y_{rh}=U_h/K+U_i+U_s \quad \text{and} \quad \hat{Y}_{rh}=\hat{U}_h/\hat{K}+\hat{U}_i+U_s=U_h/\hat{K}+\hat{U}_i+U_s . \tag{62}$$

Finally, $U_i$ and $\hat{U}_i$ are connected with maximum and minimum aggregate sizes at the start of shrinkage ($X_{min}$, $X_m$) and after the aggregate destruction at the start of swelling ($\hat{X}_{min}=X_{min}$, $\hat{X}_m$) [24,39] as

$$U_i=G(x_n/X_m, P_h, U_h) \quad \text{and} \quad \hat{U}_i=G(x_n/\hat{X}_{mz}, P_z, U_z) \tag{63a}$$

where $x_n \cong X_{min} = \hat{X}_{min}$ is the mean size of soil solids; $P_h$ and $P_z$ are the structural porosity at shrinkage close to $W=W_h$ and $W=W_z$ as

$$P_h=U_s/Y_h=U_s/(U_h+U_s) \quad \text{and} \quad P_z=U_s/Y_{rz}=U_s/(U_z/K+U_i+U_s) , \tag{63b}$$

and the $G(\alpha,\beta,\chi)$ function is defined as

$$G(\alpha,\beta,\chi)=\chi\left\{1-(1-\alpha)^3+3[\alpha/(1-\alpha)]\int_0^1 \frac{\eta^2 F(\eta,\beta)d\eta}{[\eta+\alpha/(1-\alpha)]^4}\right\} \tag{63c}$$

with $F(\eta,\beta)$ distribution from [42] as

$$F(\eta,\beta)=(1-\beta^{I_o(\eta)/8.4})/(1-\beta) , \quad I_o(\eta)=\ln(6)(4\eta)^4\exp(-4\eta), \quad 0\le\eta\le 1 . \tag{63d}$$

In case of the negligible structural porosity ($\beta\to 0$)

$$F(\eta, 0)=\{1-\exp[-I_o(\eta)]\}/[1-\exp(-8.4)] . \tag{63e}$$

Eqs.(63a)-(63e) are used in Section 4.3.3 when considering the effects of loading.

(2) The soil water content at the *first reference primary* shrinkage ($W$) and swelling ($\hat{W}$) (Fig.9, curves 1 and 2) is presented as

$$W(w') = w' + \omega(w'), \qquad 0 \le w' \le w'_h, \qquad (64a)$$

$$\hat{W}(\hat{w}') = \hat{w}' + \hat{\omega}(\hat{w}'), \qquad 0 \le \hat{w}' \le \hat{w}'_h. \qquad (64b)$$

Here $w'$ and $\hat{w}'$ are connected with $K$ and $\hat{K}$, and with $U_i$ and $\hat{U}_i$ (Eqs.(61a) and (61b)) through $w'K = \hat{w}'\hat{K}$ (see Eqs.(60a) and (60b)). $\omega(w')$ and $\hat{\omega}(\hat{w}')$ [24] are the contributions of the interface layer (Fig.1) at reference primary shrinkage and swelling, respectively. The $\hat{\omega}(\hat{w}')$ contribution at swelling (Eq.(64b)) is divided into contributions of the interface layer part that has existed at shrinkage before the aggregate destruction ($\omega(\hat{w}')$) (the part with the specific volume, $U_i$) and of the additional interface layer part that comes into being after the destruction ($\Delta\hat{\omega}(\hat{w}')$) (the part with the specific volume, $\Delta\hat{U}_i = \hat{U}_i - U_i$). The $\omega(w')$, $\omega(\hat{w}')$, and $\Delta\hat{\omega}(\hat{w}')$ contributions are connected with the maximum internal sizes, $R(\zeta)$ [25] and $\hat{R}(\zeta)$ [24] of the water-filled clay pores in the intra-aggregate matrix at reference primary shrinkage and swelling, respectively (see below point (4)) and calculated as [24]

$$\omega(w') = \begin{cases} 0, & 0 \le w' < w'_s \\ \rho_w U_i \Pi_h F_i(\eta(R(w'), R_{min}, R_m), \Pi_h), & w'_s \le w' < w'_h, \end{cases} \qquad (65a)$$

$$\omega(\hat{w}') = \begin{cases} 0, & 0 \le \hat{w}' < \hat{w}'_b \\ \rho_w U_i \Pi_h F_i(\eta(\hat{R}(\hat{w}'), R_{min}, R_m), \Pi_h), & \hat{w}'_b \le \hat{w}' < \hat{w}'_h, \end{cases} \qquad (65b)$$

$$\Delta\hat{\omega}(\hat{w}') = \begin{cases} \rho_w \Delta\hat{U}_i \Pi_z F_i(\eta(\hat{R}(\hat{w}'), \hat{R}_{min}, \hat{R}_m), \Pi_z), & 0 \le \hat{w}' \le \hat{w}'_e \\ \rho_w \Delta\hat{U}_i \Pi_z, & \hat{w}'_e \le \hat{w}' \le \hat{w}'_h \end{cases} \qquad (65c)$$

where $\Pi_h$ and $\Pi_z$ are the clay porosity of the interface layer part of the $U_i$ volume and $\Delta\hat{U}_i$ volume, respectively, as

$$\Pi_h = 1 - (u_s + u_{lph})/u_h \qquad \text{and} \qquad \Pi_z = 1 - (u_s + u_{lpz})/u_z. \qquad (65d)$$

The $F_i(\eta, \Pi)$ function ($\Pi \equiv \Pi_h$ or $\Pi_z$) is defined as (for $I_o(\eta)$ see Eq.(63d))

$$F_i(\eta, \Pi) = (1 - (1-\Pi)^{I_o(\eta)/8.4})/\Pi. \qquad (65e)$$

The $\eta(x, y, z)$ function is defined as

$\eta(x, y, z) = (x-y)/(z-y)$ . (65f)

$R_{\min}$ and $R_{\mathrm{m}}$ are the minimum and maximum sizes of the non-shrinking clay pores in the interface layer part of the $U_{\mathrm{i}}$ volume at shrinkage (for more detail see [25]) and connected with $r_{\mathrm{m}}(v)$ (Eq.(6a)) as

$R_{\min} = (r_{\mathrm{m}}(v_{\mathrm{n}}) + r_{\mathrm{m}}(v_{\mathrm{h}}))/2$    and    $R_{\mathrm{m}} = r_{\mathrm{m}}(v_{\mathrm{h}})$ . (65g)

$\hat{R}_{\min}$ and $\hat{R}_{\mathrm{m}}$ are the minimum and maximum sizes of non-shrinking and non-swelling clay pores in the interface layer part of the $\Delta \hat{U}_{\mathrm{i}}$ volume at swelling (for more detail see [24]) and connected with $r_{\mathrm{m}}(v)$ (Eq.(6a)) and $r_{\mathrm{o}}(v)$ (Eq.(6b)) as

$\hat{R}_{\min} = r_{\mathrm{o}}(v_{\mathrm{z}})$    and    $\hat{R}_{\mathrm{m}} = r_{\mathrm{m}}(v_{\mathrm{z}})$ . (65h)

$w'_{\mathrm{s}}$, $\hat{w}'_{\mathrm{b}}$, and $\hat{w}'_{\mathrm{e}}$ meet the following conditions [24]

$R(w'_{\mathrm{s}}) = R_{\min}$,    $\hat{R}(\hat{w}'_{\mathrm{b}}) = R_{\min}$,    and    $\hat{R}(\hat{w}'_{\mathrm{e}}) = \hat{R}_{\mathrm{m}}$ . (65i)

Eqs.(65a)-(65i) are used in Section 4.3.4 when considering the effects of loading. According to Eqs.(64a), (64b), and (65a)-(65c) the found $\omega(w')$, $\omega(\hat{w}')$, and $\Delta\hat{\omega}(\hat{w}')$ show that $\Delta W_{\mathrm{h}} = W_{\mathrm{h}} - \hat{W}_{\mathrm{h}} \equiv W(w'_{\mathrm{h}}) - \hat{W}(\hat{w}'_{\mathrm{h}}) > 0$ [24] (Fig.9). With that [24]

$W_{\mathrm{h}} = w'_{\mathrm{h}} + \rho_{\mathrm{w}} U_{\mathrm{i}} \Pi_{\mathrm{h}}$    and    $\hat{W}_{\mathrm{h}} = w'_{\mathrm{h}} K/\hat{K} + \rho_{\mathrm{w}} U_{\mathrm{i}} \Pi_{\mathrm{h}} + \rho_{\mathrm{w}} \Delta \hat{U}_{\mathrm{i}} \Pi_{\mathrm{z}}$ , (66)

(3) Eqs.(59a), (64a) and Eqs.(59b), (64b) give the parametrical determination of the *first reference primary shrinkage* curve, $Y_{\mathrm{r}}(W)$ at $0 \leq W \leq W_{\mathrm{h}}$ (Fig.9, curve 1) and the *first reference primary swelling* curve, $\hat{Y}_{\mathrm{r}}(\hat{W})$ at $0 \leq \hat{W} \leq \hat{W}_{\mathrm{h}}$ (Fig.9, curve 2).

(4) The $R(\zeta)$ size entering Eq.(65a) at $\zeta_{\mathrm{s}} \leq \zeta < \zeta_{\mathrm{h}}$ ( $w'_{\mathrm{s}} \leq w' < w'_{\mathrm{h}}$ ) is calculated as [25]

$R(\zeta) = r_{\mathrm{m}}(v(\zeta))$ ,    $\zeta_{\mathrm{n}} \leq \zeta < \zeta_{\mathrm{h}}$ . (67)

($r_{\mathrm{m}}(v(\zeta))$ is from Eq.(6a)). It is worth noting the equation that is important for the calculation of $\omega(\hat{w}')$ (Eq.(65b)) and $\Delta\hat{\omega}(\hat{w}')$ (Eq.(65c)) and determines the maximum internal size, $\hat{R}(\zeta)$ of the water-filled clay pores in the intra-aggregate matrix at the first reference primary swelling [24] to be

$f(\eta(\hat{R}(\zeta)), \hat{P}) = (1-v_{\mathrm{s}})\zeta/(\hat{v}(\zeta)-v_{\mathrm{s}}) + (\hat{v}(\zeta)-v(\zeta))/(\hat{v}(\zeta)-v_{\mathrm{s}})$,    $0 \leq \zeta \leq \zeta_{\mathrm{h}}$ . (68a)

The terms in the right part give the summary volume fraction of adsorbed water and entrapped air. The left part gives the volume fraction of the clay pores of the maximum size, $\hat{R}(\zeta)$. The $f(\eta, \hat{P})$ function is written as [42] (for $I_{\mathrm{o}}(\eta)$ see Eq.(63d))

$f(\eta, \hat{P}) = (1-(1-\hat{P})^{I_{\mathrm{o}}(\eta)/8.4})/\hat{P}$ ;    $0 \leq \eta \leq 1$ (68b)

where $\hat{P}$ is clay porosity at swelling (see Eq.(5) at $v(\zeta) \to \hat{v}(\zeta)$) and

$$\eta(\hat{R}(\zeta)) = (\hat{R}(\zeta) - \hat{r}_o(\zeta))/(\hat{r}_m(\zeta) - \hat{r}_o(\zeta)) \tag{68c}$$

(see Eqs.(6a) and (6b) at $v(\zeta) \to \hat{v}(\zeta)$). Eqs.(67), (68a)-(68c)) will be needed in Section 4.3.4 when considering the $\omega(w')$ and $\omega(\hat{w}')$ variation under loading.

Based on the above, one can consider the *second* and *following reference primary cycles* (Fig.9, curves 3 and 4). For the second cycle the initial maximum volume and water content are $\hat{Y}_{rh}$ instead of $Y_{rh}$ and $\hat{W}_h$ instead of $W_h$ (Fig.9). It is obvious that such cycles, with water content variation in the *maximum possible* range and, correspondingly, with *aggregate destruction* can only be realized in the *upper* part of the soil profile. The estimates [24] show that the cycles lead to quick destruction of aggregates. Thereby the model, at least qualitatively, also explains and describes soil crusting (more detail consideration of the latter is beyond the scope of the work). For the physical parameters of the contributive clay as well as soil texture and structure, that in the *single-valued* manner determine the *reference primary* shrinkage and swelling curves see [28,24].

*4.1.2. Reference scanning shrinkage and swelling curves of a soil*

At sufficiently large depths of soil profile the water content usually varies in a relatively *small* range, and shrinkage-swelling does not lead to aggregate destruction. In terms of the reference primary shrink-swell cycle it means that $\hat{U}_i = U_i$ ($\Delta \hat{U}_i = 0$) and $\hat{K} = K$ (Eq.(61b)). Correspondingly, $Y_{rh} = \hat{Y}_{rh}$ ($\Delta Y_{rh} = 0$; Eq.(62)) and $W_h = \hat{W}_h$ ($\Delta W_h = 0$; Eq.(66)). That is, in the consideration of the reference *scanning* shrinkage and swelling curves (i.e., curves that vary in a range $\Delta W < W_h$ and $\Delta \hat{W} < \hat{W}_h$), we should use the *quasi primary* shrink-swell cycle that includes the *reference quasi primary* swelling curve, $\hat{\underline{Y}}_r(\hat{W})$ corresponding to $\hat{U}_i = U_i$ and $\hat{K} = K$ (Fig.9, dashed line 2' of the first cycle; see also Figs.10 and 11). Similar to the reference *primary* shrinkage curve, $Y_r(W)$ (Eqs.(59a) and (64a)) the reference *scanning* shrinkage curve, $\underline{Y}_r(\underline{W}, \underline{W}_o)$ (Fig.10, curve 3) includes the contributions of the intra-aggregate matrix, interface layer, and structural pores (Fig.1) to the specific soil volume, $\underline{Y}_r(w', w'_o)$ and the two first contributions to the soil gravimetric water content, $\underline{W}(w', w'_o)$ that can be presented as (accounting for $\hat{U}_i = U_i$, $\hat{K} = K$, and $\hat{w}' = w'$)

$$\underline{Y}_r(w', w'_o) = \underline{U}'(w', w'_o) + U_i + U_s, \qquad 0 \le w' \le w'_o \le w'_h, \tag{69}$$

$$\underline{W}(w', w'_o) = w' + \underline{\omega}(w', w'_o), \qquad 0 \le w' \le w'_o \le w'_h. \tag{70}$$

Here the "o" subscript marks the starting point of the $\underline{Y}_r(\underline{W}, \underline{W}_o)$ shrinkage curve (Fig.10, curve 3) in terms of $w'_o$, $w_o$, or $\underline{W}_o$; $U_i$ and $U_s$ were indicated above [24]. The $\underline{U}'(w', w'_o)$ contribution in Eq.(69) is expressed through the corresponding specific volume, $\underline{U}(w, w_o)$ of the intra-aggregate matrix (Section 3.1.2) as (cf. Eq.(60a))

$\underline{U}'(w',w'_o) = \underline{U}(w,w_o)/K$,   $0 \leq w' \leq w'_o \leq w'_h$   ($w=w'K$, $w_o=w'_oK$) . (71)

By analogy to the expression for $\omega(w')$ [24] in Eq.(65a) the $\underline{\omega}(w',w'_o)$ contribution to the water content in Eq.(70) can be written as (for $F_i(\eta,\Pi_h)$ see Eqs.(65d) and (65e))

$$\underline{\omega}(w',w'_o) = \begin{cases} 0, & 0 \leq w' < \underline{w}'_s(w'_o) < w'_o \\ \rho_w U_i \Pi_h F_i(\eta(\underline{R}(w',w'_o)),\Pi_h), & \underline{w}'_s(w'_o) \leq w' < w'_o < w'_h \end{cases} \quad (72)$$

(for $I_o(\eta)$ see Eq.(63d)). Here the $\eta$ parameter is determined as

$\eta(\underline{R}(w',w'_o)) = (\underline{R} - R_{min})/(R_m - R_{min})$,   $R_{min} < \underline{R}(w',w'_o) < \underline{R}(w'_o,w'_o) = \hat{R}(w'_o)$ (73)

where $\underline{R}(w',w'_o)$ (see below) is the maximum internal size of water-filled clay pores of the intra-aggregate matrix along the *reference scanning shrinkage* curve (Fig.10, curve 3). The $\underline{w}'_s(w'_o)$ point (Eq.(72)) is determined from condition, $\underline{R}(\underline{w}'_s,w'_o) = R_{min}$ (cf. $w'_s$ in [24] and Eq.(65i)). The starting point of the scanning shrinkage curve in terms of $w'_o$ in Eqs.(69) and (70) is naturally connected with $\underline{W}_o$ in $\underline{Y}_r(\underline{W},\underline{W}_o)$ (Fig.10, curve 3). Indeed, excluding $w'$ from Eqs.(69) and (70) we obtain $\underline{Y}_r(\underline{W},w'_o)$. According to Fig.10 at $\underline{W} = \underline{W}_o$ one has the condition

$\underline{Y}_r(\underline{W}_o,w'_o) = \hat{\underline{Y}}_r(\underline{W}_o)$ (74)

where $\hat{\underline{Y}}_r(\hat{W})$ is calculated as $\hat{Y}_r(\hat{W})$ (Eqs.(59b), (60b), and (61b)) at $\hat{U}_i = U_i$ and $\hat{K} = K$. Substituting for $w'_o$ in $\underline{Y}_r(\underline{W},w'_o)$ the solution, $w'_o = w'_o(\underline{W}_o)$ of Eq.(74) we obtain the $\underline{Y}_r(\underline{W},\underline{W}_o)$ presentation (Fig.10). A similar consideration of the reference *scanning swelling* curve, $\hat{Y}_r(\hat{\underline{W}},\hat{\underline{W}}_o)$ (Fig.11, curve 3) leads to the following presentation (at $\hat{U}_i = U_i$, $\hat{K} = K$, and $\hat{w}' = w'$; cf. Eqs.(69) and (70))

$\hat{\underline{Y}}_r(w',w'_o) = \hat{\underline{U}}'(w',w'_o) + U_i + U_s$,   $0 \leq w'_o \leq w' \leq w'_h$, (75a)

$\hat{\underline{W}}(w',w'_o) = w' + \hat{\underline{\omega}}(w',w'_o)$,   $0 \leq w'_o \leq w' \leq w'_h$, (75b)

where $w'_o$, or $w_o$, or $\hat{\underline{W}}_o$ is the starting point of the reference scanning swelling curve (Fig.11, curve 3); $U_i$ and $U_s$ were indicated above [24]; and

$\hat{\underline{U}}'(w',w'_o) = \hat{\underline{U}}(w,w_o)/K$,   $0 \leq w'_o \leq w' \leq w'_h$   ($w=w'K$, $w_o=w'_oK$) (76)

(cf. Eqs.(60b) and (71); for $\hat{\underline{U}}(w,w_o)$ see Section 3.1.2). By analogy to the expression for $\hat{\omega}(\hat{w}')$ [24] in Eq.(65b) the $\hat{\underline{\omega}}(w',w'_o)$ contribution to the water content in Eq.(75b) can be written as (cf. Eq.(72))

$$\hat{\underline{\omega}}(w',w'_o) = \begin{cases} 0, & w'_o \leq w' < \underline{w}'_b(w'_o) < w'_h \\ \rho_w U_i \Pi_h F_i(\eta(\hat{\underline{R}}(w',w'_o)),\Pi_h), & w'_o < \underline{w}'_b(w'_o) \leq w' < w'_h \end{cases}, \quad (77)$$

where (for $F_i(\eta,\Pi_h)$ see Eqs.(65d) and (65e))

$\eta(\hat{\underline{R}}(w',w'_o)) = (\hat{\underline{R}} - R_{min})/(R_m - R_{min})$,   $R_{min} < \hat{\underline{R}}(w',w'_o) < R_m$ . (78)

$\hat{\underline{R}}(w',w'_o)$ (see below) is the maximum internal size of water-filled clay pores of the intra-aggregate matrix along the *reference scanning swelling* curve (Fig.11, curve 3). The $\underline{w}'_b(w'_o)$ point (Eq.(77)) is determined from the condition, $\hat{\underline{R}}(\underline{w}'_b,w'_o)=R_{min}$ (cf. $w'_b$ in [24] and Eq.(65i)).

The starting point of the scanning swelling curve in terms of $w'_o$ in Eqs.(75a) and (75b) is connected with $\hat{\underline{W}}_o$ in $\hat{\underline{Y}}_r(\hat{\underline{W}},\hat{W}_o)$ (Fig.11, curve 3). Excluding $w'$ from Eqs.(75a) and (75b) we obtain $\hat{\underline{Y}}_r(\hat{\underline{W}},w'_o)$. According to Fig.11 at $\hat{\underline{W}}=\hat{W}_o$ one has

$$\hat{\underline{Y}}_r(\hat{\underline{W}}_o,w'_o)=Y_r(\hat{\underline{W}}_o) \ . \tag{79}$$

Substituting for $w'_o$ in $\hat{\underline{Y}}_r(\hat{\underline{W}},w'_o)$ the solution, $w'_o=w'_o(\hat{\underline{W}}_o)$ of Eq.(79) we obtain the $\hat{\underline{Y}}_r(\hat{\underline{W}},\hat{\underline{W}}_o)$ presentation (Fig.11, curve 3).

Now we should consider the maximum internal size, $\underline{R}(\zeta,\zeta_o)$ of water-filled clay pores along the *scanning shrinkage* curve, $\underline{v}(\zeta,\zeta_o)$ and the maximum internal size, $\hat{\underline{R}}(\zeta,\zeta_o)$ of water-filled clay pores along the *scanning swelling* curve, $\hat{\underline{v}}(\zeta,\zeta_o)$. $\underline{R}(\zeta,\zeta_o)$ and $\hat{\underline{R}}(\zeta,\zeta_o)$ determine $\underline{\omega}(w',w'_o)$ (Eq.(72)) and $\hat{\underline{\omega}}(w',w'_o)$ (Eq.(77); for interconnections $\zeta\leftrightarrow w'$, $\zeta_o\leftrightarrow w'_o$ see Eq.(48) and last paragraph of Section 3.1.2). Similar to Eqs.(67), (68a), and (68b) for $\hat{R}(\zeta)$ (see also [24]), in case of $\underline{R}(\zeta,\zeta_o)$ one can write the following equation

$$f(\eta(\underline{R}(\zeta,\zeta_o)),\underline{P}(\zeta,\zeta_o))=(1-v_s)\zeta/(\underline{v}(\zeta,\zeta_o)-v_s)+(\underline{v}(\zeta,\zeta_o)-v(\zeta))/(\underline{v}(\zeta,\zeta_o)-v_s) \ ,$$
$$0\leq\zeta<\zeta_o\leq\zeta_h \ . \tag{80}$$

The physical meaning of the terms of this equation is similar to that in Eq.(68a), but as applied to the *scanning shrinkage* curve, $\underline{v}(\zeta,\zeta_o)$ (instead of the *primary swelling* curve, $\hat{v}(\zeta)$ in Eq.(68a)). The clay porosity along the *scanning shrinkage* curve is

$$\underline{P}(\zeta,\zeta_o)=1-v_s/\underline{v}(\zeta,\zeta_o) \ , \qquad 0\leq\zeta<\zeta_o\leq\zeta_h \ . \tag{81}$$

$$f(\eta,\underline{P})=(1-(1-\underline{P})^{I_o(\eta)/8.4})/\underline{P} \ ; \qquad 0\leq\eta\leq 1 \ , \tag{82}$$

is the distribution function [42] with $I_o(\eta)$ from Eq.(63d). Finally,

$$\eta(\underline{R}(\zeta,\zeta_o))=(\underline{R}(\zeta,\zeta_o)-r_o(\zeta,\zeta_o))/(r_m(\zeta,\zeta_o)-r_o(\zeta,\zeta_o)) \ , \qquad 0\leq\zeta<\zeta_o\leq\zeta_h \tag{83}$$

(for $r_o$ and $r_m$ see Eq.(27)). Analogously one can write the equation for $\hat{\underline{R}}(\zeta,\zeta_o)$ as

$$f(\eta(\hat{\underline{R}}(\zeta,\zeta_o)),\hat{\underline{P}}(\zeta,\zeta_o))=(1-v_s)\zeta/(\hat{\underline{v}}(\zeta,\zeta_o)-v_s)+(\hat{\underline{v}}(\zeta,\zeta_o)-v(\zeta))/(\hat{\underline{v}}(\zeta,\zeta_o)-v_s) \ ,$$
$$0\leq\zeta_o<\zeta\leq\zeta_h \ . \tag{84}$$

$$\hat{\underline{P}}(\zeta,\zeta_o)=1-v_s/\hat{\underline{v}}(\zeta,\zeta_o) \ , \qquad 0\leq\zeta_o<\zeta\leq\zeta_h \tag{85}$$

is the clay porosity along the *scanning swelling* curve, $\hat{\underline{v}}(\zeta,\zeta_o)$.

$f(\eta, \underline{\hat{P}}) = (1-(1-\underline{\hat{P}})^{I_o(\eta)/8.4})/\underline{\hat{P}}$ ;     $0 \leq \eta \leq 1$ ,     (86)

with $I_o(\eta)$ from Eq.(63d). And (for $\hat{r}_o$ and $\hat{r}_m$ see Eq.(27))

$\eta(\underline{\hat{R}}(\zeta,\zeta_o)) = (\underline{\hat{R}}(\zeta,\zeta_o) - \hat{r}_o(\zeta,\zeta_o))/(\hat{r}_m(\zeta,\zeta_o) - \hat{r}_o(\zeta,\zeta_o))$ ,     $0 \leq \zeta_o < \zeta \leq \zeta_h$     (87)

One can see that the $\underline{R}(\zeta,\zeta_o)$ and $\underline{\hat{R}}(\zeta,\zeta_o)$ as functions of $\zeta$ should be arranged in Fig.4 of [24] between the $R(\zeta)$ and $\hat{R}(\zeta)$ curves as $R(\zeta) < \underline{R}(\zeta,\zeta_o) < \hat{R}(\zeta)$ at $0 \leq \zeta < \zeta_o \leq \zeta_h$ and $R(\zeta) < \underline{\hat{R}}(\zeta,\zeta_o) < \hat{R}(\zeta)$ at $0 \leq \zeta_o < \zeta \leq \zeta_h$. With that $\underline{R}(\zeta,\zeta_o)$ and $\underline{\hat{R}}(\zeta,\zeta_o)$ meet the intersection conditions with $R(\zeta)$ and $\hat{R}(\zeta)$ at $\zeta = \zeta_o$ as

$\underline{R}(\zeta,\zeta_o)|_{\zeta=\zeta_o} = \hat{R}(\zeta_o)$   and   $\underline{\hat{R}}(\zeta,\zeta_o)|_{\zeta=\zeta_o} = R(\zeta_o)$ .     (88)

Thus, the reference *scanning* shrinkage ($\underline{Y}_r(\underline{W},\underline{W}_o)$) and swelling ($\underline{\hat{Y}}_r(\underline{\hat{W}},\underline{\hat{W}}_o)$) curves *of a soil* can be, in the *single-valued* manner, predicted through the reference *primary* shrinkage ($Y_r(W)$) and *quasi primary* swelling ($\hat{Y}_r(\hat{W})$) curves *of the soil* (Section 4.1.1; the first paragraph of this section; Figs 10 and 11), *scanning* curves of the *intra-aggregate matrix* ($\underline{U}(w,w_o)$ and $\underline{\hat{U}}(w,w_o)$ entering Eqs.(71) and (76); Section 3.1.2), and *scanning* curves of the *contributive clay matrix* ($\underline{v}(\zeta,\zeta_o)$ and $\underline{\hat{v}}(\zeta,\zeta_o)$ entering Eqs.(80) and (84); Section 2.1.2). Note that, if necessary, the reference scanning curves of the soil can also be calculated for the *second quasi primary* shrink-swell cycle (Fig.9, curves 3 and 2) after the evident replacements, $Y_{rh} \rightarrow \hat{Y}_{rh}$, $W_h \rightarrow \hat{W}_h$, $K \rightarrow \hat{K}$, and $U_i \rightarrow \hat{U}_i$.

*4.1.3. The reference transitive and steady shrink-swell cycles of a soil*

The known reference *scanning* shrinkage ($\underline{Y}_r(\underline{W},\underline{W}_o)$) and swelling ($\underline{\hat{Y}}_r(\underline{\hat{W}},\underline{\hat{W}}_o)$) curves *of a soil* inside the *quasi primary* shrink-swell cycle (Figs.10 and 11) enable one to construct, in a *single-valued* manner, the reference *transitive* and *steady* shrink-swell cycles of *the soil* in a given water content range, $0 < W_1 < W < W_2 < W_h$ (for the *steady* cycle see Fig.12) in the total analogy to the consideration for a clay paste in Section 2.1.3 (Figs.4 and 5).

The qualitative picture of the *reference transitive* cycles of a soil is as in Fig.4 *after replacements*: $v(\zeta) \rightarrow Y_r(W)$, $\hat{v}(\zeta) \rightarrow \hat{Y}_r(\hat{W})$, $\underline{v}(\zeta,\zeta_o) \rightarrow \underline{Y}_r(\underline{W},\underline{W}_o)$, $\underline{\hat{v}}(\zeta,\zeta_o) \rightarrow \underline{\hat{Y}}_r(\underline{\hat{W}},\underline{\hat{W}}_o)$, $\zeta_1 = \zeta_{o1} \rightarrow W_1$, $\zeta_2 \rightarrow W_2$, $\zeta_{o2} \rightarrow W_{o2}$, $\zeta_{o3} \rightarrow W_{o3}$, $\zeta_{o4} \rightarrow W_{o4}$. Then Eq.(28) (at $\underline{W} = W_2$, $\underline{W}_o = W_{o2}$, $\underline{\hat{W}} = W_2$, and $\underline{\hat{W}}_o = W_1$) turns into condition at $\underline{W} = \underline{\hat{W}} = W_2$ as

$\underline{Y}_{r1}(W_2, W_{o2}) = \underline{\hat{Y}}_{r1}(W_2, W_1)$     (89)

to determine the starting point, $\underline{W}_o = W_{o2}$ of the reference *scanning shrinkage* curve in the *first transitive* cycle (Fig.4 after the replacements). Eq.(29) (at $\underline{\hat{W}} = W_1$, $\underline{\hat{W}}_o = W_{o3}$, $\underline{W} = W_1$, and $\underline{W}_o = W_{o2}$) turns into condition at $\underline{\hat{W}} = \underline{W} = W_1$ as

$$\hat{\underline{Y}}_{r2}(W_1,W_{o3}) = \underline{Y}_{r1}(W_1,W_{o2}) \tag{90}$$

to determine the starting point, $\hat{\underline{W}}_o = W_{o3}$ of the reference *scanning swelling* curve in the *second transitive* cycle (Fig.4 after the replacements). Eq.(30) (at $\underline{W} = W_2$, $\underline{W}_o = W_{o4}$, $\hat{\underline{W}} = W_2$, and $\hat{\underline{W}}_o = W_{o3}$) turns into condition at $\underline{W} = \hat{\underline{W}} = W_2$ as

$$\underline{Y}_{r2}(W_2, W_{o4}) = \hat{\underline{Y}}_{r2}(W_2, W_{o3}) \tag{91}$$

to determine the starting point, $\underline{W}_o = W_{o4}$ of the reference *scanning shrinkage* curve in the *second transitive* cycle (Fig.4 after the replacements).

Fig.12 shows the qualitative view of the reference *steady* shrink-swell cycle of the soil. In this case Eqs.(31) and (32) turns into conditions as

$$\underline{Y}_r(W_1, W_{o1}) = \hat{\underline{Y}}_r(W_1, W_{o2}) \tag{92}$$

(at $\underline{W} = \hat{\underline{W}} = W_1$, $\underline{W}_o = W_{o1}$, and $\hat{\underline{W}}_o = W_{o2}$) and

$$\underline{Y}_r(W_2, W_{o1}) = \hat{\underline{Y}}_r(W_2, W_{o2}) \tag{93}$$

(at $\underline{W} = \hat{\underline{W}} = W_2$, $\underline{W}_o = W_{o1}$, and $\hat{\underline{W}}_o = W_{o2}$) to determine the starting points, $\underline{W}_o = W_{o1}$ of the reference *scanning shrinkage* branch (Fig.12, curve 3) and $\hat{\underline{W}}_o = W_{o2}$ of the reference *scanning swelling* branch (Fig.12, curve 4) of the steady cycle (Fig.12).

### *4.2. Shrinkage-swelling and cracking of a soil layer with no loading*

In a soil *layer* of any thickness and sufficiently *large sample* the crack appearance and development at shrinkage-swelling is inevitable [24,28]. Below, since we are interested in shrinkage-swelling and cracking in a soil profile, only the case of a soil layer is considered. It is necessary to estimate the crack volume and cracked-soil volume along the *different* possible shrinkage and swelling curves of the soil layer. The different *reference* shrinkage and swelling curves (of a small sample with no cracking; Section 4.1) serve as an *important starting point* in such an estimation.

First (Sections 4.2.1-4.2.3), the *initial* thickness, $h$ of a layer (i.e., the thickness at the maximum soil swelling, $W=W_h$) is considered to be *known*. Then (Section 4.2.4) we show how it can be estimated in conditions of a soil profile (with no loading).

### *4.2.1. Multifold primary shrinkage-swelling of a cracked soil layer with no loading*

Below we briefly give the major points of the transition from the *reference primary* shrink-swell curves (i.e., the curves in the maximum possible water content range and with no cracking and loading; Section 4.1) to the *primary* shrink-swell curves of a *cracked* soil layer [24,28]. The points are necessary in the following.

(1) The *primary shrinkage* curve (of a soil layer) with *crack* contribution and no loading, $Y(W)$ (Fig.13) is evidently connected with the corresponding *reference primary shrinkage* curve, $Y_r(W)$ (Section 4.1.1; Fig.13) as

$$Y(W) = Y_r(W) + U_{cr}(W) - U_s, \qquad 0 < W < W_h \tag{94}$$

where $Y_r(W)$ is from Eq.(59a); the total water content, $W$ at primary shrinkage is calculated as indicated for $Y_r(W)$ (Eq.(64a)) [24]; the $U_{cr}(W) - U_s$ difference gives the additional specific volume of cracks, that develop of the structural pores at shrinkage, compared to the initial volume, $U_s$ of the pores. The specific crack volume at primary shrinkage, $U_{cr}(W)$ meets at the start of the first shrinkage the evident condition, as (see Fig.9, the start of curve 1 of the first reference primary shrinkage)

$$U_{cr}(W_h)=U_s . \tag{95}$$

To estimate $Y(W)$ from (Eq.(94)), first we should find $U_{cr}(W)$. According to its definition the crack factor, $q$ [28,24] connects the increments of the specific crack volume, $U_{cr}(W)$ of a soil layer and specific soil volume at reference shrinkage, $Y_r(W)$ (or the intra-aggregate matrix contribution, $U'$ to $Y_r$ since $dY_r=dU'$ from Eq.(59a)) as

$$dU_{cr}(W)=-q\,dY_r(W), \qquad 0<W<W_h \tag{96}$$

The crack factor in the *layer case* depends on the *initial* layer thickness, $h$ (i.e., at $W=W_h$) and characteristics of the aggregate size distribution at $W=W_h$ (the maximum and minimum aggregate sizes and inter-aggregate porosity) as [28]

$$q(h/h^*)=b_1(h/h^*)^2 , \qquad 0 \leq h/h^* \leq \delta \tag{97a}$$

$$q(h/h^*)=1-b_2/(h/h^*) , \qquad h/h^* \geq \delta \tag{97b}$$

where the universal constants are $b_1 \cong 0.15$, $b_2 \cong 1$, and $\delta \cong 1.5$. The critical layer thickness, $h^* \cong 2\text{-}5\,cm$ is determined as [28]

$$h^*=10^3(X_m/h^*_o)^3 h^*_o , \qquad h^*_o=(l_{min}l_m)^{1/2} \tag{98}$$

where $l_{min}$ and $l_m$ are the mean distances between the aggregates of the minimum, $X_{min}$ and maximum, $X_m$ size, respectively, at $W=W_h$. $l_{min}$ and $l_m$ are calculated as

$$l_{min}=(1-P_h)^{-1/3}(X_{min}+\Delta X/2)/[(dF/dX)_{|X=X_{min}+\Delta X/2}\Delta X]^{1/3} , \tag{99a}$$

$$l_m=(1-P_h)^{-1/3}(X_m-\Delta X/2)/[(dF/dX)_{|X=X_m-\Delta X/2}\Delta X]^{1/3} \tag{99b}$$

where
$$\Delta X=\max(0.05\,mm, X_{min}) \ll X_m \tag{99c}$$

and the aggregate-size distribution, $F(X, X_{min}, X_m, P_h)$ is from [42] according to Eq.(63d) at $\beta=P_h$ and

$$\eta=(X-X_{min})/(X_m-X_{min}) , \qquad X_{min}<X<X_m . \tag{99d}$$

Eqs.(99a)-(99d)) are used in Section 4.3.5 when considering the effects of loading.

Eq.(96), after integrating with the condition from Eq.(95), gives the crack volume at the first primary shrinkage, $U_{cr}(W)$ through the specific volume, $Y_r(W)$ at reference primary shrinkage (Section 4.1.1) to be

$$U_{cr}(W,h/h^*)=q(h/h^*)(Y_{rh}-Y_r(W))+U_s , \qquad 0<W<W_h . \tag{100}$$

The substitution for $U_{cr}(W)$ in Eq.(94) from Eq.(100) gives the link, $Y(W) \leftrightarrow Y_r(W)$ as

$$Y(W,h/h^*)=(1-q(h/h^*))Y_r(W)+q(h/h^*)Y_{rh}, \qquad 0<W<W_h. \qquad (101)$$

Note that using Eq.(59a) one can express $U_{cr}(W,h/h^*)$ (Eq.(100)) and $Y(W,h/h^*)$ (Eq.(101)) through $U(W)$ (Section 3.1.1) instead of $Y_r(W)$ [24]. Fig.13 illustrates the primary shrinkage curves with crack contribution, $Y(W)$ ($\equiv Y(W,h/h^*)$) and with no cracks, $Y_r(W)$. The $Y(W)$-$Y_r(W)$ difference gives the crack volume variation with water content at the first primary shrinkage (see Eq.(94)).

(2) The *primary swelling curve* (of the soil layer) with *crack* contribution and no loading, $\hat{Y}(\hat{W})$ (Fig.13) is connected with the corresponding *reference primary swelling* curve, $\hat{Y}_r(\hat{W})$ (Section 4.1.1; Fig.13) as (cf. Eq.(94))

$$\hat{Y}(\hat{W})=\hat{Y}_r(\hat{W})+\hat{U}_{cr}(\hat{W})-U_s, \qquad 0<\hat{W}<\hat{W}_h \qquad (102)$$

where $\hat{Y}_r(\hat{W})$ is from Eq.(59b); the total water content at primary swelling, $\hat{W}$ is calculated as indicated for $\hat{Y}_r(\hat{W})$ (Eq.(64b)) [24]; the $\hat{U}_{cr}(\hat{W})$-$U_s$ difference gives the resulting specific volume of cracks that develop of structural pores at shrinkage in the $0<W<W_h$ range and following swelling up to the water content $\hat{W}>0$ (Fig.13), compared to the initial volume, $U_s$. The condition for $\hat{U}_{cr}(\hat{W})$ at $\hat{W}=W=0$ (Fig.13) follows from Eq.(100) accounting for Eqs.(59a) and (60a) as

$$\hat{U}_{cr}(0)=U_{cr}(0)=q(h/h^*)(U_h-U_z)/K+U_s. \qquad (103)$$

To estimate $\hat{Y}(\hat{W})$ from Eq.(102), we should find $\hat{U}_{cr}(\hat{W})$. To this end we introduce the concept of the crack factor, $\hat{q}$ at swelling using the relations for $q$ (Eqs.(97a), (97b), (98), (99a)-(99d)), but after the replacements: $X_m \to \hat{X}_m$, $l_m \to \hat{l}_m$, $l_{min} \to \hat{l}_{min}$, $h^*_o \to \hat{h}^*_o$, $h^* \to \hat{h}^*$. These replacements reflect the aggregate destruction when transiting from shrinkage to swelling at $W \cong 0$ (in [24] the differences between $q$ and $\hat{q}$ were not taken into account for simplicity). The indicated expression for $\hat{q}(h/\hat{h}^*)$ will be needed in Section 4.3.5 when considering the $\hat{q}$ variation under loading. Relying on the definition of the crack factor, $\hat{q}(h/\hat{h}^*)$ (of the layer of initial thickness $h$) we can link the increments of the specific crack volume, $\hat{U}_{cr}(\hat{W})$ (of the soil layer) and the specific soil volume at reference swelling $\hat{Y}_r(\hat{W})$ (or the intra-aggregate matrix contribution, $\hat{U}'$ to $\hat{Y}_r$ since $d\hat{Y}_r=d\hat{U}'$ from Eq.(59b)) as

$$d\hat{U}_{cr}(\hat{W})=-\hat{q}(h/\hat{h}^*)d\hat{Y}_r(\hat{W}), \qquad 0<\hat{W}<\hat{W}_h. \qquad (104)$$

Eq.(104) after integrating with condition from Eq.(103) gives the crack volume, $\hat{U}_{cr}(\hat{W})$ at the first primary swelling following primary shrinkage (Fig.13) through the specific volume, $\hat{Y}_r(\hat{W})$ at reference primary swelling (Section 4.1.1) as

$$\hat{U}_{cr}(\hat{W},h/h^*,h/\hat{h}^*)=q(h/h^*)(U_h-U_z)/K+U_s-\hat{q}(h/\hat{h}^*)(\hat{Y}_r(\hat{W})-Y_{rz}), \qquad 0<\hat{W}<\hat{W}_h. \qquad (105)$$

The replacement of $\hat{U}_{cr}(\hat{W})$ in Eq.(102) with its expression from Eq.(105) gives the link, $\hat{Y}(\hat{W}) \leftrightarrow \hat{Y}_r(\hat{W})$ as

$$\hat{Y}(\hat{W},h/h^*,h/\hat{h}^*) = (1 - \hat{q}(h/\hat{h}^*))\hat{Y}_r(\hat{W}) + q(h/h^*)(U_h - U_z)/K + \hat{q}(h/\hat{h}^*)Y_{rz},$$
$$0 < \hat{W} < \hat{W}_h . \tag{106}$$

Note that using Eq.(59b) we can express $\hat{U}_{cr}(\hat{W},h/h^*,h/\hat{h}^*)$ (Eq.(105)) and $\hat{Y}(\hat{W},h/h^*,h/\hat{h}^*)$ (Eq.(106)) through $\hat{U}(\hat{W})$ (Section 3.1.1) instead of $\hat{Y}_r(\hat{W})$ [24]. The $\hat{Y}(\hat{W})$-$\hat{Y}_r(\hat{W})$ difference in Fig.13 ($\hat{Y}(\hat{W}) \equiv \hat{Y}(\hat{W},h/h^*,h/\hat{h}^*)$) gives the crack volume variation with water content at the first primary swelling (see Eq.(102)).

Thus, knowing the *reference primary* shrinkage and swelling curves of a soil, $Y_r(W)$ and $\hat{Y}_r(\hat{W})$ (Section 4.1.1), we can, in a *single-valued* manner, find the *primary* shrinkage and swelling curves of a *cracked* soil layer, $Y(W,h/h^*)$ and $\hat{Y}(\hat{W},h/h^*,h/\hat{h}^*)$ as well as the *crack* volume, $U_{cr}(W,h/h^*)$ and $\hat{U}_{cr}(\hat{W},h/h^*,h/\hat{h}^*)$ along the curves. Note that omission of the subscript "r" and word "reference" in Fig.9 and its caption, throughout, enables the use of Fig.9 to also illustrate the *multiple primary cycles* for curves, $Y(W,h/h^*)$ and $\hat{Y}(\hat{W},h/h^*,h/\hat{h}^*)$ with *crack* contribution.

*4.2.2. Scanning shrinkage-swelling of a cracked soil layer with no loading*

By definition, any *scanning* shrinkage and swelling curves vary in a range, $\Delta W < W_h$ and $\Delta \hat{W} < \hat{W}_h$. For this reason, similar to the case of the *reference scanning* shrinkage and swelling curves (see the first paragraph of Section 4.1.2), in consideration of the *scanning* shrinkage and swelling curves with *crack contribution*, we use the *quasi primary* shrink-swell cycle of a *cracked* soil layer (instead of the primary cycle from Section 4.2.1). That includes the *quasi primary* swelling curve, $\hat{\tilde{Y}}(\hat{W})$ (Fig.13) corresponding to the absence of the aggregate destruction when $\hat{U}_i = U_i$, $\hat{K} = K$, $\hat{X}_m = X_m$, $\hat{h}^* = h^*$, and $\hat{q} = q$ (see also Figs.9-11 after the mental transformation that was indicated at the end of Section 4.2.1 as applied to Fig.9). The $\hat{\tilde{Y}}(\hat{W})$ curve (curve 2' in Figs.9-11 after the above indicated mental transformation) can be written as ($\hat{\tilde{Y}}(\hat{W}) \equiv \hat{\tilde{Y}}(\hat{W},h/h^*)$)

$$\hat{\tilde{Y}}(\hat{W}) = \hat{\tilde{Y}}_r(\hat{W}) + \hat{\tilde{U}}_{cr}(\hat{W}) - U_s, \qquad 0 < \hat{W} < \hat{W}_h \tag{107}$$

($\hat{\tilde{Y}}_r$ is calculated as $\hat{Y}_r$ at $\hat{U}_i = U_i$ and $\hat{K} = K$ in Section 4.1.1). The crack volume, $\hat{\tilde{U}}_{cr}(\hat{W},h/h^*)$ (of the soil layer) along the quasi primary swelling curve, $\hat{\tilde{Y}}(\hat{W},h/h^*)$ and this curve itself are obtained from Eqs.(105) and (106) at $\hat{U}_{cr} \to \hat{\tilde{U}}_{cr}$, $\hat{Y} \to \hat{\tilde{Y}}$, $\hat{Y}_r \to \hat{\tilde{Y}}_r$, and $\hat{q} \to q$ as

$$\hat{\tilde{U}}_{cr}(\hat{W},h/h^*) = q(h/h^*)(U_h - U_z)/K + U_s - q(h/h^*)(\hat{\tilde{Y}}_r(\hat{W}) - Y_{rz}), \qquad 0 < \hat{W} < \hat{W}_h, \tag{108}$$

$$\hat{\tilde{Y}}(\hat{W},h/h^*) = (1 - q(h/h^*))\hat{\tilde{Y}}_r(\hat{W}) + q(h/h^*)(U_h - U_z)/K + q(h/h^*)Y_{rz}, \qquad 0 < \hat{W} < \hat{W}_h. \tag{109}$$

The *scanning shrinkage* curves of the *quasi primary* shrink-swell cycle with *crack* contribution, $\underline{Y}(\underline{W},\underline{W}_o)$ (see curve 3 in Fig.10 after the above indicated mental transformation) and with *no cracking*, $\underline{Y}_r(\underline{W},\underline{W}_o)$ (Section 4.1.2; curve 3 in Fig.10) evidently differ by the crack contribution as (cf. Eqs.(94) and (102))

$$\underline{Y}(\underline{W},\underline{W}_o)=\underline{Y}_r(\underline{W},\underline{W}_o)+\underline{U}_{cr}(\underline{W},\underline{W}_o)-U_s, \qquad 0<\underline{W}<\underline{W}_o<W_h \qquad (110)$$

where $\underline{Y}_r(\underline{W},\underline{W}_o)$ is from Eq.(69); $\underline{W}$ and $\underline{W}_o$ are again calculated from Eqs.(70), (72)-(73), (80)-(83) as in the case of the reference scanning shrinkage curve (Section 4.1.2); $\underline{U}_{cr}(\underline{W},\underline{W}_o)-U_s$ gives the crack volume (minus its initial value, the structural pore volume, $U_s$) along the scanning shrinkage curve, $\underline{Y}(\underline{W},\underline{W}_o)$ (curve 3 in Fig.10 with $\underline{Y}_r \to \underline{Y}$, $Y_r \to Y$, and $\hat{\underline{Y}}_r \to \hat{\underline{Y}}$) that starts at $\underline{W}=\underline{W}_o$. To estimate $\underline{Y}(\underline{W},\underline{W}_o)$ (Eq.(110)) for a soil layer we should first find $\underline{U}_{cr}(\underline{W},\underline{W}_o)$. Again, using the definition of the crack factor, $q(h/h^*)$ (of the soil layer of the initial thickness, $h$) we have the differential relation for $\underline{U}_{cr}$ as

$$d\underline{U}_{cr}(\underline{W},\underline{W}_o)=-q(h/h^*)d\underline{Y}_r(\underline{W},\underline{W}_o), \qquad 0<\underline{W}<\underline{W}_o<W_h. \qquad (111)$$

The crack volume, $\underline{U}_{cr}(\underline{W},\underline{W}_o,h/h^*)$ along the scanning shrinkage curve, $\underline{Y}(\underline{W},\underline{W}_o,h/h^*)$ (curve 3 in Fig.10 with replacements: $\underline{Y}_r \to \underline{Y}$, $Y_r \to Y$, and $\hat{\underline{Y}}_r \to \hat{\underline{Y}}$) and crack volume, $\hat{U}_{cr}(\hat{W},h/h^*)$ (Eq.(108)) along the quasi primary swelling curve, $\hat{\underline{Y}}(\hat{W},h/h^*)$ (Eq.(109); curve 2' in Fig.10 with the replacements) should evidently coincide at $\underline{W}=\hat{W}=\underline{W}_o$ as

$$\underline{U}_{cr}(\underline{W},\underline{W}_o,h/h^*)|_{\underline{W}=\underline{W}_o}=\hat{U}_{cr}(\hat{W},h/h^*)|_{\hat{W}=\underline{W}_o}, \qquad 0<\underline{W}_o<W_h. \qquad (112)$$

Here $\hat{U}_{cr}(\hat{W},h/h^*)$ is given by Eq.(108) with $\hat{Y}_r(\hat{W})$ from Eqs.(59b), (60b), and (61b) at $\hat{U}_i=U_i$ and $\hat{K}=K$. Integrating Eq.(111) and accounting for the condition given by Eq.(112) as well as the similar condition, $\underline{Y}_r(\underline{W}_o,\underline{W}_o)=\hat{\underline{Y}}_r(\underline{W}_o)$ for the specific soil volumes with no cracks at $\underline{W}=\hat{W}=\underline{W}_o$, we obtain the crack volume, $\underline{U}_{cr}(\underline{W},\underline{W}_o,h/h^*)$ to be

$$\underline{U}_{cr}(\underline{W},\underline{W}_o,h/h^*)=q(h/h^*)(\hat{\underline{Y}}_r(\underline{W}_o)-\underline{Y}_r(\underline{W},\underline{W}_o))+\hat{U}_{cr}(\underline{W}_o,h/h^*),$$
$$0<\underline{W}<\underline{W}_o<W_h. \qquad (113)$$

Here $\hat{U}_{cr}(\underline{W}_o,h/h^*)$ in the right part is the crack volume that has *already existed* in the starting point, $\underline{W}=\underline{W}_o$ of the scanning shrinkage curve, $\underline{Y}(\underline{W},\underline{W}_o,h/h^*)$ (curve 3 in Fig.10 with the replacements). The term with $q$ in the right part of Eq.(113) gives the positive *additional contribution* to the crack volume, $\underline{U}_{cr}(\underline{W},\underline{W}_o,h/h^*)$ (at a given $\underline{W} < \underline{W}_o$) in the process of scanning shrinkage. Substitution for $\underline{U}_{cr}(\underline{W},\underline{W}_o,h/h^*)$ in Eq.(110) from Eq.(113) leads to $\underline{Y}(\underline{W},\underline{W}_o,h/h^*)$ as

$$\underline{Y}(\underline{W},\underline{W}_o,h/h^*)=(1-q(h/h^*))\underline{Y}_r(\underline{W},\underline{W}_o)+q(h/h^*)\hat{\underline{Y}}_r(\underline{W}_o)+\hat{\underline{U}}_{cr}(\underline{W}_o,h/h^*)-U_s,$$
$$0<\underline{W}<\underline{W}_o<W_h. \qquad (114)$$

Using Eq.(69) one can express $\underline{U}_{cr}(\underline{W},\underline{W}_o,h/h^*)$ and $\underline{Y}(\underline{W},\underline{W}_o,h/h^*)$ through $\underline{U}(\underline{W},\underline{W}_o)$ instead of $\underline{Y}_r(\underline{W},\underline{W}_o)$. According to Eqs.(113), (114), and (108) at $q\to 0$ (when there are no cracks) $\hat{\underline{U}}_{cr}\to U_s$, $\underline{U}_{cr}\to U_s$, and $\underline{Y}\to\underline{Y}_r$ as it should be.

The *scanning swelling* curves of the *quasi primary* shrink-swell cycle with *crack contribution*, $\hat{\underline{Y}}(\hat{\underline{W}},\hat{\underline{W}}_o)$ (see curve 3 in Fig.11 with $\hat{\underline{Y}}_r\to\hat{\underline{Y}}$, $Y_r\to Y$, and $\hat{\hat{Y}}_r\to\hat{\hat{Y}}$) and with *no cracking*, $\hat{\underline{Y}}_r(\hat{\underline{W}},\hat{\underline{W}}_o)$ (Section 4.1.2; curve 3 in Fig.11) differ by the crack contribution as (cf. Eqs.(94), (102), (107), and (110))

$$\hat{\underline{Y}}(\hat{\underline{W}},\hat{\underline{W}}_o)=\hat{\underline{Y}}_r(\hat{\underline{W}},\hat{\underline{W}}_o)+\hat{\underline{U}}_{cr}(\hat{\underline{W}},\hat{\underline{W}}_o)-U_s, \qquad 0<\hat{\underline{W}}_o<\hat{\underline{W}}<W_h \qquad (115)$$

where $\hat{\underline{Y}}_r(\hat{\underline{W}},\hat{\underline{W}}_o)$ is from Eq.(75a); $\hat{\underline{W}}$ and $\hat{\underline{W}}_o$ are again found from Eqs.(75b), (77), (78), (84)-(87) as in the case of the reference scanning swelling curve (Section 4.1.2); $\hat{\underline{U}}_{cr}(\hat{\underline{W}},\hat{\underline{W}}_o)-U_s$ is the crack volume (minus its initial value, $U_s$) along the scanning swelling curve, $\hat{\underline{Y}}(\hat{\underline{W}},\hat{\underline{W}}_o)$ (curve 3 in Fig.11 with replacements: $\hat{\underline{Y}}_r\to\hat{\underline{Y}}$, $Y_r\to Y$, and $\hat{\hat{Y}}_r\to\hat{\hat{Y}}$) that starts at $\hat{\underline{W}}=\hat{\underline{W}}_o$. To find $\hat{\underline{Y}}(\hat{\underline{W}},\hat{\underline{W}}_o)$ (Eq.(115)) for a soil layer one should first estimate $\hat{\underline{U}}_{cr}(\hat{\underline{W}},\hat{\underline{W}}_o)$. Using the definition of the crack factor, $q(h/h^*)$ (of the soil layer of the initial thickness, $h$) we have for $\hat{\underline{U}}_{cr}$ (cf. Eq.(111))

$$d\hat{\underline{U}}_{cr}(\hat{\underline{W}},\hat{\underline{W}}_o)=-q(h/h^*)d\hat{\underline{Y}}_r(\hat{\underline{W}},\hat{\underline{W}}_o), \qquad 0<\hat{\underline{W}}_o<\hat{\underline{W}}<W_h \qquad (116)$$

One can also use $\hat{\underline{U}}'$ instead of $\hat{\underline{Y}}_r$ since $d\hat{\underline{Y}}_r = d\hat{\underline{U}}'$ from Eq.(75a). The crack volume, $\hat{\underline{U}}_{cr}(\hat{\underline{W}},\hat{\underline{W}}_o,h/h^*)$ along the scanning swelling curve, $\hat{\underline{Y}}(\hat{\underline{W}},\hat{\underline{W}}_o,h/h^*)$ (curve 3 in Fig.11 with the above indicated replacements) should evidently coincide with the crack volume, $U_{cr}(W,h/h^*)$ from Eq.(100) along the primary shrinkage curve, $Y(W,h/h^*)$ (Eq.(101); curve 1 in Fig.11 with the above indicated replacements) at $\hat{\underline{W}}=W=\hat{\underline{W}}_o$ as

$$\hat{\underline{U}}_{cr}(\hat{\underline{W}},\hat{\underline{W}}_o,h/h^*)|_{\hat{\underline{W}}=\hat{\underline{W}}_o}=U_{cr}(W,h/h^*)|_{W=\hat{\underline{W}}_o}, \qquad 0<\hat{\underline{W}}_o<W_h. \qquad (117)$$

Integrating Eq.(116) and accounting for the condition from Eq.(117) as well as the similar condition, $\hat{\underline{Y}}_r(\hat{\underline{W}}_o,\hat{\underline{W}}_o)=Y_r(\hat{\underline{W}}_o)$ for the specific soil volumes with no cracks at $\hat{\underline{W}}=W=\hat{\underline{W}}_o$, we obtain the crack volume, $\hat{\underline{U}}_{cr}(\hat{\underline{W}},\hat{\underline{W}}_o,h/h^*)$ as

$$\hat{\underline{U}}_{cr}(\hat{\underline{W}},\hat{\underline{W}}_o,h/h^*)=-q(h/h^*)(\hat{\underline{Y}}_r(\hat{\underline{W}},\hat{\underline{W}}_o)-Y_r(\hat{\underline{W}}_o))+U_{cr}(\hat{\underline{W}}_o,h/h^*),$$

$$0 < \underline{\hat{W}}_o < \underline{\hat{W}} < W_h \ . \tag{118}$$

$U_{cr}(\underline{\hat{W}}_o, h/h^*)$ in the right part of Eq.(118) is the crack volume that *already existed* in the starting point, $\underline{\hat{W}} = \underline{\hat{W}}_o$ of the scanning swelling curve, $\underline{\hat{Y}}(\underline{\hat{W}}, \underline{\hat{W}}_o, h/h^*)$ (curve 3 in Fig.11 with the above indicated replacements). The term with $q(h/h^*)$ in the right part gives the negative *addition* to the crack volume, $\underline{\hat{U}}_{cr}(\underline{\hat{W}}, \underline{\hat{W}}_o, h/h^*)$ (at a given $\underline{\hat{W}} > \underline{\hat{W}}_o$) in the process of scanning swelling. Substitution for $\underline{\hat{U}}_{cr}(\underline{\hat{W}}, \underline{\hat{W}}_o, h/h^*)$ in Eq.(115) from Eq.(118) gives $\underline{\hat{Y}}(\underline{\hat{W}}, \underline{\hat{W}}_o, h/h^*)$ as

$$\underline{\hat{Y}}(\underline{\hat{W}}, \underline{\hat{W}}_o, h/h^*) = (1-q(h/h^*)\underline{\hat{Y}}_r(\underline{\hat{W}}, \underline{\hat{W}}_o) + q(h/h^*)Y_r(\underline{\hat{W}}_o) + U_{cr}(\underline{\hat{W}}_o, h/h^*) - U_s ,$$
$$0 < \underline{\hat{W}}_o < \underline{\hat{W}} < W_h \tag{119}$$

where $\underline{\hat{Y}}_r(\underline{\hat{W}}, \underline{\hat{W}}_o, h/h^*)$ is from Eq.(75a), $Y_r(\underline{\hat{W}}_o)$ is from Eq.(59a), and $U_{cr}(\underline{\hat{W}}_o, h/h^*)$ is from Eq.(100). Using Eq.(75a) we can express $\underline{\hat{U}}_{cr}(\underline{\hat{W}}, \underline{\hat{W}}_o, h/h^*)$ (Eq.(118)) and $\underline{\hat{Y}}(\underline{\hat{W}}, \underline{\hat{W}}_o, h/h^*)$ (Eq.(119)) through $\underline{\hat{U}}(\underline{\hat{W}}, \underline{\hat{W}}_o)$ instead of $\underline{\hat{Y}}_r(\underline{\hat{W}}, \underline{\hat{W}}_o)$. According to Eqs.(118), (119), and (100) at $q \to 0$ (when there are no cracks) $U_{cr} \to U_s$, $\hat{U}_{cr} \to U_s$, and $\underline{\hat{Y}} \to \underline{\hat{Y}}_r$ as it should be. Thus, knowing the *reference scanning* shrinkage and swelling curves, $\underline{Y}_r(\underline{W}, \underline{W}_o)$ (curve 3 in Fig.10) and $\underline{\hat{Y}}_r(\underline{\hat{W}}, \underline{\hat{W}}_o)$ (curve 3 in Fig.11) we can, in the *single-valued* manner, find the *scanning* shrinkage and swelling curves of a *cracked* layer, $\underline{Y}(\underline{W}, \underline{W}_o, h/h^*)$ and $\underline{\hat{Y}}(\underline{\hat{W}}, \underline{\hat{W}}_o, h/h^*)$ and the *crack* volume $U_{cr}(\underline{W}, \underline{W}_o, h/h^*)$ and $\hat{U}_{cr}(\underline{\hat{W}}, \underline{\hat{W}}_o, h/h^*)$ along the curves.

*4.2.3. Shrink-swell cycles and crack volume hysteresis in a soil layer with no loading*

Passing from the *scanning* curves with *crack* contribution, $\underline{Y}(\underline{W}, \underline{W}_o, h/h^*)$ (Eq.(114)) and $\underline{\hat{Y}}(\underline{\hat{W}}, \underline{\hat{W}}_o, h/h^*)$ (Eq.(119)) to the corresponding *transitive* and *steady shrink-swell cycles* of the *cracked* soil layer, one should note that in this case (of a cracked layer) both Eqs.(89)-(91) which determine the characteristics of the *successive reference transitive* shrink-swell cycles (cf. Fig.4) and Eqs.(92) and (93) which determine the characteristics of the *reference steady* cycle (Fig.12) are totally kept, obviously omitting the "r" subscript.

Let us consider in more detail the *steady* shrink-swell cycle of the *cracked* layer and the *volume* variation of accompanying *cracks*. Accepting in Eq.(113) $\underline{W} \equiv W$ and $\underline{W}_o \equiv W_{o1}$, we can write the specific *crack* volume, $\underline{U}_{cr}(W, W_{o1}, h/h^*)$ along the *shrinkage stage* of the *steady* shrink-swell cycle (curve 3 in Fig.12) as

$$\underline{U}_{cr}(W, W_{o1}, h/h^*) = Y_o(W_{o1}, h/h^*) - q(h/h^*)\underline{Y}_r(W, W_{o1}) , \quad 0 < W_1 < W < W_2 < W_{o1} < W_h \tag{120}$$

where

$$Y_o(W_{o1}, h/h^*) \equiv q(h/h^*)\hat{\underline{Y}}_r(W_{o1}) + \hat{\underline{U}}_{cr}(W_{o1}, h/h^*) . \tag{121}$$

Similarly, accepting in Eq.(118) $\hat{W} \equiv W$ and $\hat{W}_o \equiv W_{o2}$, we can write the specific *crack* volume, $\hat{U}_{cr}(W, W_{o2}, h/h^*)$ along the *swelling stage* of the *steady* shrink-swell cycle (curve 4 in Fig.12) as

$$\hat{\underline{U}}_{cr}(W,W_{o2},h/h^*) = \hat{Y}_o(W_{o2},h/h^*) - q(h/h^*)\hat{\underline{Y}}_r(W,W_{o2}), \quad 0<W_{o2}<W_1<W<W_2<W_h \quad (122)$$

where

$$\hat{Y}_o(W_{o2},h/h^*) \equiv q(h/h^*)Y_r(W_{o2}) + U_{cr}(W_{o2},h/h^*) . \quad (123)$$

Since, by definition and physical meaning of the steady cycle (Fig.12), $\underline{U}_{cr}(W,W_{o1},h/h^*)$ and $\underline{Y}_r(W,W_{o1})$ entering Eq.(120) coincide with $\hat{\underline{U}}_{cr}(W,W_{o2},h/h^*)$ and $\hat{\underline{Y}}_r(W,W_{o2},h/h^*)$ entering Eq.(122), respectively, at $W=W_1$ (as well as at $W=W_2$) (see Fig.12), the $Y_o$ term (Eqs.(120) and (121)) and $\hat{Y}_o$ term (Eqs.(122) and (123)) should also coincide as

$$Y_o(W_{o1},h/h^*) = \hat{Y}_o(W_{o2},h/h^*) . \quad (124)$$

The expressions for $\underline{U}_{cr}(W,W_{o1},h/h^*)$ (Eq.(120)) and $\hat{\underline{U}}_{cr}(W,W_{o2},h/h^*)$ (Eq.(122)) (accounting for Eqs.(119), (123), and (124)) through $\underline{Y}_r(W,W_{o1})$ and $\hat{\underline{Y}}_r(W,W_{o2})$, respectively, enable us to present in Fig.14 the qualitative view of the crack volume evolution during the *steady* shrink-swell cycle in the *cracked* soil layer, using the *reference steady* shrink-swell cycle in Fig.12. The c*rack* volume at the *shrinkage stage*, $\underline{U}_{cr}(W,W_{o1},h/h^*)$ (curve 3 in Fig.14) corresponds to the *reference scanning shrinkage* curve, $\underline{Y}_r(W,W_{o1})$ (curve 3 in Fig.12; Eq.(120)) and to the *scanning shrinkage* curve of *cracked* soil, $\underline{Y}(W,W_{o1},h/h^*)$ (see below and curve 3 in Fig.12 with mental omission of the "r" subscript throughout). Similarly, the c*rack* volume at the *swelling stage*, $\hat{\underline{U}}_{cr}(W,W_{o2},h/h^*)$ (curve 4 in Fig.14) corresponds to the *reference scanning swelling* curve, $\hat{\underline{Y}}_r(W,W_{o2})$ (curve 4 in Fig.12; Eq.(122)) and to the *scanning swelling* curve of *cracked* soil, $\hat{\underline{Y}}(W,W_{o2},h/h^*)$ (see below and curve 4 in Fig.12 with mental omission of the "r" subscript throughout).

Thus, the *hysteretic steady* shrink-swell cycle of the specific *crack* volume (Fig.14) in the *single-valued* manner flows out of the *reference* (i.e., with no cracks) *steady* shrink-swell cycle of soil volume (Fig.12; Eqs.(120)-(124)) accounting for the $q$ factor value. The minimum crack volume is obtained from Eqs.(120) or (122) at $W=W_2$ (Fig.14) as

$$\underline{U}_{cr\,min} = \underline{U}_{cr}(W_2,W_{o1},h/h^*) = \hat{\underline{U}}_{cr}(W_2,W_{o2},h/h^*)$$

$$= Y_o(W_{o1},h/h^*) - q(h/h^*)\underline{Y}_r(W_2,W_{o1}) = Y_o(W_{o1},h/h^*) - q(h/h^*)\hat{\underline{Y}}_r(W_2,W_{o2}) . \quad (125)$$

The maximum crack volume is obtained from the same equations at $W=W_1$ (Fig.14) as

$$\underline{U}_{cr\,max} = \underline{U}_{cr}(W_1,W_{o1},h/h^*) = \hat{\underline{U}}_{cr}(W_1,W_{o2},h/h^*)$$

$$= Y_o(W_{o1},h/h^*) - q(h/h^*)\underline{Y}_r(W_1,W_{o1}) = Y_o(W_{o1},h/h^*) - q(h/h^*)\hat{\underline{Y}}_r(W_1,W_{o2}) . \quad (126)$$

The *hysteretic* variation of the *crack* volume, $\Delta\underline{U}_{cr}(W)$ at $W_1<W<W_2$ (Fig.14) (and at given $W_{o1}$ and $W_{o2}$ in Fig.12) is found from Eqs.(120), (122), and (124) as

$$\Delta \underline{U}_{cr}(W) = \underline{U}_{cr}(W,W_{o1},h/h^*) - \hat{\underline{U}}_{cr}(W,W_{o2},h/h^*)$$

$$= q(h/h^*)(\hat{\underline{Y}}_r(W,W_{o2}) - \underline{Y}_r(W,W_{o1})), \qquad W_1 < W < W_2. \qquad (127)$$

Besides $\underline{U}_{cr\,min}$ (Eq.(125); Fig.14) and $\underline{U}_{cr\,max}$ (Eq.(126); Fig.14), the *maximum hysteretic* variation of the crack volume, $\Delta \underline{U}_{cr}(W_m)$ (Eq.(127); Fig.14) that can be found at each depth of the soil profile can essentially influence the soil hydraulic conductivity of the depth. The above stated allows one to estimate the crack volume in the soil profile based on the physical parameters of the soil (for the parameters see [24,28] and the end of Section 6).

The scanning shrinkage and swelling curves of a *cracked* soil layer, $\underline{Y}(W,W_{o1},h/h^*)$ and $\hat{\underline{Y}}(W,W_{o2},h/h^*)$ from Eqs.(114) and (119) in the case of a *steady* shrink-swell cycle can be written accounting for Eqs.(121), (123), and (124) as

$$\underline{Y}(W,W_{o1},h/h^*) = Y_o(W_{o1},h/h^*) - U_s + (1-q(h/h^*))\underline{Y}_r(W,W_{o1}), \quad 0 < W < W_{o1} < W_h, \qquad (128)$$

$$\hat{\underline{Y}}(W,W_{o2},h/h^*) = Y_o(W_{o1},h/h^*) - U_s + (1-q(h/h^*))\hat{\underline{Y}}_r(W,W_{o2}), \quad 0 < W_{o2} < W < W_h. \qquad (129)$$

These relations show that (as noted above) the *steady* shrink-swell cycle of a *cracked* soil layer ($\underline{Y}(W,W_{o1},h/h^*), \hat{\underline{Y}}(W,W_{o2},h/h^*)$) in the $W_1 < W < W_2$ range qualitatively repeats the *reference steady* shrink-swell cycle ($\underline{Y}_r(W,W_{o1}), \hat{\underline{Y}}_r(W,W_{o2})$) in Fig.12.

Thus, above mentioned shows that in the course of the *steady* shrink-swell cycle of a *cracked* soil layer in the water content range, $0 < W_1 < W < W_2 < W_h$ (Fig.12 with omission of the "r" subscript): (i) the *crack* volume is always *more* than *zero* (Fig.14); (ii) the *crack* volume varies between the *minimum* at $W=W_2$ and *maximum* at $W=W_1$ (Fig.14); and (iii) inside the water content range, $W_1 < W < W_2$ the *crack* volume at *drying* (see Fig.12, curve 3 and Fig.14, curve 3) is always *higher* than at *wetting* (see Fig.12, curve 4 and Fig.14, curve 4), that is the crack volume *hysteresis* exists.

Note that in connection with the possible water content variations in *different* ranges, $\Delta W < W_h$ at *different* depths, the "non-natural" situations are possible when the crack volume *grows* with the depth increase in some range. It is also noteworthy that possible erratic variations of the water content boundaries, $W_1$ and $W_2$ (see also the note after Eq.(32)) can complicate the hysteretic behavior of the crack volume (beyond the scope of this work).

If some previous soil drainage leads to $W=0$, and the aggregate destruction occurs at following wetting (the possible situation at a depth of soil profile), one should consider the *second* quasi primary shrink-swell cycle (Fig.9, curves 3 and 2) instead of the *first* one (Fig.9, curves 1 and 2'; see the last sentence of Section 4.1.2) with corresponding replacements of the reference and scanning shrinkage and swelling curves in Sections 4.2.2 and 4.2.3 as well as $h^* \to \hat{h}^*$ and $q \to \hat{q}$. All the above results, (i)-(iii) relative to crack volume hysteresis are retained. It is essential that in the case of the *second* quasi primary shrink-swell cycle the *residual* cracks exist even at $W = \hat{W}_h$ (Fig.9) [24]. For this reason, in the course of a *steady scanning* shrink-swell cycle (inside the *second* quasi primary shrink-swell cycle), the *residual* crack volume should keep even at $W_2$ close to $\hat{W}_h$. This result is in agreement with observations [6].

Finally, it should be emphasized that the *origin* of the *above* considered *shrink-swell cracks at any depth* is connected with the seasonal and other *variations* of water content at a *given depth* in the $W_1<W<W_2$ range inside the maximum possible range, $0<W<W_h$, unlike the shrink-swell *cracks* at the *soil surface* (that can, nevertheless, extend to several meters keeping a micro width [4,22,43]). The origin of the latter shrink-swell *cracks* is connected with the vertical *gradient* of the mean (at a given depth) water content [4,22,43] close to the soil surface.

*4.2.4. Estimating the initial thickness* (*h*) *of a cracked layer with no loading*

In Sections 4.2.1-4.2.3, regarding the different shrinkage and swelling curves of a horizontal soil layer, we considered that the *initial* thickness, $h$ of the layer at $W=W_h$ is *known*. However, in fact, each layer under consideration should be separated from the soil profile based on the criterion of water content homogeneity within the limits of the layer. With that the water content, $W$ of the layer can be less than $W_h$, and its thickness, $z$ at this water content is less than $h$. Thus, after dividing the soil profile into the set of horizontal layers with water content being (approximately) homogeneous within each layer, one only knows the *current* layer thickness, $z<h$ at $W<W_h$, but not the *initial* thickness, $h$ at $W=W_h$. Therefore before using the results of Sections 4.2.1-4.2.3 for a layer one should estimate the $h$ value of the layer from its *known* thickness, $z$ at the *known* water content, $W$ and corresponding specific volume, $Y$ with cracks.

Let $Y$ symbolizes the current specific volume of a cracked soil layer along any primary/scanning shrinkage/swelling curve that is possible for the soil, and $z$ symbolizes the corresponding current thickness of the cracked soil layer. Then, $Y$ and $z$ are simply connected with the initial (maximum) thickness, $h$ of the soil layer and the maximum specific volume, $Y_h$ of the soil at $W=W_h$ (for $W_h$ and $Y_h$ see Fig.13) as

$$z/h=Y/Y_h, \qquad (130)$$

even though the curve under consideration is a scanning one and does not include the point $W=W_h$. Eq.(130) is immediately evident from the definition of the values entering it. One can also derive this relation from the equality between the two different presentations of the current layer volume (per $h \times h$ surface area of the layer), $h^2 z = Y h^3/Y_h$ (note that $h^3/Y_h$ is the solid mass of the layer per $h \times h$ surface area). As noted we can substitute for $Y$ in Eq.(130) the primary shrinkage curve, $Y(W, h/h^*)$ (Eq.(101)), or the primary swelling curve, $\hat{Y}(\hat{W}, h/h^*, h/\hat{h}^*)$ (Eq.(106)), or the quasi primary swelling curve $\hat{\hat{Y}}(\hat{W}, h/h^*)$ (Eq.(109)), or any scanning shrinkage curve, $\underline{Y}(\underline{W}, \underline{W}_o, h/h^*)$ (Eq.(114) at any possible $\underline{W}_o$), or any scanning swelling curve, $\underline{\hat{Y}}(\underline{\hat{W}}, \underline{\hat{W}}_o, h/h^*)$ (Eq.(119) at any possible $\underline{\hat{W}}_o$). In each case $z$ gives the corresponding layer thickness as a function of all the indicated physical parameters.

We are especially interested in the substitution for $Y$ in Eq.(130) the two scanning curves that correspond to the shrinkage and swelling stages of the *steady* shrink-swell cycle in the cracked soil layer (Fig.12, curves 3 and 4; Eqs.(128) and (129)). Indeed, we believe that the layer parameters ($z$ and $W$) in the natural conditions vary along either the shrinkage or swelling branch of the steady shrink-swell cycle. Thus, to estimate the initial layer thickness, $h$ we can use one of the following two equations

$$z/h=\underline{Y}(W,W_{o1},h/h^*)/Y_h, \qquad 0<W<W_{o1}<W_h, \qquad (131)$$

$$z/h=\underline{\hat{Y}}(W,W_{o2},h/h^*)/Y_h, \qquad 0<W_{o2}<W<W_h \qquad (132)$$

where $\underline{Y}$ ($W,W_{o1},h/h^*$) and $\hat{\underline{Y}}$ ($W,W_{o2},h/h^*$) are from Eqs.(128) and (129) accounting for all the previous derivations. $W_{o1}$ and $W_{o2}$ are the functions of $W_1$ and $W_2$ (Fig.12; see Eqs.(92) and (93)) that are determined by the particular site conditions. Knowing $z$ at a given $W_1<W<W_2$ (Fig.12) for shrinkage (or swelling) stage we can solve Eq.(131) (or (132)) to find $h$ as some function of $z/h^*$ and $W$

$$h/h^* = f(z/h^*, W, W_1, W_2, Y_h) . \tag{133}$$

Then this $h$ can be used as stated in Sections 4.2.1-4.2.3. As a result of variations in the water content profile a layer with initially homogeneous water content should be divided after some time into two thinner layers or, on the contrary, this layer can enter a thicker layer. In such cases the recalculation of $h$ is needed at another $z$ thickness of a new layer with new homogeneous water content, $W$.

### *4.3. Shrinkage-swelling and cracking of a soil under loading*

The objective of this Section is to *extend* the results of Sections 4.1 and 4.2 to the case of soil layer *loading*, $L$. In general, as shown in Section 2, the variation of the *clay* shrinkage and swelling behavior with increase in $L$ is determined by dependences on $L$ of two clay characteristics, $v_z(L)$ (Eq.(36)) and $v_h(L)$ (Eq.(40)) (or $\zeta_h(L)$, Eq.(39)). In other words all other clay paste characteristics change with $L$ as a result of their links with $v_z$ and $v_h$ (or $\zeta_h$). As shown in Section 3 the variation of shrinkage-swelling behavior of the *intra-aggregate matrix* (of a soil) with increase in $L$ is also determined by the, $v_z(L)$ and $v_h(L)$ (or $\zeta_h(L)$) dependences of the *contributive clay*. That is, all relevant characteristics of the intra-aggregate matrix can only change with $L$ as functions of two *basic* values, $v_z$ and $v_h$ (or $\zeta_h$) of the contributive clay. In this section we intend to show that the third *basic* value (besides $v_z$ and $v_h$) that determines *soil* shrinkage and swelling behavior with loading increase (including crack *volume* behavior with the $L$ variation) is the maximum aggregate size (at the shrinkage limit), $X_{mz}$ or (at maximum swelling) $X_m$. This means that all other relevant soil characteristics only change with $L$ through the dependences, $v_z(L)$ and $v_h(L)$ (or $\zeta_h(L)$) of the contributive clay, and possible dependence, $X_{mz}(L)$ (or $X_m(L)$). Based on the results of Sections 4.1 and 4.2 we first indicate the relevant soil characteristics that are added, compared to the case of the intra-aggregate matrix, and can change with $L$ (Section 4.3.1). Then we look after their links with $v_z$, $v_h$ (or $\zeta_h$), and $X_{mz}$ (or $X_m$) (Sections 4.3.2-4.3.5). The possible dependence of $X_{mz}$ on $L$ is discussed in Section 4.3.6. Section 4.3.7 contains some final remarks relative to the loading case. Below we also mention the maximum aggregate size, $\hat{X}_{mz}$ after aggregate destruction at swelling at $W\cong 0$. It should be taken into account, however, that in the conditions of the water content variation when $W_{min}>0$ such destruction is lacking, and $\hat{X}_{mz}=X_{mz}$ (cf. the first paragraphs of Sections 4.1.2 and 4.2.2).

### *4.3.1. Soil characteristics that can change with loading increase*

The *scanning* shrinkage-swelling curves, both in the case of reference shrinkage-swelling (Section 4.1) and in the case of shrinkage-swelling with cracks (Section 4.2), are in the single-valued manner connected with the *primary* ones. For this reason, below we only turn to the latter (as in Sections 4.1.1 and 4.2.1 in no loading case). According to Eqs.(59a), (59b), (60a), (60b), (64a), and (64b), the *reference* primary shrinkage and swelling curves of a soil are determined (besides the volume of the intra-aggregate matrix, $U$ and $\hat{U}$; see Section 3) by the following values: the specific initial volume of structural (inter-aggregate) pores, $U_s$ (Fig.1); the interface layer contribution (Fig.1) to the specific soil volume at shrinkage and swelling, $U_i$ and $\hat{U}_i$; aggregate/intra-aggregate mass ratio at shrinkage and swelling, $K$ and $\hat{K}$; and the interface layer contribution (Fig.1) to the total water content at shrinkage and swelling, $\omega$ and $\hat{\omega}$. According to Eqs.(94) and (102) the transition from *reference* primary shrinkage-swelling to primary shrinkage-swelling *with cracks* is accompanied with the addition of the specific crack volume at shrinkage, $U_{cr}$ and swelling, $\hat{U}_{cr}$. From Eqs.(100) and (105), besides the values that are connected with reference shrinkage and swelling, the volumes $U_{cr}$ and $\hat{U}_{cr}$ are only determined by the crack factor at shrinkage, $q$ and swelling, $\hat{q}$. The links between the $U_s$, $U_i$, $\hat{U}_i$, $K$, $\hat{K}$, $\omega$, $\hat{\omega}$, $q$ and $\hat{q}$ values and the three above basic dependences, $v_z(L)$, $v_h(L)$, and $X_{mz}(L)$ are considered below.

*4.3.2. Variation of the structural pore volume ($U_s$) with loading increase*

One can consider that at the maximum swelling point, $W=W_h$ the aggregated soil is in the *visco-plastic* state or close to that. In *such* a state the pore volumes of different types change with $L$ proportionally to each other. In other words, it is natural to consider that the *inter-aggregate* (structural) pore volume at maximum swelling, $U_s(L)$ (Fig.1) varies with loading, $L$ proportionally to the volume, $U_h(L)-1/\rho_s$ of the *intra-aggregate* pores (lacunar and clay matrix pores; Fig.1) at $W=W_h$. Then, accounting for $U_s(L=0)\equiv U_s$ and $U_h(L=0)\equiv U_h$ ($u_h(L=0)\equiv u_h$), as well as Eq.(48), we estimate the sought $U_s(L)$ dependence to be (cf. the paragraph after Eq.(55c))

$$U_s(L)=U_s(U_h(L)-1/\rho_s)/(U_h-1/\rho_s)=U_s(u_h(L)-u_s)/(u_h-u_s), \qquad L>0 \qquad (134)$$

where $u_s=v_s[c+v_s(1-c)]^{-1}$ [26]. According to Eq.(134) $U_s(L)=0$ follows $U_s=0$, and at $U_s>0$ $U_s(L)<U_s$ as it should be from physical considerations. $u_h(L)$ is estimated from Eq.(57a) at $\zeta=\zeta_h(L)$ and, as with all characteristics of the intra-aggregate matrix, depends on $L$ through $v_z(L)$ and $v_h(L)$ (or $\zeta_h(L)$) (Section 3.2.2). Thus $U_s(L)$ is also expressed through the same functions of $L$.

*4.3.3. Variation of the interface layer volume ($U_i$ and $\hat{U}_i$) and aggregate/intra-aggregate mass ratio ($K$ and $\hat{K}$) with loading increase*

The above considered $U_s(L)$ (Section 4.3.2) and $U_h(L)$ (Section 3.2.2) show that $P_h(L)$ (Eq.(63b)) is expressed through $v_z(L)$ and $v_h(L)$. Hence, $U_i$ (Eqs.(63a)-(63e)) is expressed through $v_z(L)$, $v_h(L)$, and $X_m(L)$ (the mean size of soil solids, $x_n\equiv X_{min}$ does not depend on $L$). Then $U_z(L)$ (Section 3.2.2) and $U_i(L)$ show that $P_z(L)$ (Eq.(63b)) and $\hat{U}_i(L)$ (Eqs.(63a)-(63e)) are also expressed through $v_z(L)$, $v_h(L)$, and $\hat{X}_{mz}(L)$ ($\hat{X}_{mz}(L)=X_{mz}(L)$ at $W_{min}>0$). After that Eqs.(61a) and (61b) show that $K(L)$ and $\hat{K}(L)$ are also expressed through the same three functions of $L$.

*4.3.4. Variation of the interface layer water content ($\omega$, $\hat{\omega}$) with loading increase*

Eqs.(65a)-(65i), (66), (67), (68a)-(68c) show that all values determining $\omega$ and $\hat{\omega}$ (Eqs.(65a)-(65c)), except for $K$, $\hat{K}$, $U_i$ and $\Delta\hat{U}_i$ in Eq.(66), relate to either the contributive clay or the intra-aggregate matrix. Therefore, all these values with the loading increase are expressed through $v_z(L)$ and $v_h(L)$. In addition, according to Section 4.3.3 $K$, $\hat{K}$, $U_i$ and $\Delta\hat{U}_i=\hat{U}_i-U_i$ entering Eq.(66) also lead to the dependence of $\omega(L)$ and $\hat{\omega}(L)$ through $X_m(L)$.

*4.3.5. Variation of the crack factor ($q$ and $\hat{q}$) with loading increase*

Eqs.(97a)-(97b), (98), (99a)-(99d) and the same equations after replacements, $X_m \rightarrow \hat{X}_m$, $l_m \rightarrow \hat{l}_m$, $l_{min} \rightarrow \hat{l}_{min}$, $h^*_o \rightarrow \hat{h}^*_o$, $h^* \rightarrow \hat{h}^*$ show that the dependences of $q$ and $\hat{q}$ on $L$ are expressed through $v_z(L)$ and $v_h(L)$ (at the expense of the structural porosity, $P_h$, see section 4.3.3) as well as $X_m(L)$ and $\hat{X}_m(L)$. Note that $X_m = x_n + (X_{mz} - x_n)(u_h/u_z)^{1/3}$ [39] and $\hat{X}_m = x_n + (\hat{X}_{mz} - x_n)(u_h/u_z)^{1/3}$ [24].

*4.3.6. Possible dependence of the maximum aggregate size on loading*

Shrinkage with *no loading* does not change the maximum aggregate size, both $X_{mz}$ and $X_m$. In the beginning of swelling with *no loading* (at $W \cong 0$) the maximum aggregate size decreases to $\hat{X}_{mz} < X_{mz}$ since aggregates are destructed under the action of increasing pore pressure inside them at the expense of air entrapped and compressed in the pores of the contributive clay at soil wetting [40,41,21,24]. The maximum aggregate sizes at shrinkage, $X_{mz}$ and following swelling, $\hat{X}_{mz}$ with *no loading* can be considered as independently measured characteristics. However, *loading* during swelling should, in part or totally, offset the internal pressure in aggregates and prevent their destruction. This will be checked in Section 7.3 (point 3). In general, sufficiently *high* loading at *shrinkage* can lead to the aggregates sticking together and increasing $X_{mz}$. However, because $X_{mz}$ relates to the dry state with Young's modulus ~$10^2$-$10^3$MPa (see Section 2.2.2), such sufficiently high loading at shrinkage should be no less than 0.1-1MPa. Thus, at least at $L < 0.1$-1MPa the maximum aggregate size, $X_{mz}$ should not change under loading, and one can usually accept $X_{mz}(L) =$ const. This will be checked in Section 7.3 (point 3).

*4.3.7. Shrinkage-swelling, cycling, and crack volume hysteresis of a loaded layer*

Thus, knowing $v_z(L)$, $v_h(L)$ (or $\zeta_h(L)$), $X_{mz}(L)$, and $\hat{X}_{mz}(L)$ and using them successively throughout Sections 2, 3, 4.1, and 4.2 instead of $v_z$, $v_h$ (or $\zeta_h$), $X_{mz}$, and $\hat{X}_{mz}$, one can find any shrinkage and swelling curve, steady cycling, and crack volume hysteresis of a soil layer at a given loading, $L$. In particular, it is clear that all the equations of Section 4.2.3 (Eqs.(120)-(129)) are also retained in the case of *loading*. All the results of Section 4.2.3, relative to the shrink-swell cycles and crack volume *hysteresis* of a cracked soil layer, are also *qualitatively kept* in case of loading (see Figs.12 and 14) with a natural *quantitative* correction at the expense of $v_z \rightarrow v_z(L)$, $v_h \rightarrow v_h(L)$, $X_{mz} \rightarrow X_{mz}(L)$, $\hat{X}_{mz} \rightarrow \hat{X}_{mz}(L)$. Finally, in the above consideration of the *loading* case aspects, the *initial* thickness, $h(L)$ of a layer (i.e., the thickness at maximum soil swelling, $W = W_h(L)$) was suggested to be *known* at a *given* loading, $L$. At the same time, similar to what is stated in Section 4.2.4 (for a no loading case), in real conditions we can only know the thickness, $z$ of the layer in the soil profile at water content, $W < W_h(L)$ and loading, $L$. However, we can find the initial thickness, $h(L)$ that corresponds to these $z$, $W$, and $L$ in the total analogy to Section 4.2.4 after the following replacements in Eqs.(130)-(133): $Y \rightarrow Y(L)$, $Y_h \rightarrow Y_h(L)$, $W_h \rightarrow W_h(L)$, $\underline{Y}(W,W_{o1},h/h^*) \rightarrow \underline{Y}(W,W_{o1},h/h^*(L),L)$, $\underline{\hat{Y}}(W,W_{o2},h/h^*) \rightarrow \underline{\hat{Y}}(W,W_{o2},h/h^*(L),L)$, $h^* \rightarrow h^*(L)$, and $f(z/h^*,W,W_1,W_2,Y_h) \rightarrow f(z/h^*(L),W,W_1,W_2,Y_h(L),L)$. All the necessary functions of $L$ are calculated as considered above, but with $v_z(L)$, $v_h(L)$ (or $\zeta_h(L)$), $X_{mz}(L)$, and $\hat{X}_{mz}(L)$.

**5. Analysis of available data to check the basic model relations**

Experimental checking of the shrinkage curves of the contributive clay, intra-aggregate matrix, soil without and with cracks in the *maximum* possible water content range (the *primary shrinkage* curves) with *no loading* have been conducted based on the available data [30-32,25-28]. The same relates, to some extent, to the *primary swelling* curves of the soil media [24]. In addition, according to the above theory, the *scanning shrinkage and swelling* curves of these media with *no loading*, as well as the transitive and steady shrink-swell cycles and crack volume hysteresis in a soil, flow out of the primary shrinkage and swelling curves in the *single-valued* manner. It follows the importance of checking the relations for the primary and scanning shrinkage and swelling curves, shrink-swell cycles, and crack volume hysteresis *under loading*, in other words, the checking of the *transition* from the relations with *no loading* to those *under loading*. Among all the relations of interest in the *loading* case the *functional* dependences of the *primary shrinkage* curves of the above soil media *on loading*, $L$ are the most important because the primary swelling curves, scanning shrinkage and swelling curves, shrink-swell cycles, and crack volume variation *under loading* are then constructed in the *single-valued* manner. The major functions of $L$ to be checked are $v_z(L)$ (Eq.(36)), $v_h(L)$ (Eq.(40)), $\zeta_h(L)$ (Eq.(39)), and $X_m(L) \cong$ const (see Section 4.3).

The available data for soil *layers* in field conditions (large depths) have a *qualitative* character, i.e., they indicate the *fact itself* of the crack volume occurrence and variation with a water content profile [6]. The necessary quantitative data are only available as applied to the soil *sample* case (see Section 5.1). The above theory, as a whole, is also applicable in this case with the expression for the crack factor, $q(h/h^*)$ that is modified compared to Eqs.(97a) and (97b) as [28] ($h$ is the sample height that should be of the same order of magnitude as the sample diameter)

$$q_s(h/h^*)=0, \qquad 0<h/h^*\leq 1 \qquad (135a)$$

$$q_s(h/h^*)=b_1(h/h^*-1)^2, \qquad 1\leq h/h^*\leq 1+\delta \qquad (135b)$$

$$q_s(h/h^*)=1-b_2/(h/h^*-1), \qquad h/h^*\geq 1+\delta \qquad (135c)$$

with the same $b_1 \cong 0.15$, $b_2 \cong 1$, and $\delta \cong 1.5$. The data for the samples under loading that are obtained in an oedometer and a triaxial test apparatus (see [44-46], among others) do not take cracking into account. In addition, such data are oriented to estimating the swelling pressure, widely use fitting procedures, and for this reason do not usually contain the physical soil characteristics that are necessary for checking the model.

*5.1. Data used*

The only data on the primary shrinkage curves of soil samples under loading that we could use were obtained in works [20,47] (note that other data from [20] on the shrinkage of soils that were *compacted* using different loads *before* shrinkage are not of interest in this work). These data were presented in the form of $e(\theta)$ (void ratio vs. moisture ratio). For this reason, predicting the shrinkage curves at different $L$, we first transform the $(\theta,e)$ variables to $(W,Y)$ ones and then, after predictions, return to $(\theta,e)$ presentation of the predicted shrinkage curves ($W=(\rho_w/\rho_s)\theta$, $Y=(e+1)/\rho_s$). In particular, the moisture ratio, $\theta_h$ at a given $L$ (Figs.15 and 16) is replaced with the corresponding $W_h$ (Fig.13) and the void ratio, $e_z$ at a given $L$ (Figs.15 and 16) with the corresponding $Y_z$ (Fig.13) as input values (see below).

*5.1.1. Peng et al.'s [20] data*

Peng et al. [20] experimentally observed shrinkage of the soil samples with a small content of organic matter under five loads including $L$=0 (Table 1; Fig.15). The samples 4.1 cm in height and 5.6 cm in diameter were prepared from a Dystric Gleysol from glacial sediment by sieving (a 2 mm mesh), wetting, and repacking with compaction (the stress reached 150 kPa and was released immediately after compacting). Each constant load was imposed at the beginning of the previous swelling and acted up to the maximum swelling and then up to the end of shrinkage. Table 1 shows the primary data ($\theta_h$ and $e_z$) and input parameters from [20] (including $W_h$ and $Y_z$ corresponding to $\theta_h$ and $e_z$) that were used in the model prediction, as well as in estimating the maximum sizes of sand grains, $x_m$ and aggregates, $X_m$ that are also needed as input in the prediction (see Sections 5.2 and 7.3). The lacunar pore volume at maximum swelling, $U_{lph}$ was accepted as $U_{lph}$=0 for different $L$ (Table 1) since the experimental shrinkage curve points at $\theta=\theta_h$ (Fig.15) are on the saturated line. We accepted the initial structural pore volume for different $L$ to be $U_s$=0 (Table 1) since the experimental shrinkage curves (Fig.15) do not have the horizontal part at the intersection with the saturation line (cf. [28]). In addition, we used the experimental points of the five loadings from [20] (see Fig.15) in the data analysis or in the comparison with the model prediction (Section 5.2).

*5.1.2. Talsma's [47] data*

Talsma [47] observed the shrinkage behavior of clay paste samples under four loads (Table 1; Fig.16). The samples 2 cm in height and 7 cm in diameter were prepared from Black Earth (NSW, Australia). Unlike Peng et al.'s [20] experiments, each constant load was imposed at the beginning of shrinkage (as soon as the sample became detached from the containing ring) and worked up to the end of shrinkage. Table 1 also shows the data from [47]. Two remarks are necessary in connection with the data of Black Earth texture ($c$, $s$; Table 1) and structure ($x_m$, $X_m$; Table 2). As for the texture, Talsma [47] only indicates the clay content $c$=0.6. However, the available data on Black Earth (NSW, Australia) (see [48,49], among others) give $c+s$=0.7-0.73 and $s$=0.14-0.17. For the model predictions we took the $c$ and $s$ values indicated in Table 1. As for the structure, Talsma [47] speaks about the samples of clay soil *paste*, i.e., the soil with such large aggregates that the latter do not lead to structural shrinkage. However, his data on shrinkage curves (Fig.16) show behavior that is similar to that of aggregated soils with the *structural* shrinkage range and the slope in the *basic* shrinkage range that is essentially *less* than unity. For this reason we tried to estimate the $x_m$ and $X_m$ values for Talsma's [47] samples, suggesting that the $X_m$ values are sufficiently large (see Section 7.3, point 3). The initial lacunar pore volume, $U_{lph}$=0 and initial structural pore volume, $U_s$=0 were taken for Talsma's [47] data (Table 1) from the same considerations as for Peng et al.'s [20] data. Finally, the experimental points of the four loadings from [47] (Fig.16) were also used in the data analysis or in the comparison with the model prediction (Sections 5.2 and 7.3).

*5.2. Data analysis*

*5.2.1. Estimating the maximum sand grain and aggregate sizes*

The data [20,47] do not include the maximum sand grain size, $x_m$ and maximum aggregate size, $X_{mz}$ that are needed for the prediction of a shrinkage curve [24,28]. By its physical meaning $x_m$ is constant when $L$ increases. We assume that at shrinkage at the $L$ values from Table 1, $X_{mz}$ also does not change with $L$ (see Section 4.3.6). For this reason we find $x_m$ and $X_{mz}$ below only using data from [20] at $L$=0 and 1.8kPa (the use here of two $L$ values is explained in Section 7.3) and from [47] at $L$=0. Then, the resulting $x_m$ and $X_{mz}$ participate together with data from Table 1 in the model prediction of the shrinkage curves for data [20,47] at other $L$ values. $x_m$ and $X_{mz}$ are estimated using the approach that was recently regarded in detail [24,28]. At a given $L$, and with input data from Table 1, $x_m$ and $X_{mz}$ are practically independent from each other and change in small ranges with $\Delta x_m/x_m \cong 0.05$, $\Delta X_{mz}/X_{mz} \cong 0.01$-0.02. $X_{mz}$ is considered to be the only fitting parameter in the calculation of the shrinkage curves using the algorithm from [28,24] and experimental points for $L=0$ and 1.8kPa in Fig.15 and $L=0$ in Fig.16. The only modification of the algorithm is the following replacement at any $L$ (note that $\zeta_h(L=0)\cong 0.5$; see Section 2.2.3)

$$u_s=1/(1+2\rho_s W_h) \rightarrow u_s=1/(1+\rho_s W_h(L)/\zeta_h(L)) \,. \tag{136}$$

Since $u_s$ in Eq.(136), by physical meaning, does not change with $L$, after finding $u_s$ for case $L=0$ we can find $\zeta_h(L)$ at $L>0$ knowing $W_h(L)$ for $L>0$ (Table 1) as

$$\zeta_h(L)=\rho_s W_h(L)/(1/u_s-1) \,. \tag{137}$$

The estimates of $x_m$, $X_{mz}$, and other major physical characteristics of the soil as a whole, intra-aggregate matrix, and contributive clay are presented in Tables 2-4 at $L=0$ and 1.8kPa for Peng et al.'s [20] soil and at $L=0$ for Talsma's [47] soil. These estimates are used in Section 5.2.2 as well as 7.1-7.3.

*5.2.2. Model prediction of shrinkage curves at different loadings*

Assuming $X_{mz}(L)=\text{const}$ (see Section 4.3.6) and using $x_m$ and $X_{mz}$ (Table 2) that were found above for Peng et al.'s [20] soil (Table 1) at $L=1.8$kPa, we predict the three shrinkage curves for the soil at $L>1.8$kPa based on the algorithm [28,24] with Eq.(137) and input data from Table 1. Fig.15 shows the predicted curves at $L>1.8$kPa. Tables 2-4 show the predicted characteristics (as well as $x_m$ and $X_{mz}$ that were used in the prediction) of the soil, intra-aggregate matrix, and contributive clay at $L>1.8$kPa. Similar predictions for Talsma's [47] soil at $L>0$ obtained using $X_{mz}$ found at $L=0$ and $x_m$ from Table 2 as well as input data (Table 1, $L>0$), are shown in Fig.16 (curves) and Tables 2-4 (characteristics). In addition, for the control, we estimated the $x_m$ and $X_{mz}$ values for Peng et al.'s [20] soil at $L>1.8$kPa and Talsma's [47] soil at $L>0$ (Table 1) using the *fitting* approach as in Section 5.2.1 ($X_{mz}$ is the only fitting parameter). All the characteristics in Tables 2-4 (including $x_m$ and $X_{mz}$) at the indicated $L$ values were found to remain the same. Table 2 also shows the corresponding goodness of fit, $r_e^2$ and estimated standard errors $\sigma_e$ of the shrinkage curve data [20,47] in Figs.15 and 16. The fitted curves in Figs.15 and 16 and those predicted also coincide. The predicted characteristics (Tables 2-4) are used in Sections 7.1-7.3.

*5.2.3. Presentation of the estimated values of the relative maximum water content of clay ($\zeta_h$) at different loading values ($L$) using the theoretical curve*

We fitted the theoretical curve, $\zeta_h(L/L^*)$ (Eq.39) to the five points, $\zeta_h(L_i)$ (i=1,…,5) of Peng et al.'s [20] soil (Table 4) using $L^*$ as the only fitting parameter. The five points, fitted curve, $L^*$ value, goodness of fit, $r_\zeta^2$, and estimated standard error, $\sigma_\zeta$ of the $\zeta_h(L_i)$ points are shown in Fig.17 and in the caption to it. Using $v_s$ and the *mean* $v_z$ value by the five loading values for the soil (Table 4), we also estimated the characteristic universal loading, $L_u$ (Eq.(41); Fig.17) (the *mean* $v_z$ value was used because, in the agreement with Eq.(36) and the statement about $v_z \cong \text{const}$ at $L</\cong 100$kPa at the end of Section 2.2.2, $v_z(L)$, at different $L$ in Table 4, does not show any trend to decreasing with $L$ increase, and the differences between $v_z$ at different $L$ in Table 4 are only stipulated by the experimental spread of the input $Y_z$ values in Table 1). The similar procedure and presentations of results were then utilized to the four points, $\zeta_h(L_i)$ (i=1,…,4) of Talsma's [47] soil (Table 4; Fig.17). These estimates are discussed in Section 7.1.

## 6. Theoretical results and discussion

(1) The *classification* of the volumetric and water content values that characterize the layer of a clayey vadose zone was *suggested* (Table 5) based on the considerations of this work and previous results [24-28]. The classification is conducted according to a number of transitions (Table 5). The latter reflect the soil structure features (Fig.1) and conditions within the limits of the layer of the clayey vadose zone during the cycles of drainage-wetting. The classification essentially facilitates the presentation of the different relevant volumetric and water content values and finding the interconnections between them.

(2) The *primary* shrinkage and swelling curves with *no loading* determine the corresponding *scanning* curves in a *single-valued* manner ($V \to \underline{V}$, $\hat{V} \to \underline{\hat{V}}$, $U \to \underline{U}$, $\hat{U} \to \underline{\hat{U}}$, $Y_r \to \underline{Y}_r$, $\hat{Y}_r \to \underline{\hat{Y}}_r$, $Y \to \underline{Y}$, $\hat{Y} \to \underline{\hat{Y}}$).

(3) The *single-valued* links between the *primary* curves of clay, intra-aggregate matrix, aggregated soil with no cracks, and the soil with cracks [24,28] (for shrinkage: $V \to U \to Y_r \to Y, U_{cr}$; and for swelling: $\hat{V} \to \hat{U} \to \hat{Y}_r \to \hat{Y}, \hat{U}_{cr}$) are transformed *under loading* to *similar* links ($V(L) \to U(L) \to Y_r(L) \to Y(L), U_{cr}(L)$ and $\hat{V}(L) \to \hat{U}(L) \to \hat{Y}_r(L) \to \hat{Y}(L), \hat{U}_{cr}(L)$).

(4) The combination of results (2) and (3) shows the possibility of the *single-valued* prediction of the different *scanning* curves under *loading*, based on the *primary* curves and *loading* conditions ($V \to V(L) \to \underline{V}(L)$, $\hat{V} \to \hat{V}(L) \to \underline{\hat{V}}(L)$, $U \to U(L) \to \underline{U}(L)$, $\hat{U} \to \hat{U}(L) \to \underline{\hat{U}}(L)$, $Y_r \to Y_r(L) \to \underline{Y}_r(L)$, $\hat{Y}_r \to \hat{Y}_r(L) \to \underline{\hat{Y}}_r(L)$, $Y \to Y(L) \to \underline{Y}(L)$, $\hat{Y} \to \hat{Y}(L) \to \underline{\hat{Y}}(L)$).

(5) Result (4) means the possibility of the *single-valued* quantitative prediction of the *transitive* and *steady* shrink-swell cycles in a given layer of a clayey vadose zone at a given range, $W_1<W<W_2$ of the water content variation.

(6) Finally, the model explains the *origin* of the crack volume and quantitatively predicts the *crack volume hysteresis* in a given layer of a clayey vadose zone, at a given range, $W_1<W<W_2$ of the water content variation. It should be noted especially, in connection with result (6), that the observations themselves of the cracks and crack volume variation in the layers of a clayey vadose zone [6] gives strong qualitative experimental evidence in favor of the model feasibility.

Results (2)-(6) show that the derivations in this work with respect to the *scanning* shrinkage and swelling curves and cycles as well as the *crack volume hysteresis* under loading in the layer of a clayey vadose zone, essentially are the extraction of the *single-valued* consequences from the *primary* shrinkage and swelling curves, i.e., the curves in the maximum possible water content range. In turn, the physical modeling, quantitative prediction, and its validation as applied to the *primary* curves, were considered in detail in recent works [24-28] based on the physical properties of the soil layer. The *input* data for the prediction include those on [24,28]: (i) initial soil layer thickness, $h$ before shrinkage; (ii) soil solid density, $\rho_s$; *texture*: $c$ being clay content, $s$ being silt content (sand content being 1-$c$-$s$), and $x_m$ being the maximum sand grain size; and *initial structure*: $X_{mz}$ being the maximum aggregate size at the shrinkage limit and $P_z$ being the structural porosity at the shrinkage limit; and (iii) soil *shrinkage*: $Y_z$ being the specific volume at the shrinkage limit, in general, with cracks (Fig.13); $W_h$ being the water content at maximum swelling before shrinkage (Fig.13); and $U_{lph}$ being the specific volume of lacunar pores at maximum swelling before shrinkage (or $Y_h$ in Fig.13 instead of $U_{lph}$).

## 7. Data analysis results and discussion

*7.1. Dependences of the contributive-clay characteristics on loading*

1. The joint analysis (Section 5.2.3) of Eq.(39) and $\zeta_h$ values for different $L$ in Table 4 for both soils (see Fig.17) shows that the theoretical curve $\zeta_h(L/L^*)$ (Eq.(39)) is in agreement with the $\zeta_h$ values (Table 4) from the viewpoint of both the fitting criterion (high $r_\zeta^2$ in Fig.17) and the standard physical criterion (the difference between the $\zeta_h(L/L^*)$ curve at each $L$ and corresponding $\zeta_h$ in Table 4 does not exceed the two standard errors, $2\sigma_\zeta$ in Fig.17). According to Section 2.2.3 this result is also evidence in favor of the feasibility of Eq.(40) for $v_h(L/L^*)$.

2. The values found for $L^*$ (Fig.17) together with the $v_s$ value and mean $v_z$ value (for the mean value see Section 5.2.3) for each soil in Table 4, lead to the choice of $L^*=L_u/[(v_z-v_s)/(1-v_s)]^2$ from two the possible variants in Eq.(41) because only in this case does $L_u$ coincide (within the limits of obvious errors) for both soils (see Fig.17) and can be considered as the *universal* characteristic loading. In view of the importance of such a characteristic, additional data are desirable to estimate $L_u$.

3. The constancy of $v_s$ (Table 4) for each soil at different $L$ just follows from its physical meaning. The constancy of the $v_z$ and $\zeta_z$ estimates (Table 4) for each soil (within the limits of the spread connected with that of $Y_z$ in Table 1), confirms Eq.(36) at indicated sufficiently small $L$ values (see the end of Section 2.2.2). Finally, the estimates of the saturated degree at the clay shrinkage limit, $F_z$ (Eq.(4)) are likely to show the weak trend to increase with $L$ growth for both soils.

*7.2. Dependences of the characteristics of the intra-aggregate matrix on loading*

1. The constancy of porosity $p$ (Table 3) flows out of the constancy of $x_m$ (Table 2), $c$, and $s$ (Table 1) [28].

2. The constancy of the critical clay content, $c^*$ (Table 3; Eq.(55c)) flows out of the constancy of $v_s$, $v_z$ (Table 4; Section 7.1), and $p$. The spread of the $c^*$ values (Table 3) is connected with the spread of $v_z$ (see Section 5.2.3).

3. The $k$ values (Table 3) flow out of Eqs.(55a) and (55b), the $c$ (Table 1) and $c^*$ (Table 3) values. $k>0$ leads to the lacunar pore volume at $W=W_z$, $U_{lpz}>0$ (Table 3) albeit $U_{lph}=0$ for both soils (Table 1).

4. By their physical meaning $u_s$ and $u_S$ (Table 3) do not change with $L$.

5. The appreciable decrease of the specific volume, $U_h$ with $L$ growth for both soils (Table 3) is only kept for the specific volumes, $U_z$ and $U_n$ (Table 3) of Talsma's [47] soil when $k>0$ (Table 3). In this case the lacunar pore volumes, $U_{lpz}>0$ and $U_{lpn}>0$ (Table 3) also show the obvious decrease with $L$ growth. In the case of Peng et al.'s [20] soil when $k=0$, $U_{lpz}=0$, and $U_{lpn}=0$ (Table 3), the estimates of $U_z$ and $U_n$ (Table 3) are constant (within the limits of the spread connected with that of $Y_z$ in Table 1). Note that in the case of Talsma's [47] soil the differences, $U_z-U_{lpz}$ and $U_n-U_{lpn}$ are also constant within the limits of the spread. That is, in this case the dependences, $U_z(L)$ and $U_n(L)$ are totally stipulated by the occurrence of the lacunar pore volume, $U_{lpz}(L)>0$ and $U_{lpn}(L)>0$ at $W=W_z$ and $W=W_n$.

*7.3. Dependences of the soil characteristics and shrinkage curves on loading*

1. The $r_e^2$ and $\sigma_e$ values (Table 2) show that the predicted shrinkage curves, $e(\theta)$ (Figs.15 and 16) are in agreement with the data [20,47] (Figs.15 and 16) at all $L$ values, from the viewpoint of both the fitting criterion (high $r_e^2$) and physical criterion (discrepancies between each predicted curve and experimental points at a given $L$ do not exceed the two standard errors, $2\sigma_e$ in Figs.15 and 16). Thus, $r_e^2$ and $\sigma_e$ (Table 2) evidence in favor of the model feasibility.

2. $x_m$ and $x_n$ (Table 2) have reasonable values for clay soils and do not depend on $L$ in the accordance with their physical meaning.

3. The difference between $X_{mz}$ (Table 2) at $L=0$ and 1.8kPa for Peng et al.'s [20] soil (and the similar difference for $X_m$ in Table 2) is connected with the features of the sample preparation before shrinkage under loading. Preliminary sample swelling at $L=0$ in [20] leads to some destruction of aggregates [24] (Section 4.3.6). Preliminary sample swelling at $L>0$ in [20] is not accompanied by aggregate destruction (Section 4.3.6). The similar $X_{mz}$ estimates at different $L>0$ (Table 2) evidence in favor of that. For this reason in Section 5.2.1 we estimated $X_{mz}$ by fitting for Peng et al.'s [20] soil not only at $L=0$, but also at $L=1.8$kPa. In [47] the samples were loaded only at shrinkage. For this reason the estimates of $X_{mz}$ for Talsma's [47] soil were similar at all $L≥0$ (Table 2). The peculiarity of Talsma's [47] soil is the relatively large $X_{mz}$ values and, correspondingly, those $X_m$ (Table 2). This feature is in the accordance with the fact that Talsma's [47] soil is close to the paste-like state. Note that the differences between the $X_m$ and $X_{mz}$ values in Table 2 at a given $L$ illustrate the variation of the maximum aggregate size at shrinkage. Finally, note that the similar $X_{mz}$ values at $L>0$ for Peng et al.'s [20] soil and at $L≥0$ for Talsma's [47] soil also confirm the statement about the constancy of $X_{mz}$ at shrinkage under sufficiently small loading (Section 4.3.6).

4. The jump-like change of $K$ (Eq.(61a)), $U_i$ (Eq.(63a)), and $h^*$ (Eqs.(98), (99a)-(99d)) (Table 2) for Peng et al.'s [20] soil between $L=0$ and 1.8kPa is explained by the same reason (different $X_{mz}$ values for these $L$). At $L>0$ for Peng et al.'s [20] soil and at $L≥0$ for Talsma's [47] soil $K$, $U_i$, and $h^*$ practically do not change with $L$. Note that in the case of Talsma's [47] soil the $K$ values (Table 2) that are close to unity and the small $U_i$ values (Table 2) correspond to the large $X_{mz}$ size. Note also that the $h^*$ values (Table 2) are in the agreement with the estimates from [24,28] (several centimeters).

5. The sufficiently *small* sample size, $h$ (Table 1) for both soils leads to $h/h^*<1$ and $q=0$ at all $L$ (Table 2) [28], i.e., to the specific volume practically without cracks.

6. Data [20,47] and predicted curves in Figs.15 and 16 show *two possible types* of soil shrinkage curve behavior under loading. In the case of Peng et al.'s [20] soil (Fig.15) the appreciable distances between the shrinkage curves for different $L$ at $\theta=\theta_h(L)$ are reduced to very small (if any) at $\theta=0$. In the case of Talsma's [47] soil (Fig.16) the curves for different $L$ are approximately parallel in the total water content range, $0<\theta<\theta_h(L)$. The behavior type in Fig.15 is connected with the *absence of cracks* ($q=0$, Table 2) because of the small sample size, and *lacunar pores* ($U_{lph}=0$, Table 1; $k=0$, Table 3) because of $c>c^*$ (Tables 1 and 3) at all the $L$ values and water contents, $0<\theta<\theta_h(L)$. The contributive-clay pores only exist in Peng et al.'s [20] soil samples. The volume of the contributive-clay pores at $\theta=\theta_h(L)$ (Fig.15) is proportional to $[v_h(L/L^*)-v_s]$ and differs essentially with $L$ increase ($v_h(L/L^*)$ qualitatively follows $\zeta_h(L/L^*)$ in Fig.17; see Eqs.(39) and (40)). At the same time, the volume of the contributive-clay pores at $\theta=0$ is proportional to $v_z-v_s$ and practically does not change with $L$ growth (Table 4; Section 7.1, point 3). For this reason the curves of the different $L$ in Fig.15 converge with dewatering. Note that $q=0$ and $k=0$ for Peng et al.'s [20] soil at all $L$ values (Tables 2 and 3) means the *unit* slope, $S$ of all the curves in Fig.15 in the basic shrinkage range ($S=1-k-q$ [28]). By contrast, the behavior type in Fig.16 is connected with the development of *cracks* (which is not the case of Talsma's [47] soil because of the small sample size) and/or *lacunar pores* ($k>0$, Table 3). In the case of Talsma's [47] soil the contributive-clay pore volume changes with $L$ and $\theta$ as in the case of Peng et al.'s [20] soil. Lacunar pores in the case of Talsma's [47] soil do not exist at $\theta=\theta_h(L)$ ($U_{lph}=0$, Table 1), but develop at $\theta=0$ ($U_{lpz}(L)>0$, Table 3). Lacunar pore volume, $U_{lpz}(L)$ essentially depends on $L$ (Table 3; Section 7.2, point 5). Thus, the differences between the *clay pore* volume at different $L$ decrease with drying, but those between the *lacunar pore* volume at different $L$ increase with drying. The summary pore volume is approximately kept. As a result, the appreciable *gaps* between the curves for different $L$ in Fig.16 are retained while dewatering. Note that $q=0$ and close values of $k>0$ for Talsma's [47] soil at all $L$ values (Tables 2 and 3) mean the approximately similar slope, $S<1$ of all the curves in Fig.16 in the basic shrinkage range ($S=1-k-q$ [28]). Finally, note that the use of the sufficiently *large* samples or consideration of a *layer* for Peng et al.'s [20] soil should lead to an appreciable crack development [28,24] and to the shrinkage curve behavior with loading increase as in Fig.16.

**8. Conclusion**

The *observations* [6] show the crack existence and variation in their volume at sufficiently large depths of a clayey vadose zone where the possible contribution of the customary shrinkage cracks that originate from the vertical gradient of the water content close to the soil surface, is negligible. The *objective* of this work is to physically explain the origin of the cracks that were observed in [6] and to suggest an approach to the physical quantitative prediction of the variation of the crack volume within the limits of a soil layer, accounting for the natural conditions in a clayey vadose zone: (i) overburden; (ii) multifold shrinkage-swelling; and (iii) relatively small water content range. The *approach* is based on: (1) the inter- and intra-aggregate soil structure (Fig.1); (2) consecutive consideration of the four increasingly complex and entering each other soil media - contributive clay, intra-aggregate matrix, aggregated soil with no cracks, and soil with cracks [24] in conditions with no loading as well as under loading; (3) classification of the volumetric and water content values according to the soil structure features (Fig.1; Table 5) and the above conditions in the clayey vadose zone (Table 5); and (4) generalization of the results of recent works [24,28,39,42]. The major *theoretical results* present the single-valued links between the different volumetric and water content values of the four increasingly complex soil media, and predict these values from: (a) the primary shrinkage and swelling curves of a contributive clay; (b) effects of the soil inter- and intra-aggregate structure; and (c) effects of the soil water content profile. Eventually, this work, in combination with [24-28] enables the totally physical prediction (i.e., without fitting) of shrink-swell cycling and crack volume hysteresis in the layers of a clayey vadose zone from the above physical soil properties, overburden, range of the water content variation, and layer thickness with a homogeneous water content. The *analysis results* of available data evidence in favor of the *basic* model relations that reflect the variation of the contributive-clay characteristics and maximum aggregate size with loading increase, and substantiate the model prediction relative to the soil shrinkage curves at different loading values. The *results can be used* for a number of aims, in particular in the frame of an approach [50,51] to develop the totally physical prediction of the water retention of a clay soil (including the scanning water retention curves at drainage and wetting), as well as in the frame of an approach [43] to predict the crack width distribution of the crack volume that was considered in this work. This distribution is important for the physical consideration of the hydraulic conductivity of a crack network in a clayey vadose zone.

**Notation**

$a, d, e$   coefficients that determine $\underline{v}(\zeta,\zeta_o)$ (dimensionless)

$\hat{a}, \hat{b}$   coefficients that determine $\hat{\underline{v}}^u(\zeta,\zeta_o)$ (dimensionless)

$c, c_*$   soil clay content and its critical value (dimensionless)

$E$   Young's modulus of dry clay matrix (MPa)

$F, \hat{F}, \underline{F}, \hat{\underline{F}}$   saturation degree of clay matrix along the curves of primary shrinkage, primary swelling, scanning shrinkage, and scanning swelling (dimensionless)

$F(\eta,\beta)$   aggregate-size distribution at structural porosity, $\beta$ (dimensionless)

$F_i$   volume fraction of water-filled interface clay pores (dimensionless)

$F_z$   $F$ value at $\zeta=\zeta_z$ (dimensionless)

$f(\eta,\varphi)$   clay pore-size distribution at clay matrix porosity, $\varphi$ ($\varphi$ can symbolize $P, \hat{P}, \underline{P},$ and $\hat{\underline{P}}$) (dimensionless)

$h, h^*$   initial and critical layer thickness at maximum swelling (cm)

$h^*_o$   rough approximation of $h^*$ (cm)

$\hat{h}^*$   $h^*$ value after aggregate destruction (cm)

$\hat{h}^*_o$   $h^*_o$ value after aggregate destruction (cm)

$I_o(\eta)$   function from Eq.(63d) (dimensionless)

$K, \hat{K}$   aggregate/intra-aggregate mass ratio at *shrinkage* and *swelling* (dimensionless)

$k$   lacunar factor (dimensionless)

$L$   loading (including overburden) (kPa)

$L_u$   universal characteristic loading of clays (kPa)

$L^*$   characteristic loading of clay (kPa)

$l_m, l_{min}$   mean distance between the aggregates of size $X_m$ and $X_{min}$ (mm)

$\hat{l}_m, \hat{l}_{min}$   $l_m$ and $l_{min}$ values after aggregate destruction (mm)

$P, \hat{P}, \underline{P}, \hat{\underline{P}}$   clay matrix porosity along the curves of primary shrinkage, primary swelling, scanning shrinkage, and scanning swelling (dimensionless)

$P_h, P_z$   structural porosity at shrinkage close to $W=W_h$ and $W=0$ (dimensionless)

$\hat{P}_z$   structural porosity at swelling close to $\hat{W}=0$ (dimensionless)

$p$   silt and sand porosity in the state of *imagined* contact (dimensionless)

$q$   crack factor of soil layer before aggregate destruction (dimensionless)

$\hat{q}$   crack factor of soil layer after aggregate destruction (dimensionless)

$R, \hat{R}, \underline{R}, \hat{\underline{R}}$   maximum internal size of water-filled clay pores along the curves of primary shrinkage, primary swelling, scanning shrinkage, and scanning swelling (µm)

$R_m$   maximum size of clay pores in interface layer part of the $U_i$ volume (µm)

$R_{min}$   minimum size of clay pores in interface layer part of the $U_i$ volume (µm)

$\hat{R}_m$   maximum size of clay pores in interface layer part of the $\Delta\hat{U}_i$ volume (µm)

$\hat{R}_{min}$   minimum size of clay pores in interface layer part of the $\Delta\hat{U}_i$ volume (µm)

$r_m, r_o$   maximum and minimum internal size of clay pores at shrinkage (µm)

$r_{mM}$   maximum external size of clay pores at $\zeta=1$ (µm)

$\hat{r}_m, \hat{r}_o$   maximum and minimum internal size of clay pores at swelling (µm)

$U, \hat{U}, \underline{U}, \hat{\underline{U}}$   specific volume of intra-aggregate matrix along the curves of primary shrinkage, primary swelling, scanning shrinkage, and scanning swelling (dm$^3$kg$^{-1}$) $U_{cr}, \hat{U}_{cr}, \underline{U}_{cr}, \underline{\hat{U}}_{cr}$ specific crack volume along the curves of primary shrinkage, primary swelling, scanning shrinkage, and scanning swelling (dm$^3$kg$^{-1}$)

$\hat{\hat{U}}_{cr}$ specific crack volume along the curve of *quasi* primary swelling (dm$^3$kg$^{-1}$)

$U_z, U_h$ $U$ value at $W=0$ and $W=W_h$ (dm$^3$kg$^{-1}$)

$U_i, \hat{U}_i$ interface layer contribution to the specific volume of aggregates at shrinkage and swelling (dm$^3$kg$^{-1}$)

$U_s$ specific volume of structural pores (dm$^3$kg$^{-1}$)

$U', \hat{U}', \underline{U}', \underline{\hat{U}}'$ intra-aggregate matrix contribution to the specific volume of aggregates along the curves of primary shrinkage-swelling ($U'=U/K$, $\hat{U}'=\hat{U}/\hat{K}$) and scanning shrinkage-swelling ($\underline{U}'=\underline{U}/K$, $\underline{\hat{U}}'=\underline{\hat{U}}/\hat{K}$) (dm$^3$kg$^{-1}$)

$u, \hat{u}, \underline{u}, \underline{\hat{u}}$ relative volume of soil intra-aggregate matrix along the curves of primary shrinkage-swelling and scanning shrinkage-swelling (dimensionless)

$u_h, \hat{u}_h$ $u$ and $\hat{u}$ values at the maximum swelling point ($u_h=\hat{u}_h$) (dimensionless)

$u_S, \hat{u}_S$ relative volume of non-clay solids of intra-aggregate matrix at shrinkage and swelling ($u_S=\hat{u}_S$) (dimensionless)

$u_s, \hat{u}_s$ relative volume of solid phase of intra-aggregate matrix at shrinkage and swelling ($u_s=\hat{u}_s$) (dimensionless)

$u_z, \hat{u}_z$ $u$ and $\hat{u}$ values at $\zeta=0$ ($u_z=\hat{u}_z$) (dimensionless)

$u_{cp}, \hat{u}_{cp}, \underline{u}_{cp}, \underline{\hat{u}}_{cp}$ relative volume of clay matrix pores in soil intra-aggregate matrix along the curves of primary shrinkage, primary swelling, scanning shrinkage, and scanning swelling (dimensionless)

$u_{lp}, \hat{u}_{lp}, \underline{u}_{lp}, \underline{\hat{u}}_{lp}$ relative volume of lacunar pores in soil intra-aggregate matrix along the curves of primary shrinkage, primary swelling, scanning shrinkage, and scanning swelling (dimensionless)

$u_{lph}, u_{lpz}$ $u_{lp}$ values at $\zeta=\zeta_h$ and $\zeta=\zeta_z$, respectively (dimensionless)

$V, \hat{V}, \underline{V}, \underline{\hat{V}}$ specific volume of clay matrix along the curves of primary shrinkage, primary swelling, scanning shrinkage, and scanning swelling (dm$^3$kg$^{-1}$)

$v, \hat{v}, \underline{v}, \underline{\hat{v}}$ ratio of clay volume to its maximum in the solid state (the liquid limit) along the curves of primary shrinkage, primary swelling, scanning shrinkage, and scanning swelling (dimensionless)

$v_h, v_n, v_z$ $v$ value at $\zeta=\zeta_h$, $\zeta=\zeta_n$, and $\zeta=\zeta_z$, respectively (dimensionless)

$v_s, \hat{v}_s$ relative clay solid volume at shrinkage and swelling ($v_s=\hat{v}_s$) (dimensionless)

$\underline{\hat{v}}^u$ relative clay volume along the $\underline{\hat{v}}(\zeta,\zeta_o)$ curve at $0<\zeta_o<\zeta_n$ (dimensionless)

$\underline{\hat{v}}^s$ relative clay volume along the $\underline{\hat{v}}(\zeta,\zeta_o)$ curve at $\zeta_n<\zeta_o<\zeta_h$ (dimensionless)

$W, \hat{W}, \underline{W}, \underline{\hat{W}}$ total water content of soil along the curves of primary shrinkage, primary swelling, scanning shrinkage, and scanning swelling (kg kg$^{-1}$)

$\underline{W}_o, \underline{\hat{W}}_o$ $\underline{W}$ and $\underline{\hat{W}}$ values at the start of scanning shrinkage and swelling (kg kg$^{-1}$)

$W_h$ $W$ value at the start of primary shrinkage (kg kg$^{-1}$)

$\hat{W}_h$ $\hat{W}$ value at the end of primary swelling ($\hat{W}_h<W_h$) (kg kg$^{-1}$)

$\hat{W}_h*$ water content (Fig.12) at which residual cracks after shrink-swell cycle would be water-filled (kg kg$^{-1}$)

$W_1, W_2$ range borders of the shrink-swell cycles in a soil layer (kg kg$^{-1}$)

$w$ water content of intra-aggregate matrix at shrinkage-swelling ($\hat{w}=w$) (kg kg$^{-1}$)

$w_h$, $w_n$    $w$ value at $\zeta=\zeta_h$ ($w_h=\hat{w}_h$) and $\zeta=\zeta_n$ ($w_n=\hat{w}_n$) (kg kg$^{-1}$)

$\overline{w}$    water content of clay matrix at shrinkage and swelling ($\overline{w}=\hat{\overline{w}}=w/c$) (kg kg$^{-1}$)

$\overline{w}_o$    $\overline{w}$ at the start of scanning shrinkage ($\underline{V}$)/swelling ($\underline{\hat{V}}$) (Figs.2 and 3) (kg kg$^{-1}$)

$\overline{w}_h$, $\overline{w}_n$    $\overline{w}$ value at $\zeta=\zeta_h$ ($\overline{w}_h=\hat{\overline{w}}_h$) and $\zeta=\zeta_n$ ($\overline{w}_n=\hat{\overline{w}}_n$) (kg kg$^{-1}$)

$w'$, $\hat{w}'$    contribution of intra-aggregate matrix to total water content at shrinkage and swelling ($w'>\hat{w}'$) (kg kg$^{-1}$)

$w'_h$    $w'$ value at maximum swelling (before shrinkage start) (kg kg$^{-1}$)

$w'_s$    $w'$ value at the end point of the structural range of primary shrinkage (kg kg$^{-1}$)

$w'_o$    corresponds to $\underline{W}_o$ and $\underline{\hat{W}}_o$ in terms of $w'$ (kg kg$^{-1}$)

$\hat{w}'_b$    $\hat{w}'$ value being the starting point of water filling in the interface layer part of the $U_i$ volume at primary swelling (kg kg$^{-1}$)

$\hat{w}'_h$    $\hat{w}'$ value at swelling finish (kg kg$^{-1}$)

$\hat{w}'_e$    $\hat{w}'$ value being the end point of water filling in the interface layer part of the $\Delta\hat{U}_i$ volume at primary swelling (kg kg$^{-1}$)

$\underline{w}'_s$    analogue of $w'_s$ in case of scanning shrinkage (kg kg$^{-1}$)

$\underline{\hat{w}}'_b$    analogue of $\hat{w}'_b$ in case of scanning swelling (kg kg$^{-1}$)

$X_m$, $X_{mz}$    maximum aggregate size at $W=W_h$ and $W=0$ (mm)

$\hat{X}_{mz}$, $\hat{X}_m$    $X_{mz}$ and $X_m$ values after aggregate destruction at $W=0$ and $W=\hat{W}_h$ (mm)

$X_{min}$    minimum aggregate size ($X_{min}=\hat{X}_{min}$ after aggregate destruction) (mm or μm)

$x_m$    maximum size of sand grains (μm)

$x_n$    mean size of soil solids ($x_n=X_{min}=\hat{X}_{min}$) (μm)

$Y$, $\hat{Y}$, $\underline{Y}$, $\underline{\hat{Y}}$    specific soil volume with crack contribution along the curves of primary shrinkage-swelling and scanning shrinkage-swelling (dm$^3$kg$^{-1}$)

$Y_h$    $Y$ value at the start of shrinkage (Fig.13) (dm$^3$kg$^{-1}$)

$Y_r$, $\hat{Y}_r$, $\underline{Y}_r$, $\underline{\hat{Y}}_r$    reference specific soil volume (with no cracks) along the curves of primary shrinkage-swelling and scanning shrinkage-swelling (dm$^3$kg$^{-1}$)

$Y_{rh}$    $Y_r$ value at the start of reference shrinkage (Fig.13) (dm$^3$kg$^{-1}$)

$Y_{rz}$    $Y_r$ value at the end of reference shrinkage (Fig.13) (dm$^3$kg$^{-1}$)

$\hat{Y}_h$    $\hat{Y}$ value after primary shrink-swell cycle with cracking (Fig.13) (dm$^3$kg$^{-1}$)

$\hat{Y}_{rh}$    $Y_r$ value after reference primary shrink-swell cycle (Fig.13) (dm$^3$kg$^{-1}$)

$\hat{Y}_{rz}$    $\hat{Y}_r$ value at the start of reference primary swelling (Fig.13) (dm$^3$kg$^{-1}$)

$\hat{Y}_z$    $\hat{Y}$ value at the start of primary swelling with cracking (Fig.13) (dm$^3$kg$^{-1}$)

$\hat{\hat{Y}}$    specific soil volume with crack contribution along the curve of *quasi* primary swelling (i.e., with no aggregate destruction) (dm$^3$kg$^{-1}$)

$\hat{\hat{Y}}_r$    *reference* specific soil volume (i.e., with no crack contribution) along the curve of *quasi* primary swelling (i.e., with no aggregate destruction) (dm$^3$kg$^{-1}$)

$z$    current layer thickness at a given $W$ (cm)

$\alpha$    function of $\sigma$ from Eq.(35) (dimensionless)

$\Delta\hat{U}_i$    additional interface layer volume that appears at swelling (dm$^3$kg$^{-1}$)

$\Delta v_z$    negative increment of $v_z$ under loading (dimensionless)

| | |
|---|---|
| $\Delta\hat{\omega}$ | water contribution of the $\Delta\hat{U}_i$ volume at primary swelling (kg kg$^{-1}$) |
| $\zeta$ | ratio of water content in clay or soil intra-aggregate matrix to its maximum in the solid state (the liquid limit) (dimensionless) |
| $\zeta_h$ | maximum swelling point on the $\zeta$ axis (dimensionless) |
| $\zeta_n$ | end point of basic (or normal) shrinkage (the air-entry point) (dimensionless) |
| $\zeta_z$ | shrinkage limit on the $\zeta$ axis (dimensionless) |
| $\zeta_o$ | $\zeta$ value at scanning shrinkage and swelling start (Figs.2 and 3) (dimensionless) |
| $\zeta_1, \zeta_2$ | range borders of the shrink-swell cycles of a clay volume (dimensionless) |
| $\underline{\zeta}_n, \underline{\zeta}_z$ | range borders where $\underline{v}$ is a square function of $\zeta$ (Fig.3) (dimensionless) |
| $\Pi_h$ | clay porosity of interface layer part of the $U_i$ volume (dimensionless) |
| $\Pi_z$ | clay porosity of interface layer part of the $\Delta\hat{U}_i$ volume (dimensionless) |
| $\rho_s$ | density of clay solids or mean density of soil solids (kg dm$^3$) |
| $\rho_w$ | water density (kg dm$^3$) |
| $\sigma$ | Poisson's ratio of dry clay matrix (dimensionless) |
| $\omega, \hat{\omega}, \underline{\omega}, \underline{\hat{\omega}}$ | interface layer contribution to the total water content along the curves of primary shrinkage-swelling and scanning shrinkage-swelling (kgkg$^{-1}$) |

**Figure captions**

**Fig.1.** Schematic illustration of the accepted soil structure [24-28]. Shown are (1) an assembly of many soil aggregates and inter-aggregate pores contributing to the specific soil volume, $Y$; (2) an aggregate, as a whole, contributing to the specific volume $U_a=U_i+U'$; (3) an aggregate with two parts indicated: (3a) an interface layer contributing to the specific volume $U_i$ and (3b) an intra-aggregate matrix contributing to the specific volumes $U$ and $U'=U/K$; (4) an aggregate with indicated intra-aggregate structure: (4a) clay, (4b) silt and sand grains, and (4c) lacunar pores; and (5) an inter-aggregate pore leading at shrinkage to an inter-aggregate crack contributing to the specific volume $U_{cr}$. $U$ is the specific volume of an intra-aggregate matrix (per unit mass of the oven-dried matrix itself). $U'$ is the specific volume of an intra-aggregate matrix (per unit mass of the oven-dried soil). $U_i$ is the specific volume of the interface layer (per unit mass of the oven-dried soil). $U_{cr}$ is the specific volume of cracks (per unit mass of the oven-dried soil). $U_a$ is the specific volume of aggregates (per unit mass of the oven-dried soil). $K$ is the aggregate/intra-aggregate mass ratio.

**Fig.2.** The qualitative view of the *clay scanning swelling curve*, $\hat{v}(\zeta,\zeta_o)$ (curve 3) that goes between the primary shrinkage, $v(\zeta)$ (curve 1) and swelling, $\hat{v}(\zeta)$ (curve 2) curves and starts at $\zeta=\zeta_o$ on the primary shrinkage curve. The initial water content, $\zeta_o$ can vary in the $0<\zeta_o<\zeta_h$ range. At $\zeta_n<\zeta_o<\zeta_h$ $\hat{v}(\zeta,\zeta_o)=v(\zeta)$, i.e., curves 3 and 1 coincide.

**Fig.3.** The qualitative view of the *clay scanning shrinkage curve*, $\underline{v}(\zeta,\zeta_o)$ (curve 3) that goes between the primary shrinkage, $v(\zeta)$ (curve 1) and swelling, $\hat{v}(\zeta)$ (curve 2) curves and starts at $\zeta=\zeta_o$ on the primary swelling curve. $\zeta_o$ can vary in the $0<\zeta_o<\zeta_h$ range. The point, $\zeta=\underline{\zeta}_n(\zeta_o)$ is a boundary of the initial linear range ($\underline{\zeta}_n(\zeta_o)<\zeta<\zeta_o$) of the $\underline{v}(\zeta,\zeta_o)$ curve. The point, $\zeta=\underline{\zeta}_z(\zeta_o)$ is a boundary of the final linear range ($0<\zeta<\underline{\zeta}_z(\zeta_o)$) of the $\underline{v}(\zeta,\zeta_o)$ curve. $\underline{\zeta}_z(\zeta_o)\to\zeta_z$ and $\underline{\zeta}_n(\zeta_o)\to\zeta_n$ at $\zeta_o\to\zeta_h$.

**Fig.4.** The qualitative view of the two initial consecutive transitive scanning shrink-swell cycles of a cycle set leading to steady shrink-swell cycle in the $\zeta_1<\zeta<\zeta_2$ range (Fig.5). The *first* cycle consists of the *swelling* branch, $\hat{v}_1(\zeta,\zeta_1)$ at $\zeta_o=\zeta_1\equiv\zeta_{o1}$ (curve 1) and *shrinkage* branch, $\underline{v}_1(\zeta,\zeta_{o2})$ at $\zeta_o=\zeta_{o2}$ (curve 2). The *second* cycle consists of the *swelling* branch, $\hat{v}_2(\zeta,\zeta_{o3})$ at $\zeta_o=\zeta_{o3}$ (curve 3) and *shrinkage* branch, $\underline{v}_2(\zeta,\zeta_{o4})$ at $\zeta_o=\zeta_{o4}$ (curve 4). Curve 5 shows the start of the following transitive cycle.

**Fig.5.** The qualitative view of the *steady* shrink-swell cycle (curves 1 and 2) of a clay in the $\zeta_1<\zeta<\zeta_2$ range. The cycle consists of the *shrinkage* branch, $\underline{v}(\zeta,\zeta_{o1})$ (curve 1) and *swelling* branch, $\hat{v}(\zeta,\zeta_{o2})$ (curve 2) where $\zeta_{o1}=\zeta_{o1}(\zeta_1,\zeta_2)$ and $\zeta_{o2}=\zeta_{o2}(\zeta_1,\zeta_2)$ are a solution of Eqs.(31) and (32).

**Fig.6.** The general qualitative view of transformation of the clay maximum shrink-swell cycle under loading. The indicated values are connected by the following relations: (i) $v(\zeta)\equiv v(\zeta,L=0)$, $\hat{v}(\zeta)\equiv\hat{v}(\zeta,L=0)$; (ii) $\zeta_h=\zeta_h(L=0)$, $\zeta_n=\zeta_n(L=0)$, $\zeta_z=\zeta_z(L=0)$; (iii) $v_h=v(\zeta=\zeta_h,L=0)$, $v_n=v(\zeta=\zeta_n,L=0)$, $v_z=v(\zeta=\zeta_z,L=0)$; (iv) $v_h(L)\equiv v(\zeta=\zeta_h(L),L)$, $v_n(L)\equiv v(\zeta=\zeta_n(L),L)$, $v_z(L)\equiv v(\zeta=\zeta_z(L),L)$.

**Fig.7.** The qualitative view of the $\zeta_h(L/L^*)$ dependence. Quantitatively, $\zeta_h(L/L^*)$ is described by Eqs.(39) and (41).

**Fig.8.** The qualitative view of the transformation of the maximum shrink-swell cycle for the intra-aggregate matrix under loading (at $c<c^*$ and $0<k<1$). The $u(\zeta)$ curve with the initial slope, $(1-k)(1-u_s)$ starts at the $(\zeta_h, u_h)$ point on the *pseudo* saturation (dash-dot) line (with the slope, $(1-u_s)$) that corresponds to $L=0$. The displacement, $\zeta_h*$-

$\zeta_h=u_{lph}/(1-u_s)$ of the line relative to the *true* saturation (dashed) line is connected with the lacunar pores that are non-filled in water. The $u(\zeta,L)$ curve with the same initial slope, $(1-k)(1-u_s)$ starts at the $(\zeta_h(L), u_h(L))$ point on *another pseudo* saturation (dotted) line that corresponds to the $L$ loading. The smaller displacement of the line relative to the true saturation line corresponds to the smaller lacunar pore volume at loading, $L$.

**Fig.9.** The qualitative view of the *reference primary* and *quasi-primary* shrink-swell cycles of a soil. 1-the reference primary *shrinkage* curve ($Y_r(W)$) of the *first* cycle; 2-the reference primary *swelling* curve ($\hat{Y}_r(\hat{W})$) of the *first* cycle; 2'-the reference *quasi*-primary *swelling* curve ($\hat{\hat{Y}}_r(\hat{W})$) of the *first* cycle; 3-the reference primary *shrinkage* curve of the *second* cycle; 4-the *swelling* branch of the *second* cycle. $Y_{rh}$ and $W_h$ correspond to maximum swelling before the first cycle; $Y_{rz}$ corresponds to maximum shrinkage in the course of the first and following cycles; $\hat{Y}_{rh}$ and $\hat{W}_h$ correspond to maximum swelling after the first cycle. The inclined (dashed) line is the saturation or quasi-saturation one.

**Fig.10.** The qualitative view of the *reference quasi-primary* shrink-swell cycle, ($Y_r(W), \hat{\hat{Y}}_r(\hat{W})$) (curves 1 and 2', respectively) and *reference scanning shrinkage* curve of a soil, $\underline{Y}_r(\underline{W}, \underline{W}_o)$ (curve 3). The inclined (dashed) line is the saturation or quasi-saturation one.

**Fig.11.** The qualitative view of the *reference quasi-primary* shrink-swell cycle, ($Y_r(W), \hat{\hat{Y}}_r(\hat{W})$) (curves 1 and 2', respectively) and *reference scanning swelling* curve of a soil, $\hat{\underline{Y}}_r(\hat{\underline{W}}, \hat{\underline{W}}_o)$ (curve 3). The inclined (dashed) line is the saturation or quasi-saturation one.

**Fig.12.** The qualitative view of the reference *steady* shrink-swell cycle (curves 3 and 4) of a soil in the $W_1<W<W_2$ range. The cycle consists of the *shrinkage* branch, $\underline{Y}_r(\underline{W}, W_{o1})$ (curve 3) and *swelling* one, $\hat{\underline{Y}}_r(\hat{\underline{W}}, W_{o2})$ (curve 4) [which are inside the reference *quasi-primary* shrink-swell cycle, ($Y_r(W), \hat{\hat{Y}}_r(\hat{W})$) (curves 1 and 2', respectively)] where $W_{o1}=W_{o1}(W_1,W_2)$ and $W_{o2}=W_{o2}(W_1,W_2)$ are a solution of Eqs.(92) and (93). The inclined (dashed) line is the saturation or quasi-saturation one.

**Fig.13.** The qualitative view of the four different *primary* shrink-swell cycles of a soil. ($Y_r(W), \hat{Y}_r(\hat{W})$) is the *reference primary* shrink-swell cycle. ($Y_r(W), \hat{\hat{Y}}_r(\hat{W})$) is the *reference quasi primary* shrink-swell cycle. ($Y(W), \hat{Y}(\hat{W})$) is the *primary* shrink-swell cycle with a *crack* contribution. ($Y(W), \hat{\hat{Y}}(\hat{W})$) is the *quasi primary* shrink-swell cycle with a *crack* contribution. The differences between the soil volume with and without cracks give the corresponding crack volume at a given water content minus the volume of the structural pores: $Y(W)-Y_r(W)=U_{cr}(W)-U_s$; $\hat{Y}(\hat{W})-\hat{Y}_r(\hat{W})=\hat{U}_{cr}(\hat{W})-U_s$; $\hat{\hat{Y}}(\hat{W})-\hat{\hat{Y}}_r(\hat{W})=\hat{\hat{U}}_{cr}(\hat{W})-U_s$. $\hat{W}_h^*$ is the $W$ value at which the residual cracks after the ($Y(W), \hat{Y}(\hat{W})$) shrink-swell cycle would be water-filled [24].

**Fig.14.** The qualitative view of the *hysteretic steady* shrink-swell cycle of the specific *crack* volume (at a given soil depth). Curve 3 presents the *crack volume* evolution, $\underline{U}_{cr}(W, W_{o1})$ (Eq.(120)) at the *shrinkage* stage and corresponds to the reference *scanning shrinkage curve*, $\underline{Y}_r(W, W_{o1})$ (curve 3 in Fig.12) in the $W_1<W<W_2$ range.

Curve 4 presents the *crack volume* evolution, $\hat{\underline{U}}_{cr}(W,W_{o2})$ (Eq.(122)) at the *swelling* stage and corresponds to the reference *scanning swelling curve*, $\hat{\underline{Y}}_r(W,W_{o2})$ (curve 4 in Fig.12) in the $W_1<W<W_2$ range. $W_{o1}$ and $W_{o2}$ are the starting points of the $\underline{Y}_r(W,W_{o1})$ and $\hat{\underline{Y}}_r(W,W_{o2})$ scanning curves (see Fig.12). $W_m$ is the maximum point of the *hysteretic* variation of the crack volume, $\Delta\underline{U}_{cr}(W)$ (Eq.(127)). $\underline{U}_{cr\,min}$ (Eq.(125)) and $\underline{U}_{cr\,max}$ (Eq.(126)) indicate the minimum and maximum crack volume of the hysteretic crack volume cycle in the $W_1<W<W_2$ range.

**Fig.15.** The experimental points [20] and predicted shrinkage curves at five loadings (see Tables 1-4). $\theta_h$, $\theta_s$, $\theta_n$, and $\theta_z$ correspond to the maximum swelling point, end point of structural shrinkage, end point of normal (basic) shrinkage, and shrinkage limit, respectively. The ranges of $\theta_h$, $\theta_s$, $\theta_n$, and $\theta_z$ values with loading variation are indicated by two arrows.

**Fig.16.** As in Fig.15, but for the experimental points [47] and predicted shrinkage curves at four loadings (see Tables 1-4).

**Fig.17.** The white squares and solid line indicate the five estimated $\zeta_h(L/L^*)$ points from Table 2 and fitted $\zeta_h(L/L^*)$ curve (Eq.(39)), respectively, for Peng et al.'s [20] soil and fitted characteristic loading, $L^*\cong20.5$kPa ($L_u\cong0.54$kPa). Goodness of fit is $r_\zeta^2=0.9848$. The estimated standard error of the $\zeta_h(L/L^*)$ points is $\sigma_\zeta=0.005$. The white circles and dashed line indicate the four estimated $\zeta_h(L/L^*)$ points from Table 2 and fitted $\zeta_h(L/L^*)$ curve (Eq.(39)), respectively, for Talsma's [47] soil and fitted characteristic loading, $L^*\cong40.6$kPa ($L_u\cong0.53$kPa). Goodness of fit is $r_\zeta^2=0.9994$. The estimated standard error of the $\zeta_h(L/L^*)$ points is $\sigma_\zeta=0.001$.

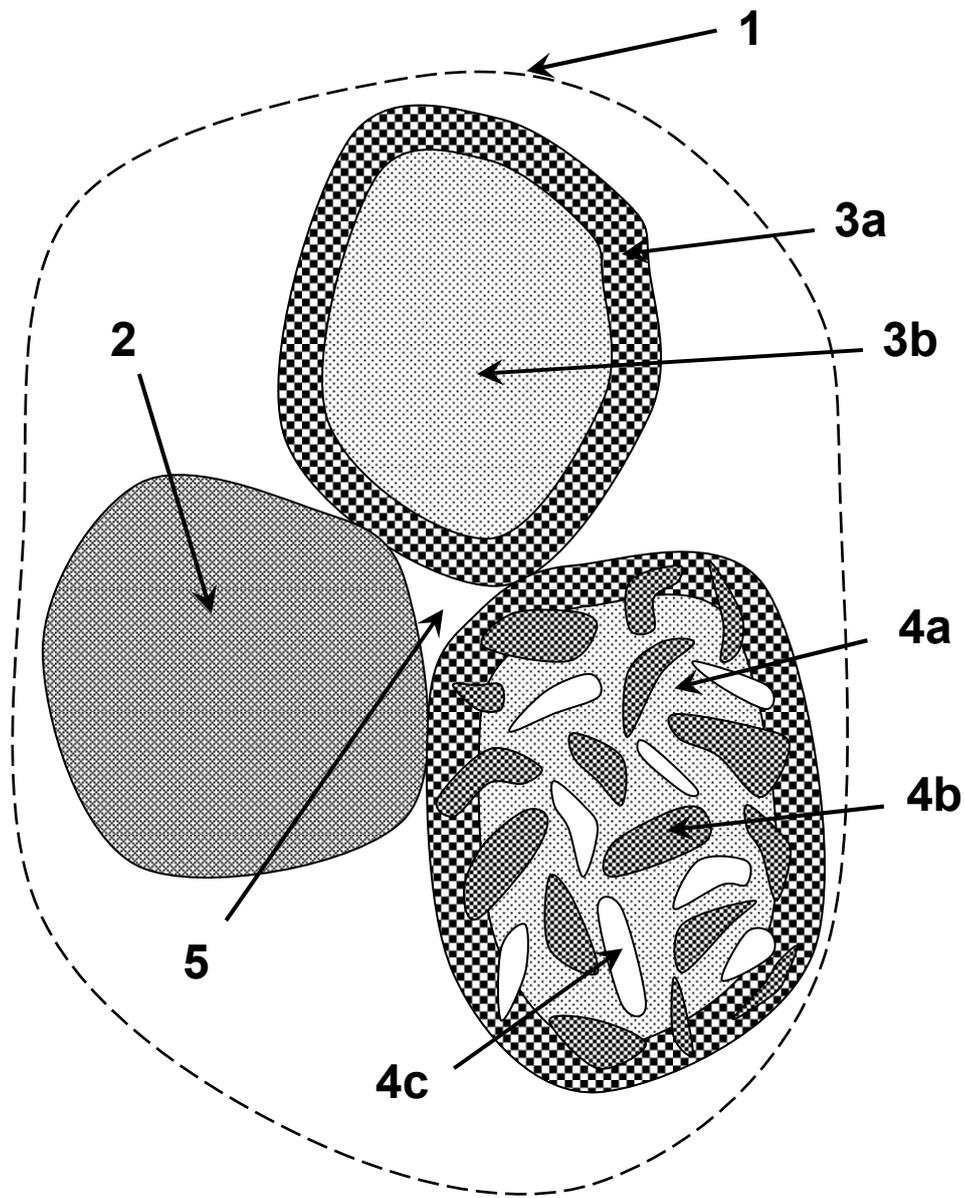

Fig.1

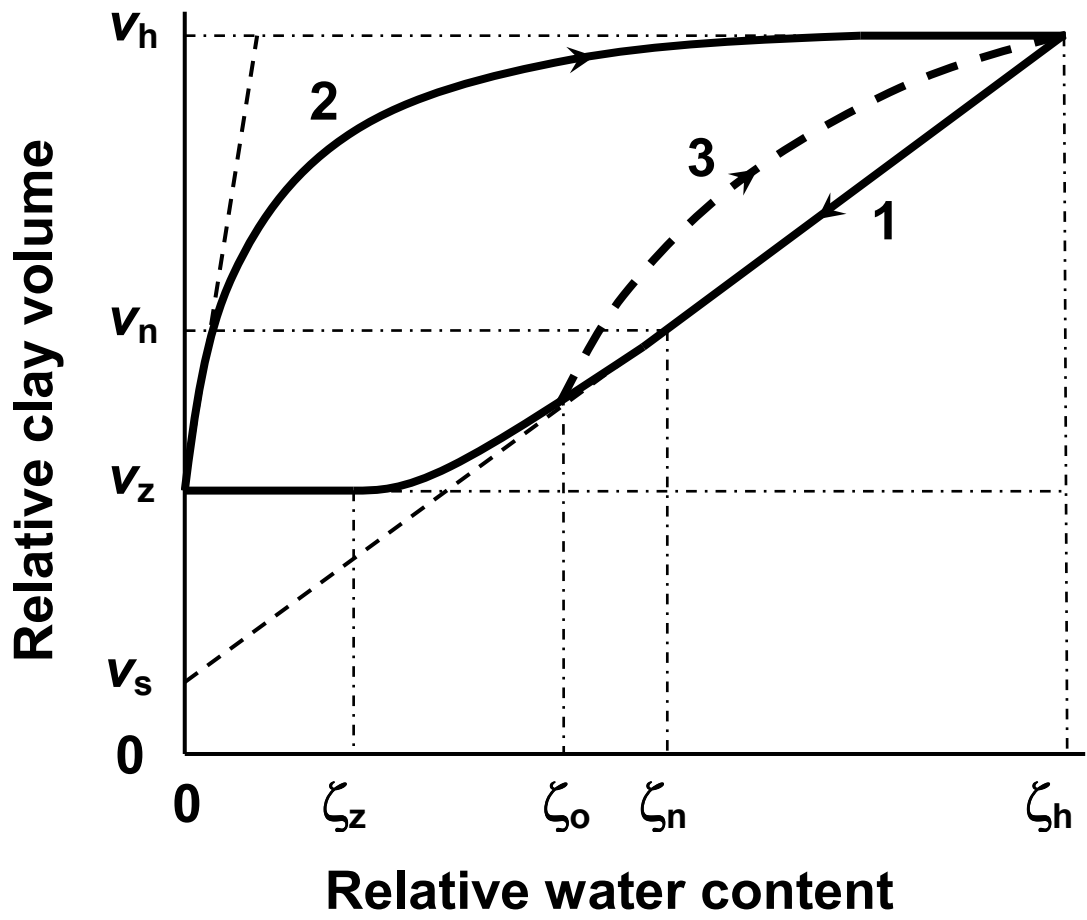

Fig.2

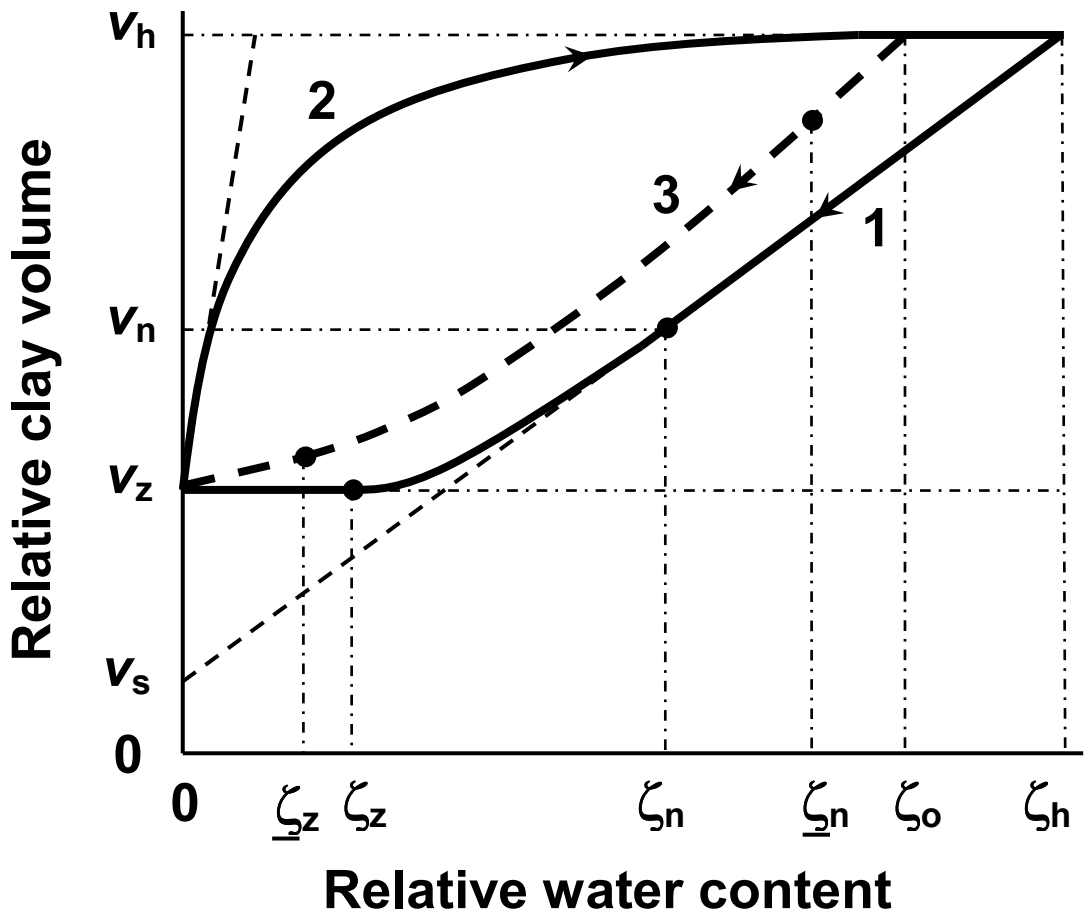

Fig.3

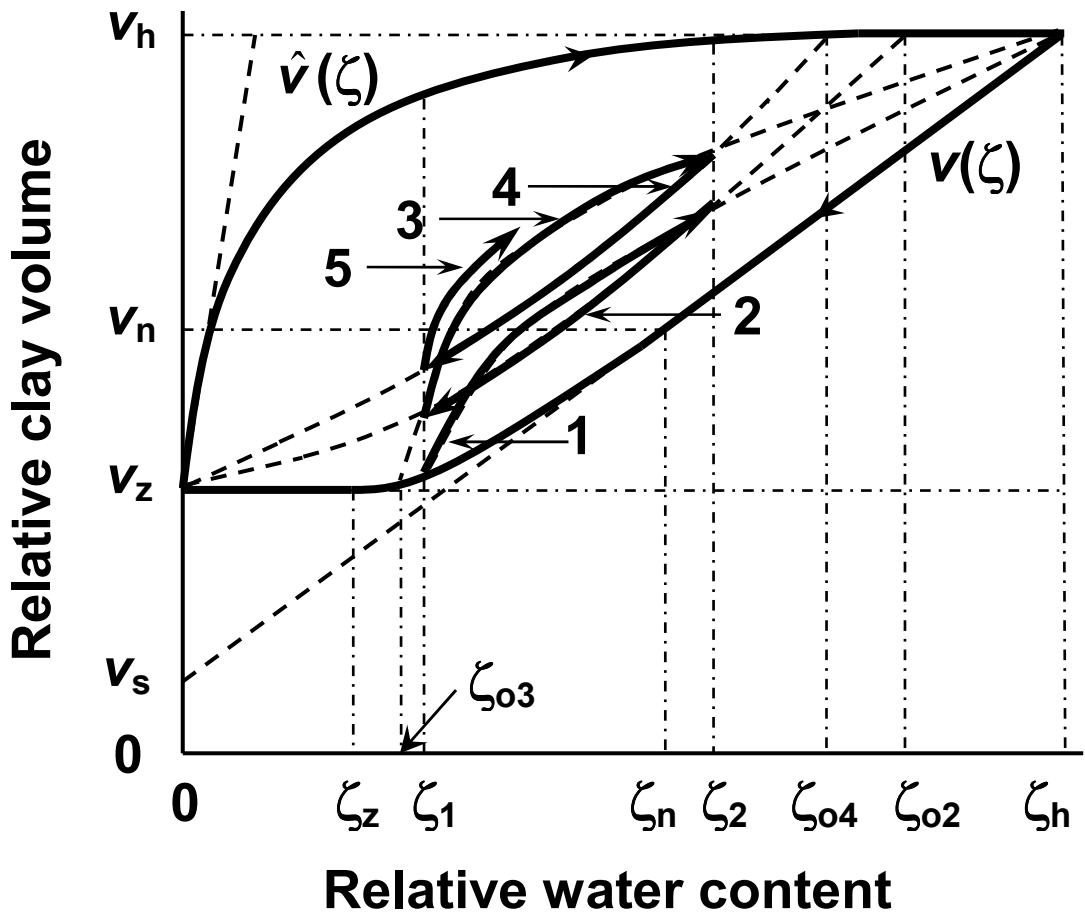

Fig.4

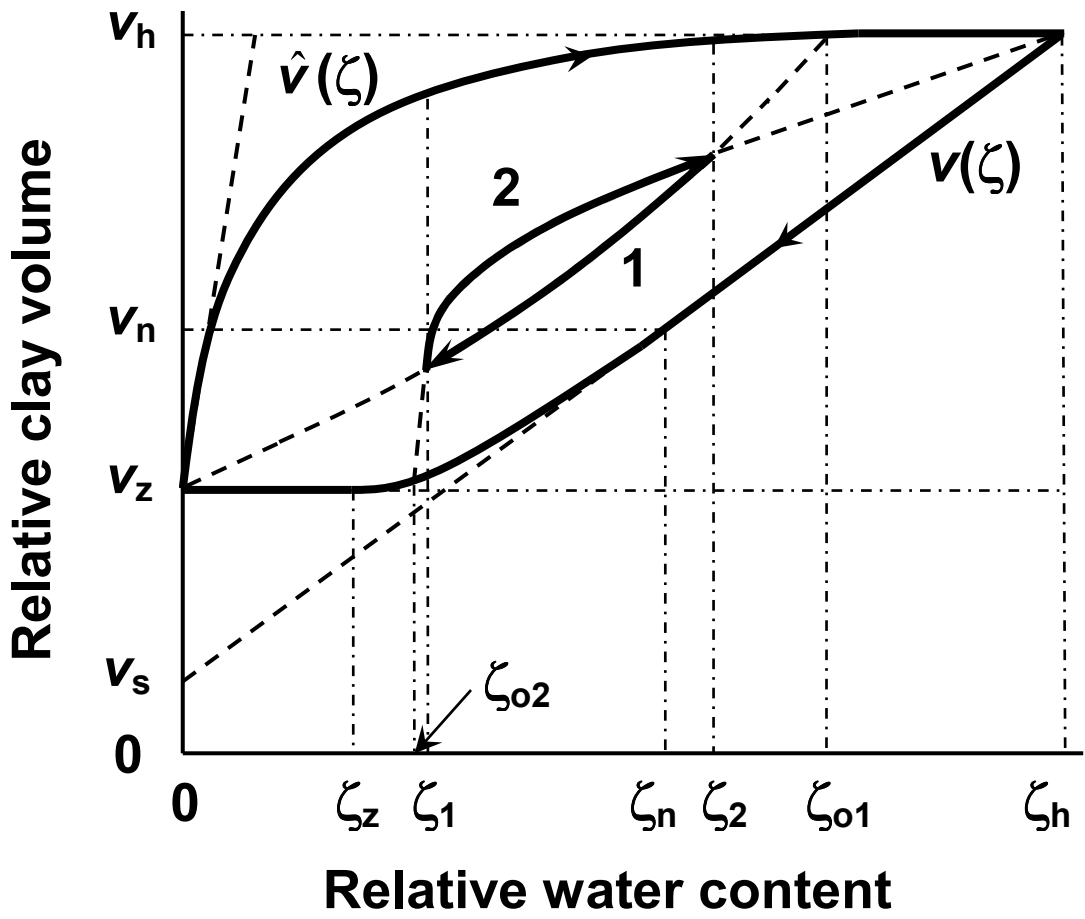

Fig.5

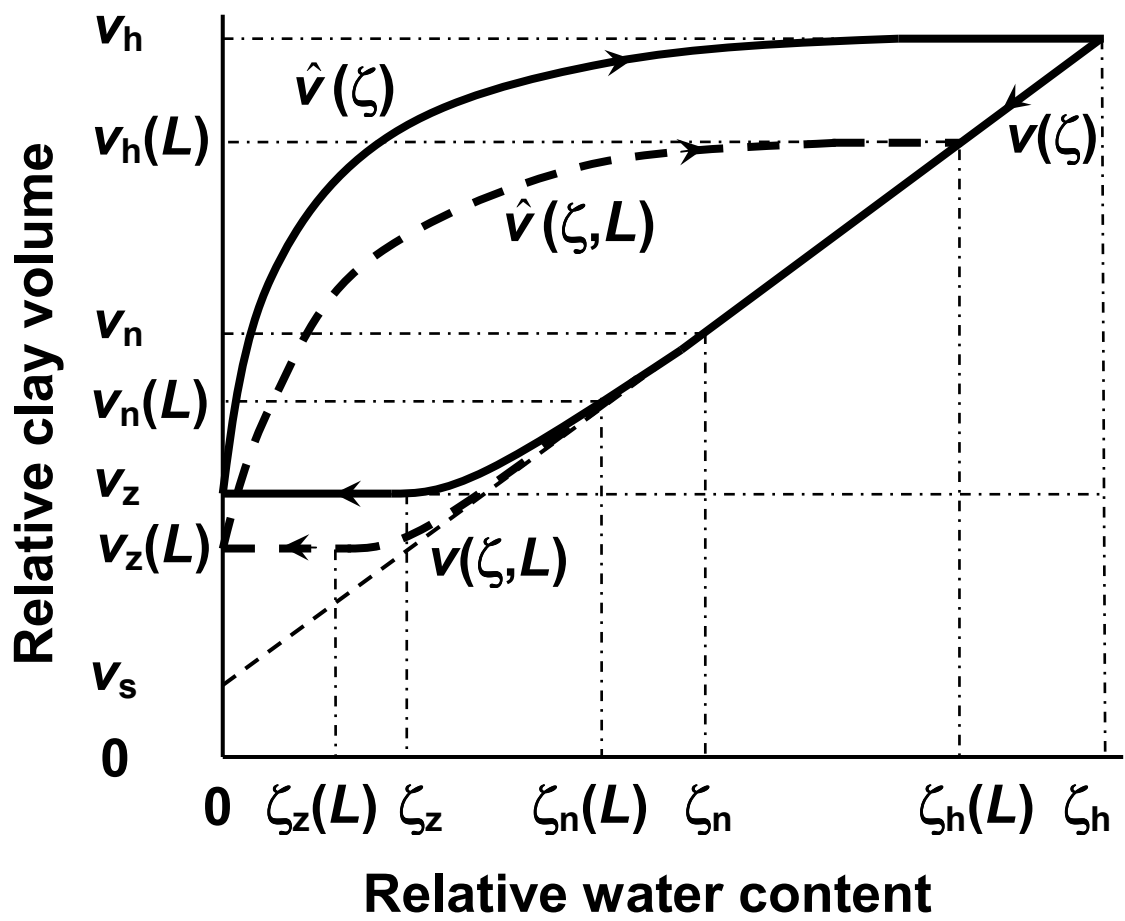

Fig.6

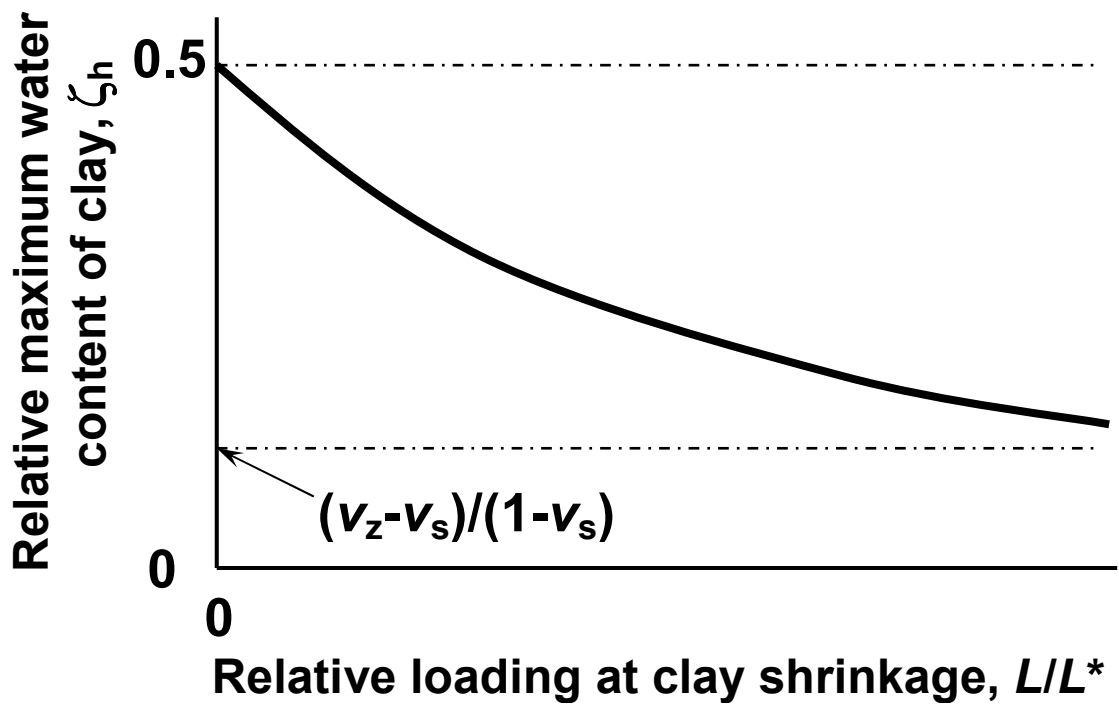

Fig.7

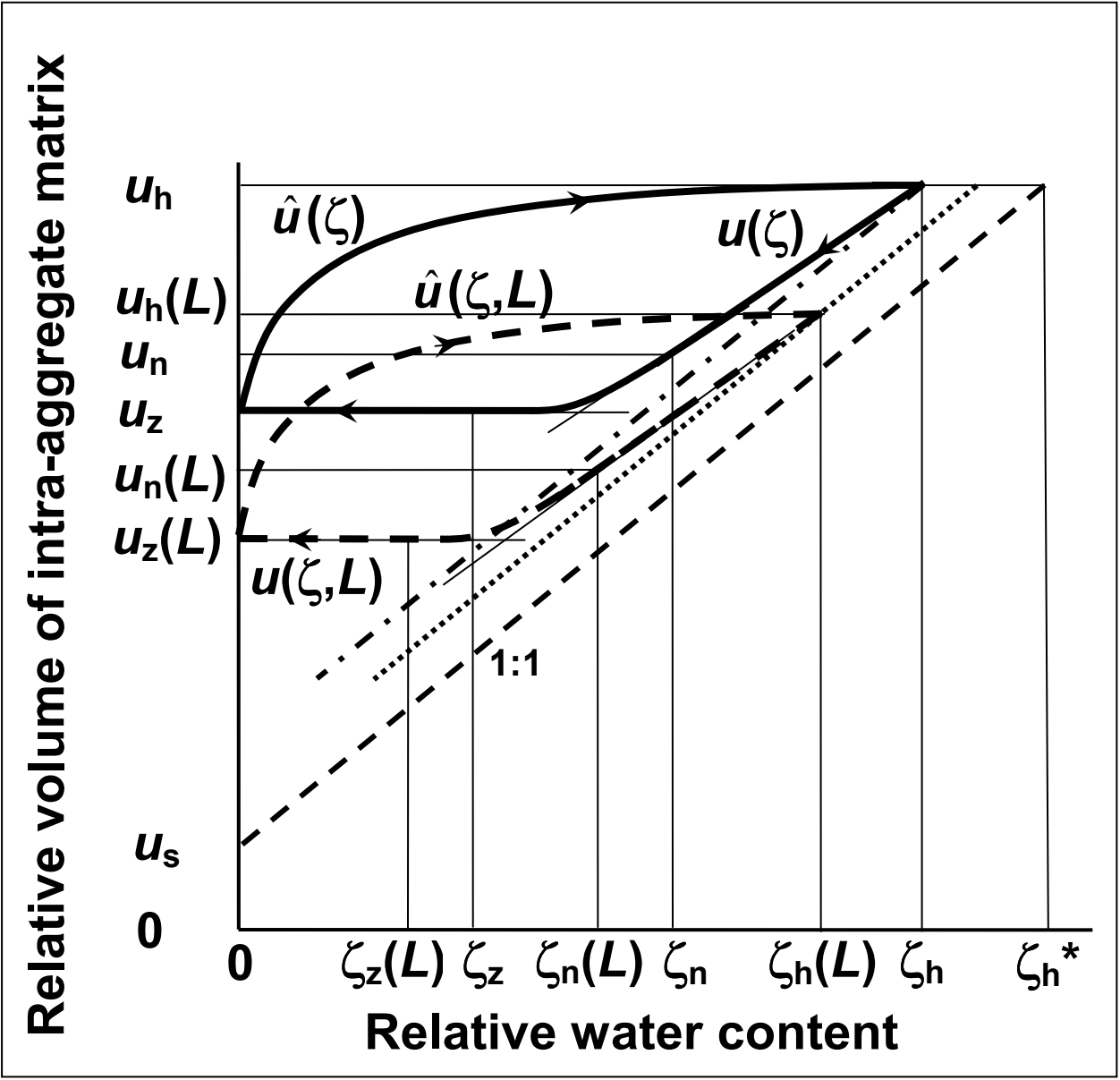

Fig.8

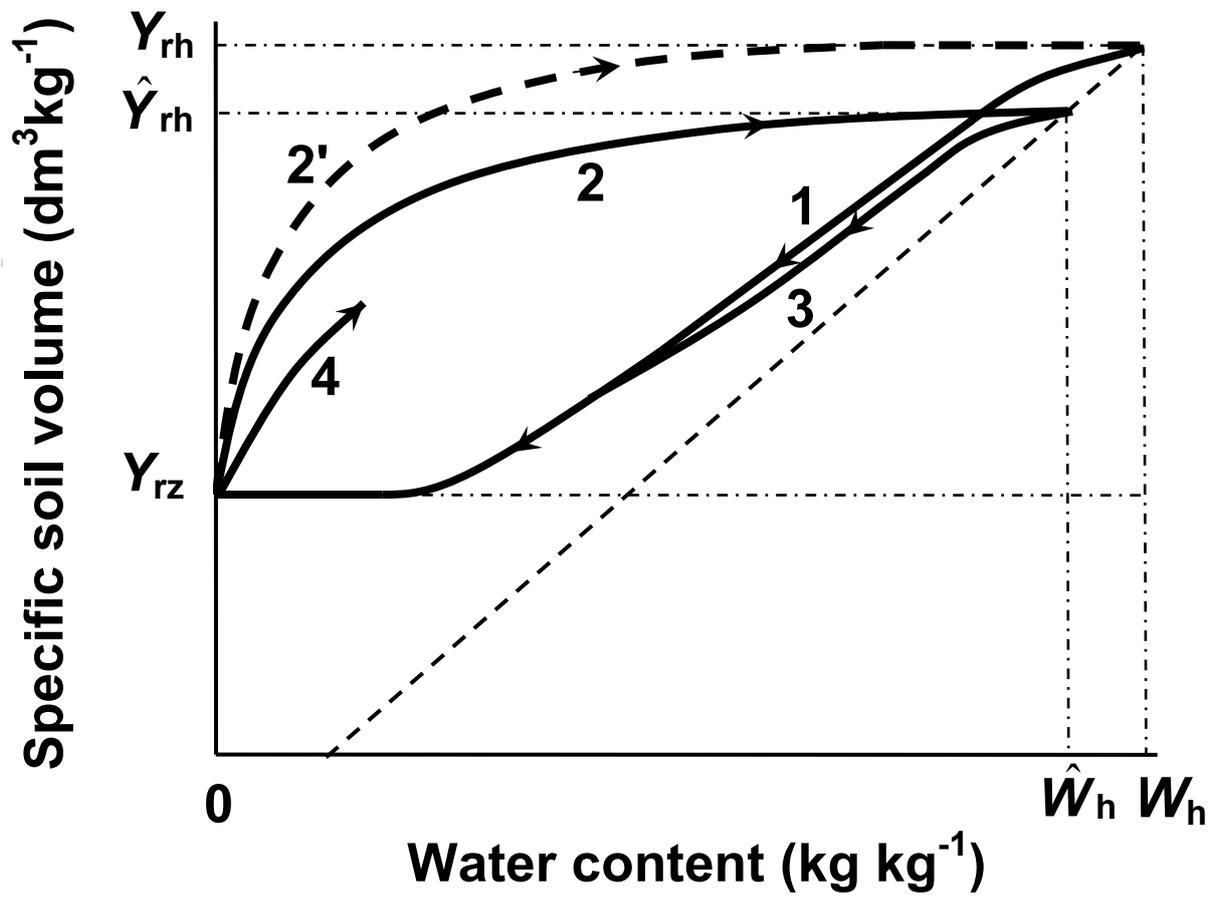

Fig.9

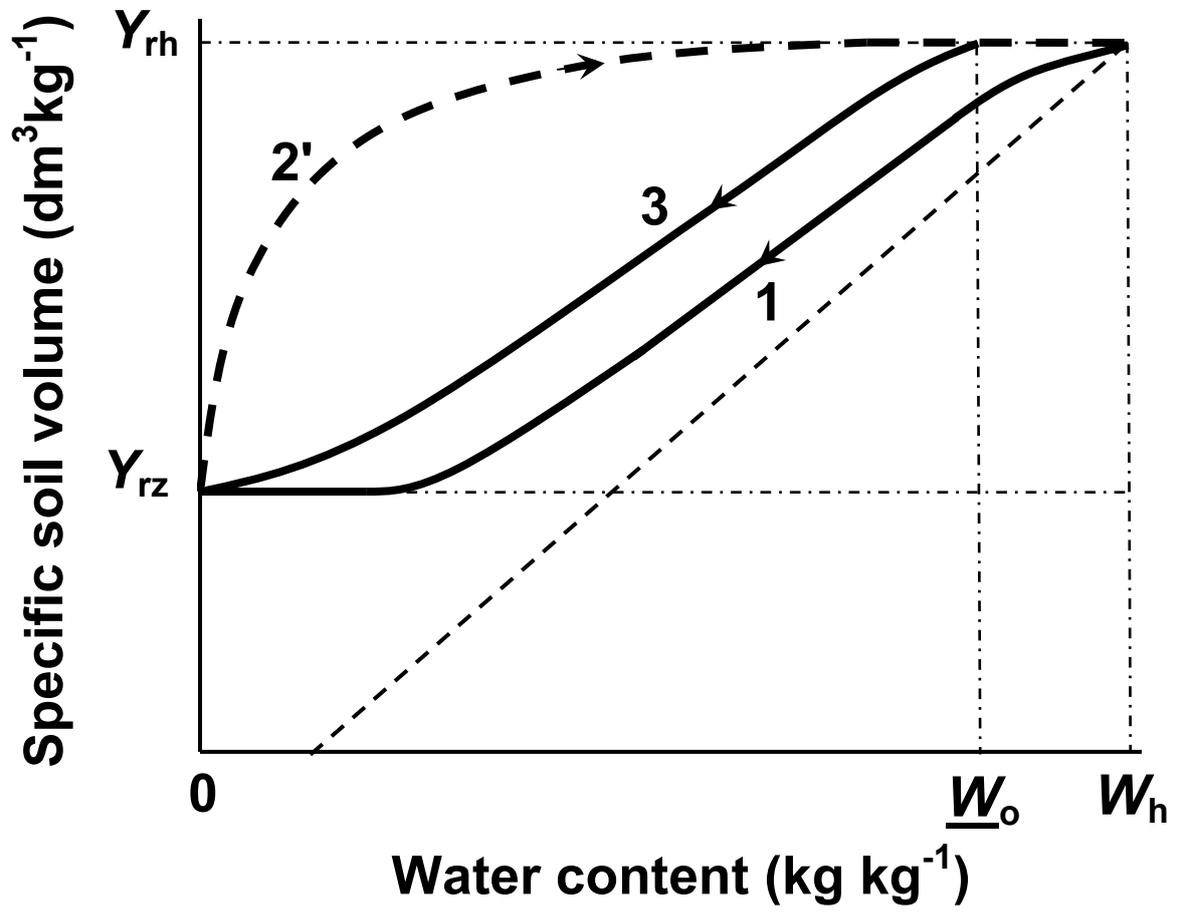

Fig.10

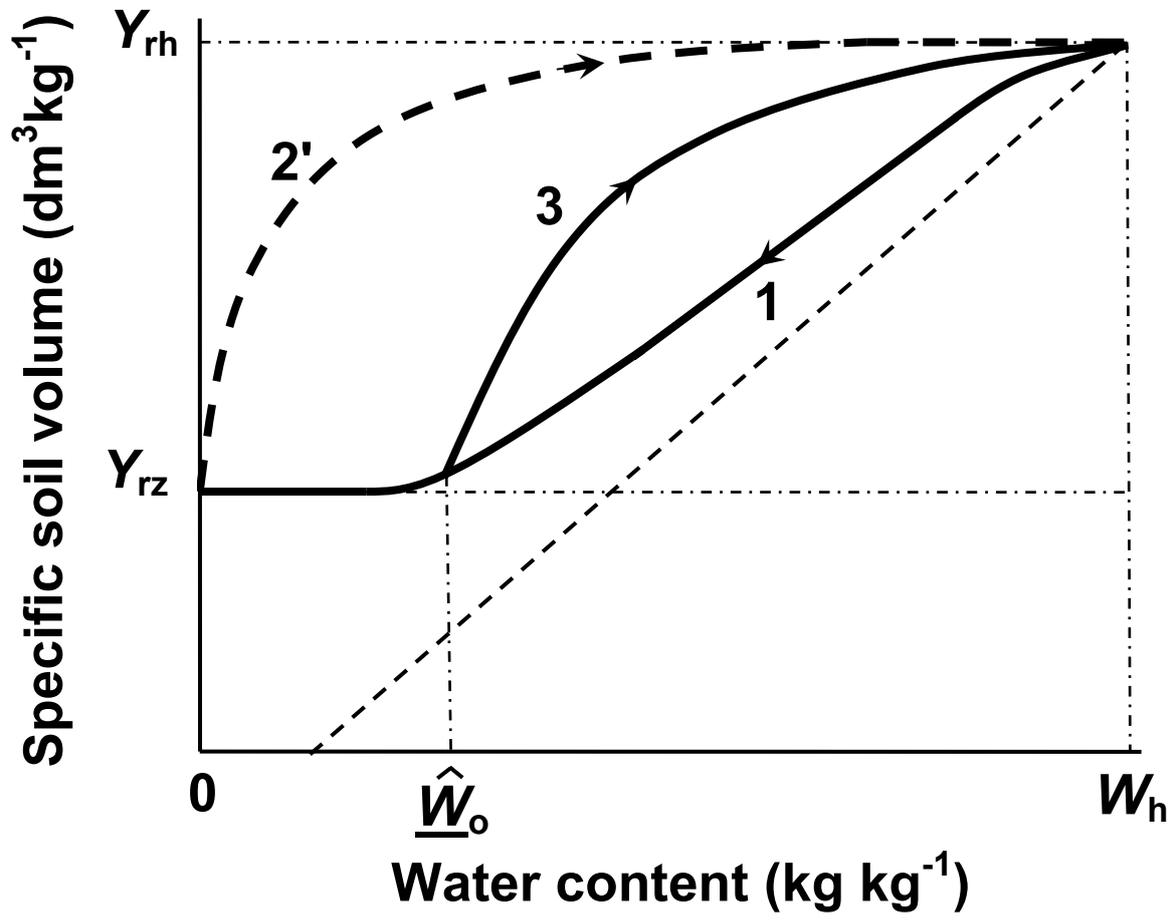

Fig.11

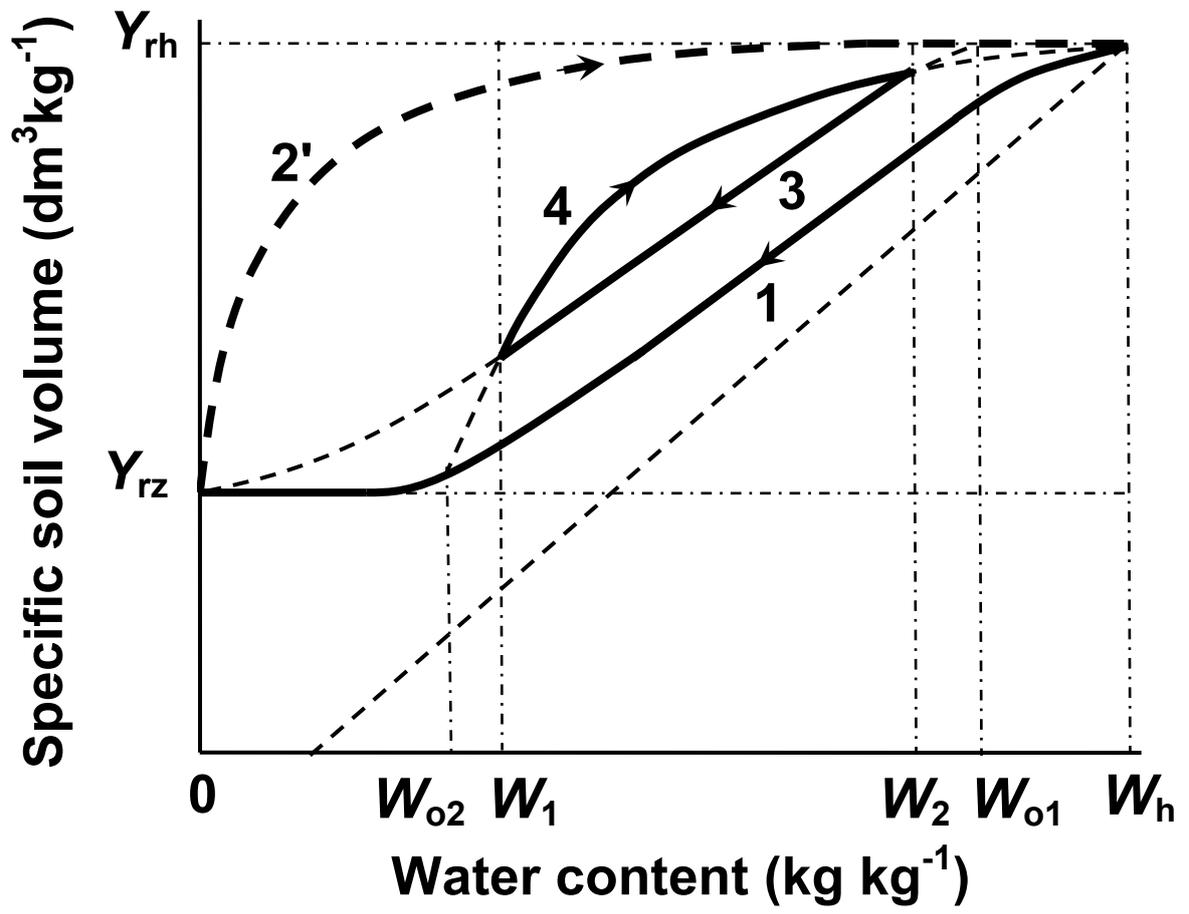

Fig.12

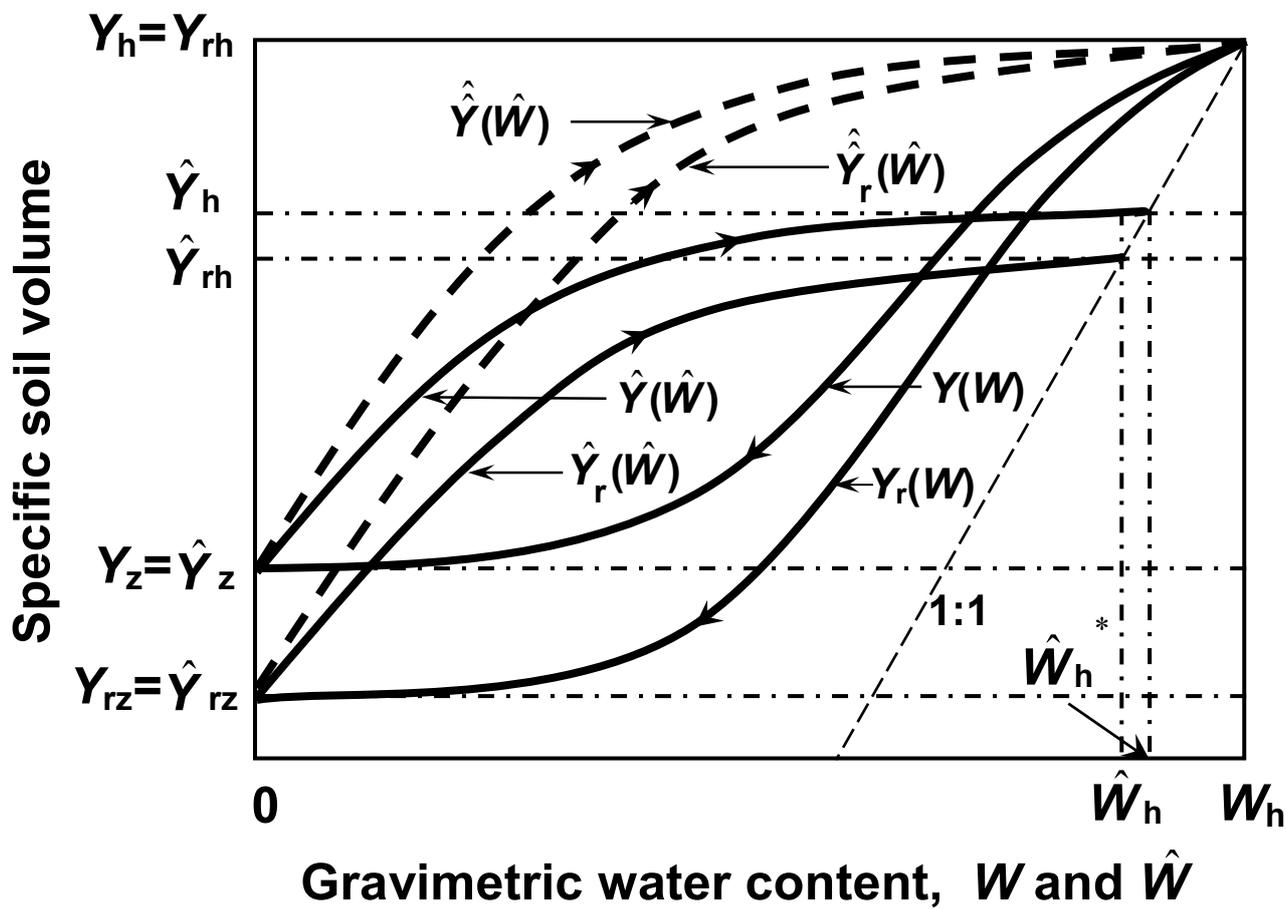

Fig.13

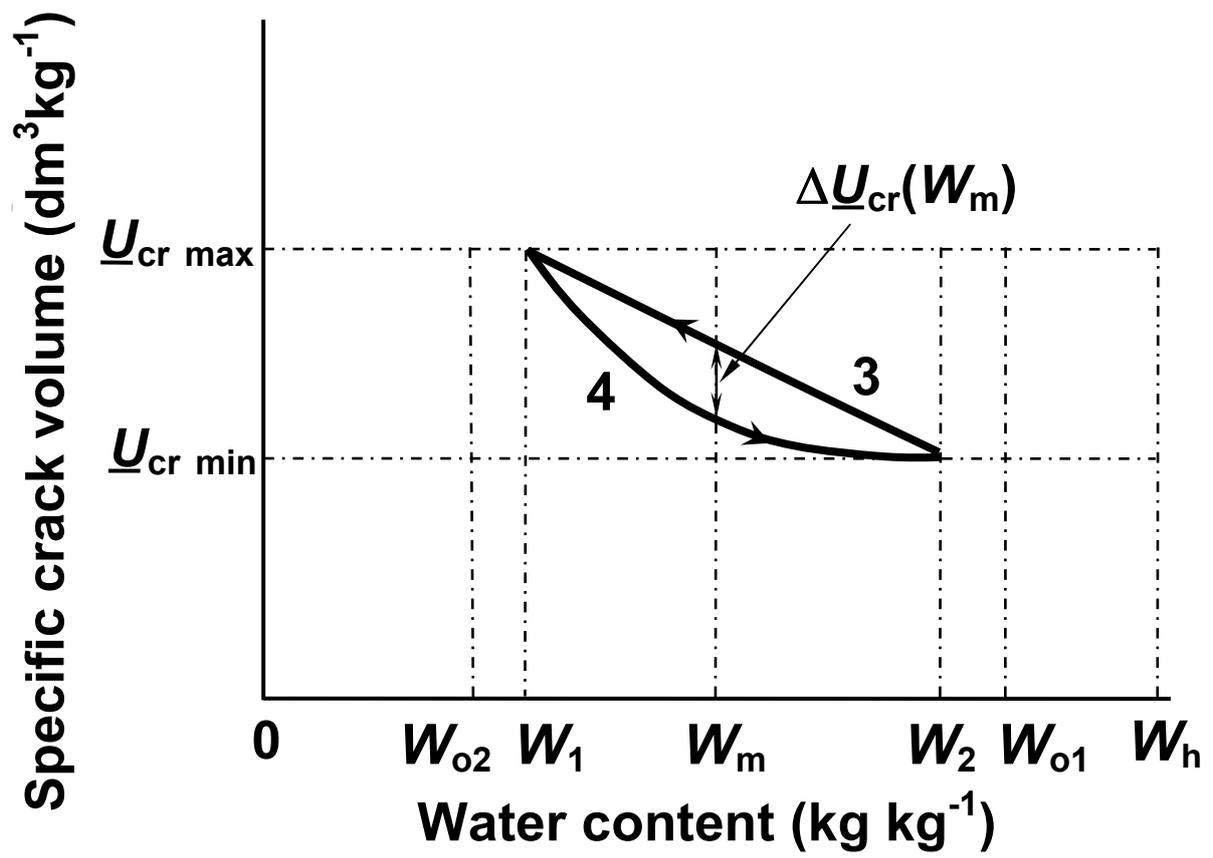

Fig.14

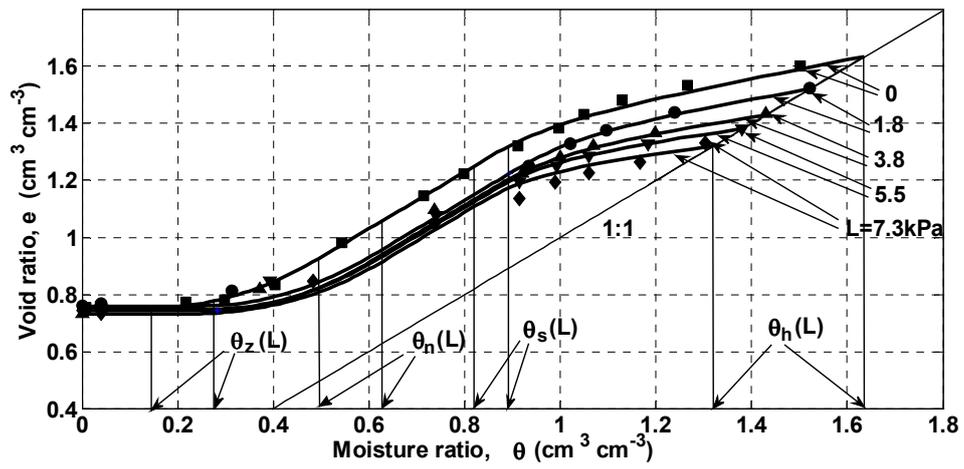

Fig.15

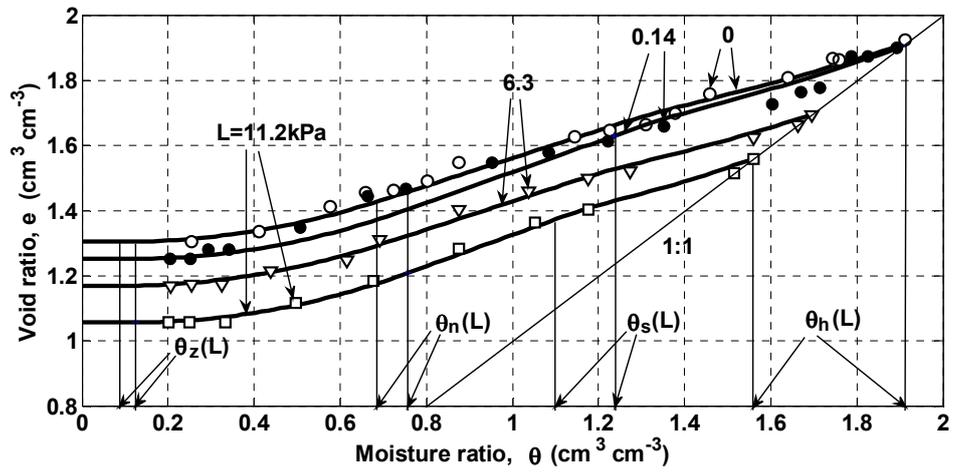

Fig.16

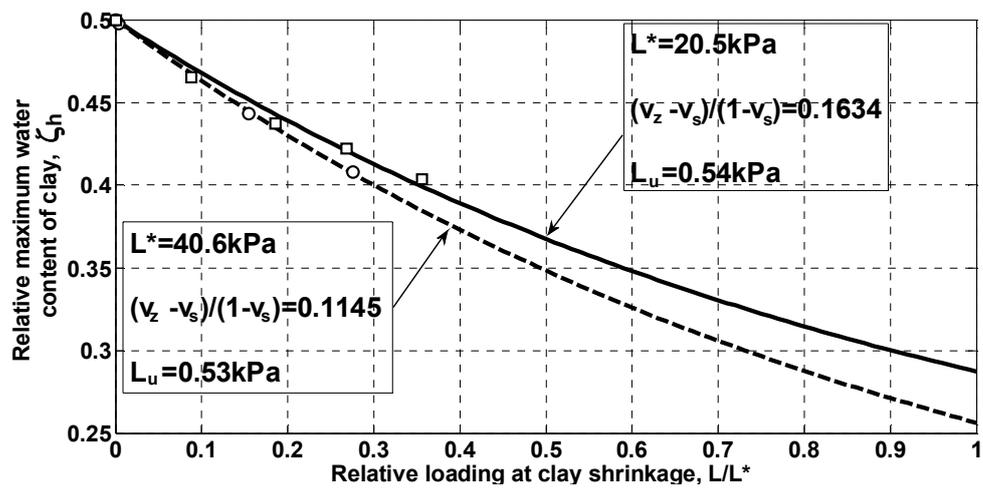

Fig.17

**Table 1**
Primary data[a] and input physical soil parameters for model prediction[b]

| Data source | Load | Primary data | | Input parameters | | | | | | | |
|---|---|---|---|---|---|---|---|---|---|---|---|
| | $L$ (kPa) | $\vartheta_h$ | $e_z$ | $\rho_s$ (kg dm$^{-3}$) | $c$ | $s$ | $U_s$ (dm$^3$kg$^{-1}$) | $U_{lph}$ (dm$^3$kg$^{-1}$) | $W_h$ (kg kg$^{-1}$) | $Y_z$ (dm$^3$kg$^{-1}$) | $h$ (mm) |
| Peng et al. [20], Fig.12 (Dystric Gleysol) | 0 | 1.6350 | 0.7541 | 2.65 | 0.649 | 0.151 | 0 | 0 | 0.6170 | 0.6619 | 41 |
| | 1.8 | 1.5211 | 0.7601 | 2.65 | 0.649 | 0.151 | 0 | 0 | 0.5740 | 0.6642 | 41 |
| | 3.8 | 1.4298 | 0.7324 | 2.65 | 0.649 | 0.151 | 0 | 0 | 0.5395 | 0.6537 | 41 |
| | 5.5 | 1.3802 | 0.7489 | 2.65 | 0.649 | 0.151 | 0 | 0 | 0.5208 | 0.6600 | 41 |
| | 7.3 | 1.3200 | 0.7429 | 2.65 | 0.649 | 0.151 | 0 | 0 | 0.4981 | 0.6577 | 41 |
| Talsma [47], Fig.1 (Black Earth from NSW, Australia) | 0[c] | 1.9123 | 1.3042 | 2.69 | 0.560 | 0.150 | 0 | 0 | 0.7109 | 0.8566 | 20 |
| | 0.14 | 1.9027 | 1.2523 | 2.69 | 0.560 | 0.150 | 0 | 0 | 0.7073 | 0.8373 | 20 |
| | 6.3 | 1.6955 | 1.1700 | 2.69 | 0.560 | 0.150 | 0 | 0 | 0.6303 | 0.8067 | 20 |
| | 11.2 | 1.5594 | 1.0573 | 2.69 | 0.560 | 0.150 | 0 | 0 | 0.5797 | 0.7648 | 20 |

[a] Soil moisture ratio at maximum swelling state before shrinkage ($\vartheta_h$), void ratio at shrinkage limit ($e_z$).
[b] Mean solid density ($\rho_s$), clay content ($c$), silt content ($s$), specific volume of structural pores at maximum swelling ($U_s$), specific lacunar pore volume at maximum swelling before shrinkage ($U_{lph}$), soil water content at maximum swelling before shrinkage ($W_h$), specific soil volume at the shrinkage limit ($Y_z$), sample height at maximum swelling before shrinkage ($h$).
[c] Actual minimum load from [47], 0.02kPa was taken to be zero.

Table 2
Predicted major physical characteristics of soil as a whole at different loading values and handling parameters

| Soil | Load $L$ (kPa) | Soil characteristics[a] | | | | | | | | Handling parameters[b] | |
|---|---|---|---|---|---|---|---|---|---|---|---|
| | | $x_m$ (mm) | $x_n(=X_{min})$ (mm) | $X_{mz}$ (mm) | $X_m$ (mm) | $K$ | $U_i$ (dm$^3$kg$^{-1}$) | $h^*$ (mm) | $q$ | $r_e^2$ | $\sigma_e$ |
| Dystric | 0 | 0.120 | 0.0216 | 0.7325 | 0.8913 | 1.3570 | 0.2616 | 53.4193 | 0 | 0.9982 | 0.0135 |
| Gleysol | 1.8 | 0.120 | 0.0216 | 0.9615 | 1.1273 | 1.2770 | 0.2064 | 52.9015 | 0 | 0.9965 | 0.0195 |
| from [20] | 3.8 | 0.120 | 0.0216 | 0.9615 | 1.1166 | 1.2799 | 0.2005 | 52.9198 | 0 | 0.9904 | 0.0283 |
| | 5.5 | 0.120 | 0.0216 | 0.9615 | 1.1016 | 1.2840 | 0.1987 | 52.9462 | 0 | 0.9865 | 0.0307 |
| | 7.3 | 0.120 | 0.0216 | 0.9615 | 1.0906 | 1.2870 | 0.1953 | 52.9659 | 0 | 0.9827 | 0.0318 |
| Black Earth | 0 | 0.118 | 0.0288 | 7.715 | 8.3744 | 1.0472 | 0.0488 | 44.5376 | 0 | 0.9884 | 0.0213 |
| from [47] | 0.14 | 0.118 | 0.0288 | 7.715 | 8.4318 | 1.0468 | 0.0483 | 44.5358 | 0 | 0.9767 | 0.0376 |
| | 6.3 | 0.118 | 0.0288 | 7.715 | 8.3236 | 1.0475 | 0.0455 | 44.5392 | 0 | 0.9940 | 0.0152 |
| | 11.2 | 0.118 | 0.0288 | 7.715 | 8.3280 | 1.0474 | 0.0431 | 44.5391 | 0 | 0.9956 | 0.0130 |

[a] Maximum sand grain size ($x_m$), mean size of soil solids ($x_n=0.001c+0.026s+(0.025+x_m/2)(1-c-s)$), minimum aggregate size ($X_{min}$), maximum aggregate size at shrinkage limit ($X_{mz}$), maximum aggregate size at maximum swelling before shrinkage ($X_m$), aggregate/intra-aggregate mass ratio at shrinkage ($K$), contribution of the interface aggregate layer to the specific volume of soil aggregates at shrinkage ($U_i$), critical sample size at shrinkage ($h^*$), sample crack factor at shrinkage ($q$).
[b] Goodness of fit of $e(\vartheta)$ ($X_{mz}$ is the only fitting parameter) to shrinkage curve data ($r_e^2$), estimated standard errors of shrinkage curve data [20,47] in Figs.15 and 16 ($\sigma_e$).

**Table 3**
Predicted major physical characteristics of intra-aggregate matrix at different loading values

| Soil | Load | Characteristics of intra-aggregate matrix[a] | | | | | | | | |
|---|---|---|---|---|---|---|---|---|---|---|
| | $L$ (kPa) | $p$ | $c^*$ | $k$ | $u_s$ | $u_S$ | $U_{lpz}$ | $U_{lpn}$ | $U_z$ (dm³kg⁻¹) | $U_n$ | $U_h(=Y_h)$ |
| Dystric | 0 | 0.2875 | 0.1940 | 0 | 0.2342 | 0.0822 | 0 | 0 | 0.5431 | 0.6365 | 0.9943 |
| Gleysol | 1.8 | 0.2875 | 0.1794 | 0 | 0.2342 | 0.0822 | 0 | 0 | 0.5844 | 0.6758 | 0.9514 |
| from [20] | 3.8 | 0.2875 | 0.1809 | 0 | 0.2342 | 0.0822 | 0 | 0 | 0.5799 | 0.6687 | 0.9169 |
| | 5.5 | 0.2875 | 0.1769 | 0 | 0.2342 | 0.0822 | 0 | 0 | 0.5921 | 0.6794 | 0.8982 |
| | 7.3 | 0.2875 | 0.1760 | 0 | 0.2342 | 0.0822 | 0 | 0 | 0.5660 | 0.6812 | 0.8755 |
| Black Earth | 0 | 0.7235 | 0.6007 | 0.5746 | 0.2073 | 0.0912 | 0.3201 | 0.2524 | 0.8458 | 0.8982 | 1.0826 |
| from [47] | 0.14 | 0.7235 | 0.5908 | 0.5295 | 0.2073 | 0.0912 | 0.2851 | 0.2202 | 0.8258 | 0.8861 | 1.0791 |
| | 6.3 | 0.7235 | 0.5991 | 0.5682 | 0.2073 | 0.0912 | 0.2694 | 0.2020 | 0.7973 | 0.8496 | 1.0020 |
| | 11.2 | 0.7235 | 0.5892 | 0.5208 | 0.2073 | 0.0912 | 0.2125 | 0.1484 | 0.7558 | 0.8153 | 0.9514 |

[a] Porosity of contributive silt and sand grains in the state of imagined contact ($p$), critical value of soil clay content ($c^*$), lacunar factor ($k$), relative volume of solid phase of intra-aggregate matrix ($u_s$), relative volume of non-clay solids of intra-aggregate matrix ($u_S$), specific volume of lacunar pores at $\zeta=\zeta_z$ and $\zeta=\zeta_n$ ($U_{lpz}$ and $U_{lpn}$, respectively), specific volume of intra-aggregate matrix at $\zeta=\zeta_z$, $\zeta=\zeta_n$, and $\zeta=\zeta_h$ ($U_z$, $U_n$, and $U_h$, respectively).

**Table 4**
Predicted major physical characteristics of contributive clay (see Fig.6) at different loading values

| Soil | Load $L$ (kPa) | Clay characteristics[a] | | | | | |
|---|---|---|---|---|---|---|---|
| | | $v_s$ | $v_z$ | $v_h$ | $\zeta_z$ | $\zeta_h$ | $F_z$ |
| Dystric Gleysol from [20] | 0 | 0.1656 | 0.2777 | 0.5828 | 0.0612 | 0.5 | 0.4558 |
| | 1.8 | 0.1656 | 0.3056 | 0.5538 | 0.0982 | 0.4652 | 0.5851 |
| | 3.8 | 0.1656 | 0.3025 | 0.5305 | 0.0938 | 0.4373 | 0.5716 |
| | 5.5 | 0.1656 | 0.3108 | 0.5178 | 0.1057 | 0.4221 | 0.6076 |
| | 7.3 | 0.1656 | 0.3128 | 0.5024 | 0.1086 | 0.4037 | 0.6158 |
| Black Earth from [47] | 0 | 0.1277 | 0.2222 | 0.5639 | 0.0254 | 0.5 | 0.2347 |
| | 0.14 | 0.1277 | 0.2314 | 0.5617 | 0.0327 | 0.4976 | 0.2753 |
| | 6.3 | 0.1277 | 0.2236 | 0.5145 | 0.0265 | 0.4434 | 0.2407 |
| | 11.2 | 0.1277 | 0.2330 | 0.4834 | 0.0341 | 0.4078 | 0.2824 |

[a] Relative clay solid volume ($v_s$), relative clay volume at $\zeta=\zeta_z$ ($v_z$), relative clay volume at $\zeta=\zeta_h$ ($v_h$), shrinkage limit on the $\zeta$ axis of relative clay water content ($\zeta_z$), maximum clay swelling point on the $\zeta$ axis ($\zeta_h$), saturation degree of clay at shrinkage limit ($F_z$).

**Table 5**
Classification and corresponding designations of volumetric and water content values (in units of specific volume and gravimetric water content)

| N | Classification according to transitions | Designation[a,b,c] | Specification | Examples |
|---|---|---|---|---|
| 1 | Contributive clay→part of soil structure→soil as a whole | $V(v) \to U(U_s, U_i, U(u), U_{lp}, U_{cr}) \to Y$ $\overline{w}(\zeta) \to w(\zeta), w', \omega \to W$ | Transitions relate to any value | — |
| 2 | Shrinkage↔swelling[d] | $...\leftrightarrow\hat{}$ | Transition relates to any value except for $U_s$, $\overline{w}$, and $w$ | $V \leftrightarrow \hat{V}$, $U_{cr} \leftrightarrow \hat{U}_{cr}$, $Y \leftrightarrow \hat{Y}$, $W \leftrightarrow \hat{W}$, $\omega \leftrightarrow \hat{\omega}$ |
| 3 | Primary curve↔scanning curve | $...\leftrightarrow\underline{\ }$ | Transition relates to any value except for $U_s$, $U_i$, and $\hat{U}_i$ | $V \leftrightarrow \underline{V}$, $\hat{U} \leftrightarrow \underline{\hat{U}}$, $Y_r \leftrightarrow \underline{Y}_r$, $Y \leftrightarrow \underline{Y}$, $\hat{\omega} \leftrightarrow \underline{\hat{\omega}}$, $\omega \leftrightarrow \underline{\omega}$, $W \leftrightarrow \underline{W}$ |
| 4 | Reference curve→curve with crack contribution | $_r \to ...$ | Transition relates only to $Y$ values | $Y_r \to Y$, $\underline{Y}_r \to \underline{Y}$, $\hat{Y}_r \to \hat{Y}$ |
| 5 | No loading→under loading | $... \to (L)$ | Transition relates to any value | $U_{cr} \to U_{cr}(L)$, $\underline{Y} \to \underline{Y}(L)$, $\omega \to \underline{\omega}(L)$, $W_h \to W_h(L)$ |

[a] See Notation
[b] $(v)$, $(u)$, and $(\zeta)$ indicate the designations of the corresponding values as relative.
[c] "..." means that the corresponding sign, "^" or "_", subscript, "$_r$", or argument, "$(L)$" is lacking.
[d] In the case of the *quasi primary swelling* curve $\hat{Y}_r \to \hat{Y}_r$, $\hat{Y} \to \hat{Y}$, and $\hat{U}_{cr} \to \hat{U}_{cr}$.